\documentclass[12pt]{article}
\usepackage{amsmath,amssymb}
\usepackage{graphicx}
\usepackage{array}
\makeatletter
\newif\if@preliminary
\@preliminaryfalse
\def\preliminary{\@preliminaryfalse}

\def\bq{\begin{equation}}
\def\eq{\end{equation}}
\def\ba{\begin{eqnarray}}
\def\ea{\end{eqnarray}}

\def\preprintno#1{\def\@preprintno{#1}}
\def\address#1{\def\@address{#1}}

\def\abstract#1{\def\@abstract{#1}}
\renewcommand\abstractname{ABSTRACT}
\newlength\preprintnoskip
\newlength\abstractwidth
\@titlepagetrue
\renewcommand\maketitle{\begin{titlepage}
  \let\footnotesize\small
  \hfill\parbox{\preprintnoskip}{
  \begin{flushright}\@preprintno\end{flushright}}\hspace*{1cm}
  \vskip 60\p@
  \begin{center}
    {\Large\bf\boldmath \@title \par}\vskip 1cm
    {\sc\@author \par}\vskip 3mm
    {\@address \par}
    \if@preliminary
      \vskip 2cm {\large\sf PRELIMINARY DRAFT \par \@date}
    \fi
  \end{center}\par
  \@thanks
  \vfill
  \begin{center}
    \parbox{\abstractwidth}{\centerline{\abstractname}
    \vskip 3mm
    \@abstract}
  \end{center}
  \end{titlepage}
  \setcounter{footnote}{0}
  \let\thanks\relax\let\maketitle\relax
  \gdef\@thanks{}\gdef\@author{}\gdef\@address{}
  \gdef\@title{}\gdef\@abstract{}\gdef\@preprintno{}
}

\def\@citex[#1]#2{\if@filesw\immediate\write\@auxout{\string\citation{#2}}\fi
  \def\@citea{}\@cite{\@for\@citeb:=#2\do
    {\@citea\def\@citea{,\penalty\@m}\@ifundefined
       {b@\@citeb}{{\bf ?}\@warning
       {Citation `\@citeb' on page \thepage \space undefined}}%
\hbox{\csname b@\@citeb\endcsname}}}{#1}}
\def\citerange{\@ifnextchar [{\@tempswatrue\@citexr}{\@tempswafalse\@citexr[]}}
\def\@citexr[#1]#2{\if@filesw\immediate\write\@auxout{\string\citation{#2}}\fi
  \def\@citea{}\@cite{\@for\@citeb:=#2\do
    {\@citea\def\@citea{--\penalty\@m}\@ifundefined
       {b@\@citeb}{{\bf ?}\@warning
       {Citation `\@citeb' on page \thepage \space undefined}}%
\hbox{\csname b@\@citeb\endcsname}}}{#1}}
\long\def\@makecaption#1#2{
  \vskip\abovecaptionskip
  \sbox\@tempboxa{#1: \emph{#2}}
  \ifdim \wd\@tempboxa >\hsize
    #1: \emph{#2}\par
  \else
    \hbox to\hsize{\hfil\box\@tempboxa\hfil}
  \fi
  \vskip\belowcaptionskip}

\def\fmslash{\@ifnextchar[{\fmsl@sh}{\fmsl@sh[0mu]}}
\def\fmsl@sh[#1]#2{
  \mathchoice
    {\@fmsl@sh\displaystyle{#1}{#2}}
    {\@fmsl@sh\textstyle{#1}{#2}}
    {\@fmsl@sh\scriptstyle{#1}{#2}}
    {\@fmsl@sh\scriptscriptstyle{#1}{#2}}}
\def\@fmsl@sh#1#2#3{\m@th\ooalign{$\hfil#1\mkern#2/\hfil$\crcr$#1#3$}}
\makeatother

\newcommand{\GeV}{{\ensuremath\rm GeV}}
\newcommand{\TeV}{{\ensuremath\rm TeV}}

\newcommand{\fb}{{\ensuremath\rm fb}}

\newcommand{\etal}{\textit{et al.}}
\newcommand{\pd}{\partial}
\newcommand{\LL}{\mathcal{L}}

\newcommand{\tr}[1]{\operatorname{tr}\left[#1\right]}
\newcommand{\hc}{\text{h.c.}}

\newcommand{\ii}{{\rm i}}
\newcommand{\I}{{\rm i}}

\renewcommand{\Re}{\text{Re}}
\renewcommand{\Im}{\text{Im}}

\renewcommand{\Re}{\operatorname{Re}}
\renewcommand{\Im}{\operatorname{Im}}

\newcommand{\MSbar}{\mbox{$\overline{\rm MS}$}}

\newcommand{\unit}{\boldsymbol{1}}

\newcommand{\vB}{\mathbf{B}}

\newcommand{\vG}{\mathbf{G}}
\newcommand{\vV}{\mathbf{V}}
\newcommand{\vT}{\mathbf{T}}
\newcommand{\vW}{\mathbf{W}}

\newcommand{\vj}{\mathbf{j}}
\newcommand{\vt}{\mathbf{a}}

\newcommand{\vw}{\mathbf{w}}

\newcommand{\vphi}{\boldsymbol{\phi}}
\newcommand{\vrho}{\boldsymbol{\rho}}

\newcommand{\cA}{\mathcal{A}}
\newcommand{\cD}{\mathcal{D}}
\newcommand{\cP}{\mathcal{P}}
\newcommand{\cS}{\mathcal{S}}
\newcommand{\cV}{\mathcal{V}}
\newcommand{\cW}{\mathcal{W}}

\newcommand{\pp}{{\prime 2}}

\newcommand{\sw}{s_{\mathrm{w}}}
\newcommand{\cw}{c_{\mathrm{w}}}
\newcommand{\eff}{{\rm eff}}
\newcommand{\lrpd}{\overset{\leftrightarrow}{\partial}}
\newcommand{\logms}{\log{\frac{M^2}{s+M^2}}}
\newcommand{\ptmax}{p_{\perp,\textrm{max}}}

\newcommand{\whizard}{\texttt{WHIZARD}}
\newcommand{\pythia}{\texttt{PYTHIA}}

\begin{document}

%\shortletter        % subdivided in paragraphs instead of sections
\preliminary        % mark on title page
%\baselineskip20pt   % stretch linespacing in main text

%%%%%%%%%%%%%%%%%%%%%%%%%%%%%%%%%%%%%%%%%%%%%%%%%%%%%%%%%%%%%%%%%%%%%%%%

\preprintno{%
  SI-HEP-2008-11\\[0.5\baselineskip]
}

\title{RESONANCES AND UNITARITY IN WEAK BOSON SCATTERING AT THE LHC
}

\author{
 A.~Alboteanu$^a$,
 W.~Kilian$^a$, and
 J.~Reuter$^b$
}

\address{\it
$^a$ University of Siegen, Fachbereich Physik, D--57068 Siegen, Germany 
\\[.5\baselineskip]
$^b$University of Freiburg, Institute of Physics, D--79104 Freiburg, Germany
}

\abstract{
A crucial test of the Standard Model
is the measurement of electroweak gauge-boson scattering. In this
paper, we describe a generic parameterization aimed at a realistic
simulation of weak-boson scattering at the LHC.  The
parameterization implements resonances of all possible spin and
isospin combinations, properly matched to the low-energy effective
(chiral) Lagrangian, includes leading higher-order effects and
contains a minimal unitarization scheme.  We implement the
parameterization in the Monte-Carlo event generator \whizard\ and
present results for complete partonic cross-section integration and
event generation.  We provide a comparison with the effective $W$
approximation that previously has been used for most $WW$ scattering
studies at hadron colliders.
}

\maketitle

%%%%%%%%%%%%%%%%%%%%%%%%%%%%%%%%%%%%%%%%%%%%%%%%%%%%%%%%%%%%%%%%%%%%%%%%
%%% Text
%%%%%%%%%%%%%%%%%%%%%%%%%%%%%%%%%%%%%%%%%%%%%%%%%%%%%%%%%%%%%%%%%%%%%%%%
%%% \begin{fmffile}{paper-graphs}

\section{Introduction}

Exploring the mechanism of electroweak symmetry breaking (EWSB) is the
primary focus of the upcoming LHC experiments ATLAS and CMS.  The
simplest explanation, the minimal Standard Model (SM), suffers from
theoretical deficiencies and does not account for all experimental
facts.  Weakly-coupled extensions of the SM such as its minimal
supersymmetric version MSSM are a possible solution.  All
weakly-coupled models contain new particles in the range between about
$100\;\GeV$ and $1\;\TeV$ that are observable at the LHC.  Among them
are light scalar states, in particular one or more neutral Higgs
bosons.

No Higgs boson has been observed so far, and the LHC will finally
decide about its existence.  If no light Higgs boson exists, we have
to consider alternatives to the familiar SM.  Models without a (light)
Higgs boson are strongly coupled, hence much less predictive and more
difficult to handle theoretically.  They need not provide new physics
below the $\TeV$ region.  While simple strongly-coupled scenarios such
as minimal technicolor tend to be at variance with known precision
data, more advanced models remain valid, and we are not even close to
a comprehensive view of the possibilities.

The theory and phenomenology of strong weak-boson scattering (for
reviews, see, e.g.,
Refs.~\cite{Chivukula:1990bc,Dobado:1997jx,Hill:2002ap,Kilian:2003pc})
has been a subject of active research for more than two decades.
Early work on a strongly interacting electroweak
sector~\cite{Chanowitz:1985hj,BESS} was motivated by the technicolor
paradigm~\cite{dynEWSB}.  In particular, Bagger
\etal~\cite{Bagger:1993zf} considered a collection of benchmark
scenarios and their observability at hadron colliders; this study was
updated for the LHC parameters in~\cite{Bagger:1995mk,Gupta:1995ru}.
Later work focused on the sub-$\TeV$ behavior and its extrapolation to
higher
energies~\cite{Dobado,ATLAS-TDR,Eboli,Butterworth:2002tt,Green:2003if,Fabbrichesi:2007ad}.
Studies of $WW$ scattering at lepton colliders are also
available~\cite{Barger:1995cn,Boos:1997gw,Gangemi:1999gt,Beyer:2006hx}.
More recently, interest in $WW$ scattering at the LHC was revived in
the context of extra-dimension
models~\cite{5D-BC,Chivukula,Davoudiasl:2003me,Birkedal:2004au}.

Since the LHC will start taking data soon, new and detailed
experimental studies are under way which prepare for the upcoming
analyses at ATLAS and CMS.  These have to operate on a solid
theoretical basis.  However, the earlier phenomenological studies
mentioned above have restricted themselves to particular benchmark
models, e.g., the SM, technicolor-inspired resonances or specific
unitary extrapolations of the low-energy behavior.  Non-SM models have
been treated using simplifying approximations, in particular the
effective $W$ approximation (EWA)~\cite{EWA}.  For the future analysis
of real LHC data, it will be crucial to get rid of approximations and
treat the problem with full generality, as far as the physics is
accessible to data analysis.

The present paper aims at a practical realization of the
strongly-interacting scenario that is suited for realistic physics
simulation and experimental analysis.  To this end, we introduce a
generic parameterization of weak-boson scattering that includes all
resonances allowed by spin and isospin with free mass and width
parameters.  We embed this in the generic effective-Lagrangian
formalism for electroweak symmetry
breaking~\cite{Weinberg:1968de,appelquistlonghitano} and properly
match the high-energy region to the low-energy expansion.  We include
the model-independent part of loop corrections to the scattering
amplitude.  For regulating the high-energy behavior, we adopt a
straightforward (K-matrix) unitarization scheme.  This approach
\emph{cum grano salis} encompasses all of the specific models studied
earlier.

The parameterization is extended off-shell in a natural way, and thus
can be implemented in a parton-level matrix element generator.  The SM
emerges as a special case.  Models can thus be studied in the context
of cross-section calculation and event generation, and there is no
need for further approximations.  The partonic simulation provides
complete six-fermion signals and irreducible background.  We have
realized this as an extension to the public Monte-Carlo simulation
package \whizard~\cite{whizard,Omega}, and we present numerical
results.

Beyond partonic cross sections and events, the implementation makes it
possible to apply parton shower, hadronization, and fast or full
detector simulation.  This should enable LHC analyses of weak-boson
scattering to derive solid conclusions from comparing simulation
results with real data, once the latter are available.

\section{Strong Weak-Boson Scattering}

In this section, we consider a generic no-(light-)Higgs scenario.  In
the absence of a light scalar resonance, weak bosons become strongly
interacting in the $\TeV$ range~\cite{Uni}, and the perturbative
expansion in the weak couplings $g,g'$ breaks down.  To the extent
that the corresponding scattering processes are observable at the LHC,
a measurement of the amplitudes is a probe of new physics in
electroweak symmetry breaking.

\subsection{The LHC Case}

\begin{figure}[hbt]
  \begin{center}
    \includegraphics[width=.4\textwidth]{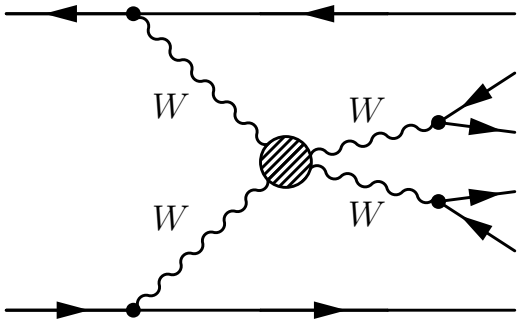}
  \end{center}
  \caption{
    Topology of vector boson scattering in proton-proton collisions.
    \label{fig:signal}}
\end{figure}

The LHC can access this kind of physics in processes of the type
$qq\to jj+4f$.  Among the Feynman diagrams there are some where the
initial quarks radiate approximately on-shell $W$ and $Z$ bosons and
become hard forward/backward (low-$p_T$) 'spectator' jets,
Fig.~\ref{fig:signal}.  The weak bosons scatter quasi-elastically and
decay into four additional fermions which appear more centrally.  This
is the strong-scattering signal that we are interested in.  It depends
on detection efficiency and background reduction, which $W/Z$ decay
modes (four leptons, semileptonic, all jets) are useful.

As an alternative to an expansion in the weak couplings $g,g'$, one can
expand in powers of $E/\Lambda$, where $E$ is the characteristic
energy scale of the subprocess, and the cutoff $\Lambda$ is loosely
defined as $4\pi v$ with the electroweak scale $v=(\sqrt2\,
G_F)^{-1/2}=246\;\GeV$.  In practice, this expansion is valid up to
about $1\;\TeV$, where scattering amplitudes approach the saturation
of unitarity limits.  The corresponding effective Lagrangian is known
as the {electroweak chiral Lagrangian}~\cite{appelquistlonghitano}.
(There is a close analogy with the chiral-Lagrangian approach to
low-energy QCD~\cite{Weinberg:1968de,CCWZ,ChPT}.)  For each scattering
process, the leading order (LO) in this expansion in $E/\Lambda$ is
completely predicted from low-energy data, while the next-to-leading
order (NLO) coefficients $\alpha_i$ have to be determined by
experiment.  Some of the parameters have been constrained by $Z$-pole
and $W$ pair production data.  LHC data, hopefully, will probe
weak-boson scattering well into the $\TeV$ range and thus provide the
information that is still missing.

A meaningful experimental analysis of a non-perturbative scenario
requires a class of models to compare with.  For each amplitude, the
low-energy region which is quantitatively described by an effective
Lagrangian, has to be matched to the region of unitarity saturation at
higher energies.  In this region, amplitudes may exhibit resonances,
or they may approach saturation only asymptotically.  There is the
actual possibility of a rich high-energy structure (like in QCD), but
we have to keep in mind the limited event rates and energy range of
the LHC: while the distinction of leading resonances from a
structureless amplitude or from each other becomes feasible, looking
further beyond and determining asymptotic behavior is quite a
challenge.

\subsection{Modeling Terra Incognita}

A comprehensive list of phenomenological models for strong EWSB
includes all types of resonances that can emerge in quasi-elastic weak
boson scattering $VV\to VV$ with $V=W,Z$.  The case $V=\gamma$ can be
ignored: the strong interactions we are interested in are a property
of the longitudinal degrees of freedom, which are absent for the
photon.  For similar reasons, we do not consider resonance couplings
to the other gauge degrees of freedom, i.e.\ transversally polarized
$W/Z$ bosons.  There is no obvious relation of such effects to
electroweak symmetry breaking.  Similarly, the couplings of a new
resonance to SM fermions may be important, but with our current
knowledge the relation to electroweak symmetry breaking is obscure, so
we do not take them into account at the present stage.  Of course, the
model may be extended to cover all of these effects as well, if
necessary.

Spin selection rules restrict $VV$ resonances to scalar, vector, and
tensor type.  In a generic approach, resonance masses and widths are
arbitrary parameters, with the limiting case $M\to\infty$ included.
For each resonance, the partial width for decay into (longitudinal)
vector bosons is determined by the couplings to the corresponding
scattering channel and sets the lower bound for the total width.  As
stated above, we neglect other couplings, so the $VV$ couplings are
directly related to the total width.  Expanding results for low
energies, each resonance contributes a calculable shift to the
chiral-Lagrangian parameters.

Low-energy weak interactions approximately respect weak isospin, also
known as custodial symmetry, $SU(2)_C$~\cite{SU2c}.  Models with
significant violation of weak isospin at high energy tend to provide a
shift to the low-energy $\rho$ parameter that is not supported by LEP
precision data.  In this paper, we therefore extend weak isospin to
high energies and consider the following resonances in $VV\to VV$
processes:
\begin{itemize}
\item scalar singlet $\sigma$, scalar quintet $\phi$,
\item vector triplet $\rho$,
\item tensor singlet $f$, tensor quintet $a$,
\end{itemize}
with arbitrary masses and widths, including $M\to\infty$.  We might
also list $\pi$ (scalar triplet), $\omega$ (vector singlet), etc., but
their couplings to weak bosons are isospin-violating and thus either
small, so we can ignore them, or require unnatural cancellations to
preserve the $\rho$ parameter.

It is straightforward to classify models of EWSB, also
weakly-interacting ones, according to their resonance content in $VV$
scattering.  For instance, a specific model with a $\sigma$ resonance
is the SM.  The vector resonance triplet $\rho$ appears in technicolor
models, but also in extra-dimension models where it is understood as a
$W/Z$ resonance~\cite{5D-BC}.  A tensor $f$ could be a graviton
resonance~\cite{Randall:1999ee}.

We should expect superpositions of resonances.  In particular,
multiplets with specific $SU(2)_L$ quantum numbers $I_L$ decompose
into superpositions of $SU(2)_C$ multiplets: for instance, the
$I_L=1/2$ Higgses of the MSSM decompose into a light singlet
$\sigma=h$ and a heavy triplet $\pi=(H^+,A,H^-)$.  With increasing
mass, the latter decouples from $VV$ scattering due to isospin.
Similarly, the Littlest Higgs model~\cite{Arkani-Hamed:2002qy}
contains a heavy complex $I_L=1$ multiplet which decomposes into a
scalar $I=2$ quintet $\phi$ and a singlet.  The parameterization that
we introduce below supports multiple resonances (one per scattering
channel).  For our numerical results, we have switched on only one
resonance at a time.

\subsection{Unitarity}

Since we are interested in strongly coupled phenomena in energy ranges
where perturbative expansions break down, phenomenological models must
have unitarity bounds explicitly built in.  For instance, the LO naive
result for the $WW\to ZZ$ on-shell amplitude yields quadratic rise
with energy, while unitarity at most allows for an asymptotically
constant value.  In a physics simulation, the naive result would
produce by far too many events at high energy, while in reality there
might be no sensitivity to this region at all.

For quasi-elastic $VV\to VV$ scattering, the unitarity requirement is
rather simple: the eigen\-amplitudes, properly normalized, must lie on
the Argand circle $|a(s)-i/2| = 1/2$.  (Strictly speaking, this is
true in the limit $g\ll E/\Lambda$ where masses are neglected, and
photon and inelastic channels are considered subleading and are
omitted.)  For $a(s)=0$, this law is trivially satisfied. A resonance
corresponds to the amplitude crossing the value $a(s)=i$.

Conservation of angular momentum implies that the eigenamplitudes have
definite angular momentum $(0,1,2,\ldots)$, and since the weak bosons
have spin $1$, at LO there is no unitarity problem for angular
momentum higher than~$2$.  Furthermore, if we keep weak isospin as a
symmetry, the eigenamplitudes also have definite isospin quantum
numbers.  The relevant channels coincide with the list of resonances
given above.

Computed at finite order in perturbation theory, a model amplitude
that rises from a small value of $a(s)$ near $s=0$, will eventually
depart from the Argand circle.  For instance, the LO higgsless SM
eigenamplitude $a_{00}^{(0)}(s)=2s/v^2$ breaks the unitarity limit
$\mathrm{Re}\;a(s)\leq 1/2$ for $E>1.2\;\TeV$, and in a perturbative
expansion this is not remedied by loop corrections in finite order.
Therefore, unitarization models have been invented.  They act as an
operator that takes a scattering amplitude and projects it onto the
Argand circle in an ad-hoc way.

For practical purposes, only gross features of the unitarization
scheme are relevant.  For instance, in ILC physics ($\sqrt{s}\leq
1\;\TeV$), unitarity saturation is not even reached, so the low-energy
expansion taken at face value is usually sufficient.  The LHC can
probe higher energies, but both quark and weak-boson effective
structure functions fall off rapidly with rising energy and strongly
suppress the impact of the multi-$\TeV$ range.  So, the most important
property of any scheme is that it does ensure unitarity, and thus
prohibits any fake $s^n$ rise of the amplitude that, in a simulation,
would produce too many events with large $VV$ invariant masses.

%%%%%%%%%%%%%%%%%%%%%%%%%%%%%%%%%%%%%%%%%%%%%%%%%%%%%%%%%%%%%%%%%%%%%%%%

\section{Basic Theory}

\subsection{Effective Lagrangian}

Without a light Higgs boson, the interactions of fermions and vector
bosons depend on an infinite number of parameters.  However, if the
$S$-matrix is expanded in a series $E/\Lambda$ with $\Lambda=4\pi v$,
at any fixed order in the expansion only a finite subset of the
parameters is relevant.  Order by order, the expansion can be
generated by a suitable low-energy effective Lagrangian.

For a useful approximation, the effective Lagrangian has to respect
the low-energy symmetries, in particular electromagnetic $U(1)$ and QCD
$SU(3)$ gauge invariance, which therefore are realized linearly on the
fields.  The electroweak symmetry $SU(2)_L\times U(1)_Y$ is broken by
fermion and boson masses, but manifest in the low-energy current
algebra as well as in the massive vector-boson couplings.  This can be
encoded in a nonlinear realization.  Grouping quarks and leptons as
left-handed and right-handed doublets $Q_{L/R}$ and $L_{L/R}$, one
introduces a matrix-valued field $\Sigma(x)$ which transforms as
\begin{equation}
  \Sigma \to U_L\, \Sigma\, U_R^\dagger
\end{equation}
under local $SU(2)_L\times U(1)_Y$ transformations, where $U_L(x)=\exp
\left(i\sum_{a=1}^3\beta^a(x)\tau^a\right)$ and $U_R(x)=\exp
\left(i\beta^0(x)\tau^3\right)$ with 
gauge parameters $\beta^a(x)$ and Pauli matrices $\tau^a$.  The $\Sigma$
matrix field is also a special unitary matrix, i.e., it can be
parameterized by
\begin{equation}
  \Sigma(x) = \exp\left(\frac{-i}{v}\vw(x)\right)
\end{equation}
with a scalar field triplet $\vw = \sum_{a=1}^3 w^a\tau^a$, cf.\
App.~\ref{app:gauge}.  The ground state for the perturbative expansion
is defined by $\Sigma=1$, i.e., $w^a\equiv 0$, and the nonlinearity
appears in the $w^a$ gauge transformations.

With these definitions, an effective Lagrangian which generates the
lowest order in $E/\Lambda$ is the chiral
Lagrangian~\cite{appelquistlonghitano,Kilian:2003pc}
\begin{equation}\label{chpt}
\begin{split}
  \LL &=  \frac{v^2}{4}\tr{(D_\mu\Sigma)^\dagger (D^\mu\Sigma)}
   - \tfrac12 \tr{\vW_{\mu\nu} \vW^{\mu\nu}} 
   - \tfrac12 \tr{\vB_{\mu\nu} \vB^{\mu\nu}}
   - \tfrac12 \tr{\vG_{\mu\nu} \vG^{\mu\nu}}.
  \\ &\quad
   + \bar Q_L \I\fmslash D Q_L + \bar Q_R \I\fmslash D Q_R
   + \bar L_L \I\fmslash D L_L + \bar L_R \I\fmslash D L_R
  \\ &\quad
        - (\bar Q_L\Sigma M_QQ_R + \bar L_L\Sigma  M_LL_R + \hc)
        - \bar L_L^c \Sigma^* M_{N_L}\frac{1+\tau^3}{2}\Sigma L_L  
        - \bar L_R^c M_{N_R}\frac{1+\tau^3}{2} L_R
\end{split}
\end{equation}
As the basis for perturbation theory in the gauge couplings $g_s$,
$g$, $g'$, and $E/\Lambda$, this Lagrangian accounts for all
particle-physics measurements that have been possible so far.

\subsection{Resonances}
\label{sec:resonances}

To describe resonances in $WW$ scattering, we add new degrees of
freedom to the chiral Lagrangian~(\ref{chpt}): scalar fields
$\sigma$ and $\vphi$, a vector field $\vrho_\mu$, and tensor
fields $f_{\mu\nu}$ and $\vt_{\mu\nu}$, represented by tensor
products of Pauli matrices in $SU(2)$ space.  In our conventions, they
all transform as matter fields under $SU(2)_L$ according to their
isospin representation,
\begin{subequations}
\begin{align}
  \sigma &\to \sigma,
& 
  \vrho &\to U_L\,\vrho\, U_L^\dagger
& 
  \vphi &\to (U_L\otimes U_L)\,\vphi\,(U_L\otimes U_L)^\dagger,
\end{align}
\end{subequations}
$f$ and $a$ analogous to $\sigma$ and $\phi$, respectively.

In terms of physical (charged) fields, the iso-singlets $\sigma,f$ are
neutral,
\begin{equation}
  \sigma = \sigma^0,
\end{equation}
the iso-triplet $\vrho$ decomposes as
\begin{equation}
  \vrho = \sqrt2\left(
  \rho^+\tau^+ + \rho^0\frac{\tau^3}{\sqrt2} + \rho^-\tau^-
  \right),
\end{equation}
and the iso-quintet fields $\phi,a$ contain doubly-charged components,
\begin{equation}
  \vphi = \sqrt2\left(\phi^{++}\tau^{++} + \phi^+\tau^+ + \phi^0\tau^0
  + \phi^-\tau^- + \phi^{--}\tau^{--}\right),
\end{equation}
where $\tau^{++}=\tau^+\otimes\tau^+$, etc. (App.~\ref{app:algebra}). 

A minimal Lagrangian for these should contain a kinetic term and the
lowest order (in a derivative expansion) of couplings to $W/Z$ pairs.
There are two possibilities: (i) couplings to transversal gauge bosons
via the field strength $\vW_{\mu\nu}$, $\vB_{\mu\nu}$, and (ii)
couplings to longitudinal gauge bosons via the covariant derivative of
the matrix field $\Sigma$.  We do not consider the first case: as
discussed above, such couplings are not directly related to EWSB.
Furthermore, transversal gauge bosons are associated with a factor $g$
or $g'$ instead of $E/\Lambda$, so the interactions of transversal
gauge bosons with a resonance are numerically subdominant.

Let us look at couplings of a heavy resonance to longitudinal gauge
bosons.  As shown by Appelquist/Longhitano
\etal~\cite{appelquistlonghitano}, all possible terms can be expressed
via the two derived fields
\begin{equation}\label{Vmu}
  \vV_\mu = \Sigma (D_\mu\Sigma)^\dagger \quad\text{and}\quad
\index{T@$T$ field}%
  \vT = \Sigma \tau^3 \Sigma^\dagger.
\end{equation}
In the unitarity gauge where $\Sigma=1$, they reduce to $\vV_\mu = -i
g\vW_\mu + ig'B_\mu$ and $T=\tau^3$.  If we insist on isospin
(custodial symmetry), the isospin-breaking spurion $\vT$ can be omitted,
and all couplings to longitudinal gauge bosons proceed via couplings
to $\vV_\mu$.  This vector field transforms under $SU(2)_L\times U(1)_Y$
as
\begin{equation}
  \vV_\mu \to U_L\,\vV_\mu\,U_L^\dagger.
\end{equation}
Each Lagrangian consists of a kinetic term for the resonance and a
linear coupling to a bosonic current.  Explicitly~\cite{Boos:1997gw},
\begin{subequations}
\begin{align}
  \label{L-sigma}
  \LL_\sigma &= 
  -\frac12 \sigma\left(M_\sigma^2 + \pd^2\right)\sigma
  + \sigma j_\sigma
\\
  \label{L-phi}
  \LL_\phi &= 
  -\frac12\left[\frac12\tr{\vphi\left(M_\sigma^2 + \pd^2\right)\vphi}
  + \tr{\vphi \vj_\phi}\right]
\\
  \label{L-rho}
  \LL_\rho &= 
  \frac12\left[
  \frac{M_\rho^2}{2}\tr{\vrho_\mu\vrho^\mu}
  -\frac14\tr{\vrho_{\mu\nu}\vrho^{\mu\nu}}
  + \tr{\vj_\rho^\mu\vrho_\mu}\right]
\\
  \label{L-f}
  \LL_f &= \LL_{\rm kin}
  -\frac{M_f^2}{2}f_{\mu\nu} f^{\mu\nu} + f_{\mu\nu} j_f^{\mu\nu}
\\
  \label{L-a}
  \LL_a &= \LL_{\rm kin} - \frac{M_t^2}{4}\tr{\vt_{\mu\nu}\vt^{\mu\nu}}
  + \frac12\tr{\vt_{\mu\nu}\vj_a^{\mu\nu}}
\end{align}
(the explicit form of the the $M=0$ kinetic term $\LL_{\rm kin}$ of
the tensor [App.~\ref{app:tensor}] is not needed) with the currents
\begin{align}
  j_\sigma &=
  \frac{g_\sigma v}{2} \tr{\vV_\mu\vV^\mu}
\\
  \vj_\phi &= 
  -\frac{g_\phi v}{2}
  \left(\vV_\mu\otimes\vV^\mu - \frac{\tau^{aa}}{6}\tr{\vV_\mu\vV^\mu}\right)
\\
  \vj_\rho^\mu &= 
  \ii g_\rho v^2\vV^\mu
\\
  j_f^{\mu\nu} &= 
  -\frac{g_f v}{2}
  \left(\tr{\vV^\mu\vV^\nu} - \frac{g^{\mu\nu}}{4}\tr{\vV_\rho\vV^\rho}\right) 
\\
  \vj_a^{\mu\nu} &=
  -\frac{g_a v}{2}\left[\frac12
    \left(\vV^\mu\otimes\vV^\nu + \vV^\nu\otimes\vV^\mu\right)
    - \frac{g^{\mu\nu}}{4}\vV_\rho\otimes\vV^\rho\right.
\nonumber
\\
  &\quad\qquad\qquad\left.
  - \frac{\tau^{aa}}{6}\tr{\vV^\mu\vV^\nu}
  + \frac{g^{\mu\nu}\tau^{aa}}{24}\tr{\vV_\rho\vV^\rho}
  \right]
\end{align}
\end{subequations}
The form of the interactions is completely determined by the
transformation laws of the fields and by the conditions of symmetry
and transversality,
\begin{subequations}
\begin{align}
  f^{\mu\nu} &= f^{\nu\mu},
&
  \vt^{\mu\nu} &= \vt^{\nu\mu},
&
  \pd^\mu\vrho_\mu &= 0,
&
  \pd^\mu f_{\mu\nu} &= 0,
&
  \pd^\mu \vt_{\mu\nu} &= 0,
\end{align}
\end{subequations}
and tracelessness with respect to $SU(2)$
\begin{equation}
  \tr{\vrho_\mu} = \tr{\vphi} = \tr{\vt_{\mu\nu}} = 0.
\end{equation}
Analogous relations hold for the currents and uniquely fix their form,
up to terms with higher powers of derivatives.

Higher-derivative terms in the amplitude can be expanded about the
resonance location.  Their on-shell values renormalize the leading
interaction terms as given above and can thus be dropped.  The
off-shell corrections are non-resonant and thus renormalize the NLO
low-energy effective Lagrangian, so they are included there and can
also be omitted.  In short, our list of resonance interactions with
longitudinal $W/Z$ bosons is exhaustive (for the vector resonance
case, see App.~\ref{app:rho}).

With the interaction Lagrangian fixed, we can evaluate the partial
widths for resonance decay into vector bosons.  Given the fact the we
do not specify couplings to transversal bosons, we can only calculate
the leading term in the electroweak coupling expansion, which is
easily computed using the Goldstone-boson equivalence theorem
(GBET)~\cite{Chanowitz:1985hj,GBET}.  The results are listed in
Table~\ref{tab:widths}.  With increasing number of spin and isospin
components, the resonance width decreases.  Furthermore, with our
normalization convention for the dimensionless couplings $g_i$, the
width of a vector resonance has a scaling behavior different from the
others.

\begin{table}[hbt]
\begin{equation*}
\begin{array}{l|ccccc}
\hline
  \text{Resonance}
  &
  \sigma
  &
  \phi
  &
  \rho
  &
  f
  &
  a
\\
\hline
  \Gamma
  &
  6
  &
  1
  &
  \frac{4}{3}(\frac{v^2}{M^2})
  &
  \frac{1}{5}
  &
  \frac{1}{30}
\\
\hline
\end{array}
\end{equation*}
\caption{Partial widths for resonance decay into longitudinally
  polarized vector bosons, computed using the GBET.  All values have
  to be multiplied by the factors $g^2/64\pi$ and $M^3/v^2$, where $g$
  is the coupling in the corresponding resonance Lagrangian.}
\label{tab:widths}
\end{table}

In a purely phenomenological approach, the couplings $g_i$ in the
interaction Lagrangian have no meaning on their own, and their
normalization is arbitrary.  Thus, it is useful to eliminate them in
favor of the resonance masses and widths which are observables, using
Table~\ref{tab:widths}.  We will do this in the following section, so
the matching to the low-energy effective theory is made free of this
ambiguity.

\subsection{Low-energy effects}
\label{sec:low-energy}

Below the first new resonance, physics is described by the chiral
Lagrangian with a double perturbative expansion in the electroweak and
strong couplings, and in $E/\Lambda$.  The LO in $E/\Lambda$ is
generated by the Lagrangian (\ref{chpt}).  The NLO in $E/\Lambda$ is
generated by one-loop corrections and by higher-order operators
$\alpha_i\LL_i$ with coefficients $\alpha_i$.  The list of NLO terms
with isospin symmetry $SU(2)_C$ consists
of~\cite{appelquistlonghitano}
\begin{subequations}
\label{eq:alphaparams}
\begin{align}
  \LL_1 &= \alpha_1 
         gg'\tr{\vB_{\mu\nu}\vW^{\mu\nu}} \label{L1},\\
  \LL_2 &= \I\alpha_2 
         g'\tr{\vB_{\mu\nu}[\vV^\mu,\vV^\nu]} \label{L2},\\
  \LL_3 &= \I\alpha_3 g\tr{\vW_{\mu\nu}[\vV^\mu,\vV^\nu]} \label{L3},\\
  \LL_4 &= \alpha_4(\tr{\vV_\mu \vV_\nu})^2 \label{L4},\\
  \LL_5 &= \alpha_5(\tr{\vV_\mu \vV^\mu})^2 \label{L5}.
\end{align}
\end{subequations}
The first two terms introduce isospin breaking in the same form as the
SM, i.e., only via the coupling to the $B_\mu$ hypercharge gauge
boson, just as the lowest order Lagrangian does.  This breaking
disappears in the limit $g'\ll g$.  $\LL_1$ corresponds to the $S$
parameter, which is well constrained by LEP data.  $\LL_2$ and $\LL_3$
affect three-boson couplings and are also constrained by LEP; these
bounds will be improved by weak-boson pair production at the LHC.  The
last two terms are observable only in weak-boson scattering and are
thus unconstrained so far.

There are several sources that contribute to the $\alpha$ parameters.
First of all, they arise as counterterms for the one-loop correction,
and therefore logarithmically depend on a renormalization scale.
Calculable contributions are generated by integrating out heavy
degrees of freedom, in particular the resonances introduced above.
Ultimately, the values of $\alpha_i$ result from matching the
underlying theory to the chiral Lagrangian; e.g., in a technicolor
model contributions to $\alpha_i$ can be estimated from technifermion
loops.  In the analogous case of low-energy QCD, such estimates are
feasible, while in the electroweak case, the underlying theory is
unknown.

Here, we consider the contributions that result from integrating out
resonances at tree level.  Formally, we can cast the interactions of
a resonance $\Phi$ in the form
\begin{equation}
  \LL_\Phi = z\left[\frac12\Phi(M^2 + A)\Phi + \Phi J\right],
\end{equation}
with a coefficient $z$ and composite operators $A$ and $J$.  The
specific formulae include sums over spin and isospin indices.

Performing the path integral over $\Phi$, we arrive at the effective
Lagrangian which we expand in powers of $1/M^2$ to obtain
\begin{equation}
  \LL_\Phi^\eff = -\frac{z}{2M^2}JJ  + \frac{z}{2M^4}JAJ + \ldots
\end{equation}
As far as this Lagrangian contains terms that are already present in
the LO chiral Lagrangian, they renormalize the LO coefficients, i.e.,
the couplings $g$ and $g'$ and the electroweak scale $v$.  Since the
values of these parameters are determined by low-energy data (in the
sub-TeV range), those shifts can be ignored.  The leading part of the
remainder can be expressed as a combination of the NLO operators
listed above.  The resulting contributions to $\alpha_4$ and
$\alpha_5$ are given in Table~\ref{tab:alpha}.  The values increase
with increasing spin and isospin, and expressed in terms of the
observable parameters $v$, $\Gamma$ and $M$ they all have the same
scaling factor $v^4/M^4$.

\begin{table}[hbt]

\begin{equation*}
\begin{array}{l|ccccc}
\hline
  \text{Resonance} &
  \sigma &
  \phi &
  \rho &
  f &
  a
\\
\hline
\\[-9pt]
  \Delta\alpha_4 &
  0 &
  \frac14 &
  \frac34 &
  \frac52 &
  -\frac58
\\[6pt]
  \Delta\alpha_5 &
  \frac{1}{12} &
  -\frac{1}{12} &
  -\frac{3}{4} &
  -\frac{5}{8} &
  \frac{35}{8}
\\[3pt]
\hline
\end{array}
\end{equation*}
\caption{Shifts in the NLO chiral Lagrangian coefficients $\alpha_4$
  and $\alpha_5$ that result from integrating out a heavy resonance at
  tree level.  All values have to be multiplied by the factors
  $16\pi\Gamma/M$ and $v^4/M^4$.}
\label{tab:alpha}
\end{table}

If a Lagrangian is used that contains a resonance explicitly, these
shifts of the $\alpha$ parameters have to be omitted since they are
replaced by the low-energy tail of the resonance.  Vice versa, if the
resonance is not explicitly included in the Lagrangian but assumed to be
present (presumably, because its mass is beyond the reach of the
experiment), the $\alpha_i$ shifts due to the resonance have to be
added to the low-energy effective Lagrangian.

\begin{figure}[hbt]
  \begin{center}
    \includegraphics[scale=1.5]{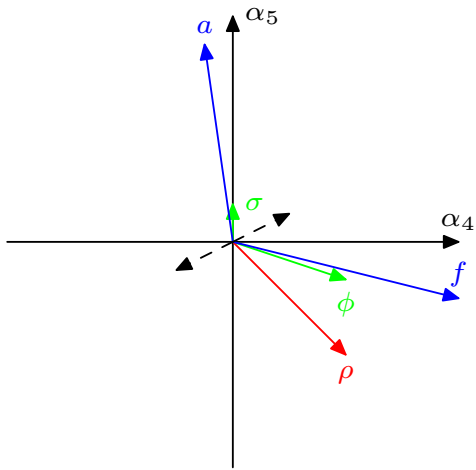}
  \end{center}
  \caption{
    Anomalous couplings $\alpha_{4/5}$ in the low-energy effective
    theory coming from the different resonances under the assumption
    of identical masses and widths (Table~\ref{tab:alpha}). The dashed
    arrow indicates the shift due to renormalization scale variation.
    (The derivations are given in the text.)  }
  \label{fig:shifts}  
\end{figure}

In Fig.~\ref{fig:shifts}, we display the directions and relative
magnitudes of these shifts in the $\alpha_4$-$\alpha_5$ plane.  We
observe that the contributions due to resonances are roughly
orthogonal to the shift which is attributed to a change of
renormalization scale in the one-loop corrections~(\ref{a45ren}),
which makes the two sources distinguishable in principle.
Furthermore, arbitrary resonance patterns induce a combined shift
which lies between the upper and lower-right directions in
Fig.~\ref{fig:shifts}.  This coincides with the region favored by
causality considerations~\cite{Fabbrichesi:2007ad}.

If there is only one important resonance, a simultaneous fit to both
$\alpha$ parameters in the low-energy region would thus enable us
first to distinguish the isosinglet case (scalar or tensor) on the one
hand from the isotriplet/-quintet case (scalar, vector or tensor) on
the other hand.  If the resonance can actually be produced, an angular
analysis of its decay products (for instance, in the \emph{golden
channel} $R\to ZZ\to 4\mu$) could then distinguish scalar from tensor.
The $\rho$ resonance multiplet has the characteristic feature that the
$ZZ$ decay channel is absent, a manifestation of the Landau-Yang
theorem.

\subsection{Reparameterizations}

In this section we discuss alternative parameterizations of the
physics we are interested in.  Due to the equivalence theorem of
quantum field theory~\cite{UET}\footnote{We may call this the
\emph{universal} equivalence theorem (UET) to distinguish it from the
Goldstone-boson equivalence theorem (GBET)~\cite{GBET} for electroweak
interactions, which is a corollary of the UET and gauge invariance.},
they can lead to different intermediate results (such as Feynman
rules), but ultimately have to yield the same observables.

(a) In the previous sections, we have chosen a particular representation
of the effective Lagrangian which manifestly exhibits $SU(2)_L\times
U(1)_Y$ gauge symmetry and $SU(2)_C$ global isospin symmetry.  While
gauging electroweak symmetry is useful for making contact with the SM
and to low-energy current algebra, and for computing loop corrections,
tree-level calculations (at least) can be done in unitarity gauge,
where weak bosons are merely heavy matter fields. The rules for
unitarity gauge are
\begin{align}
  \vw &\to 0, \\
  \Sigma &\to \unit, \\
  V_\mu &\to -ig\vW_\mu + ig'B_\mu.
\end{align}
In this gauge, the Goldstone scalars $w^a$ disappear, and only
physical degrees of freedom are present.  

(b) Alternatively, in the limit that the gauge couplings $g,g'$ can be
neglected compared with $E/\Lambda$ (gaugeless limit), one may omit
the gauge fields and study processes with external Goldstone scalars
$w^a$ only.  These calculations are particularly simple.  Due to the
GBET, in the gaugeless limit the resulting observables are identical
to observables where the Goldstone scalars are replaced by physical,
longitudinally polarized, vector bosons.

(c) The UET states that physical observables are invariant with
respect to arbitrary nonlinear field redefinitions. While manifest
symmetries should be kept in a linear realization for obvious reasons,
there is much freedom in the treatment of nonlinear symmetries.  A
simple corollary implies that all parameterizations of the unitary
matrix $\Sigma$ in terms of three scalar fields are equivalent.  For
instance, we could alternatively use 
\begin{equation}
  \Sigma(x) = \sqrt{1-\frac{\vw(x)^2}{v^2}}\,\times\,\left(\unit -
  \frac{\I}{v}\vw(x)\right) 
\end{equation}
and get new Feynman rules, but identical results for Goldstone
scattering and vector-boson scattering observables.

(d) A straightforward nonlinear reparameterization involves omitting
the $B$ field from the covariant derivative $D_\mu$ in~(\ref{Vmu}) and
expressing the couplings in terms of
\begin{equation}
  \cW_\mu^a = \tr{\vV_\mu\tau^a}
  \quad\text{and}\quad
  B_\mu,
\end{equation}
which results in vector fields that are invariant under $SU(2)_L$ but
transform nontrivially under $U(1)_Y$ instead: $\vW^\pm$ become matter
fields while $\vW^0$ behaves like a gauge field.  Analogously, by
multiplying fermion doublets with $\Sigma$ factors, fermion fields
transforming just under $U(1)_Y$ can be introduced.  This approach,
which is close to choosing unitarity gauge, has been described in
Ref~.\cite{Larios:1996ib}.

(e) The CCWZ version of the chiral Lagrangian~\cite{CCWZ}
introduces the square root of $\Sigma$,
\begin{equation}
  \Sigma = \xi\xi,
\end{equation}
so in the exponential parameterization
\begin{equation}
  \xi(x) = \exp\left(\frac{-i}{2v}\vw(x)\right).
\end{equation}
The field $\xi$ has a mixed transformation law,
\begin{equation}
  \xi \to U_L\xi U_C^\dagger = U_C\xi U_R^\dagger,
\end{equation}
which defines an $SU(2)$ matrix $U_C(x)$ as a function of the
transformations $U_L(x)$ and $U_R(x)$ and of the field $\xi(x)$.  The
matrix $U_C(x)$ can be interpreted as a local isospin
transformation, $U_C\in SU(2)_C$.  

Using $\xi$, the chiral fermion multiplets $Q_{L/R}$ and $L_{L/R}$ can
be promoted to Dirac spinor multiplets,
\begin{align}
  Q &= \begin{pmatrix} \xi Q_R \\ \xi^\dagger Q_L\end{pmatrix},
&
  L &= \begin{pmatrix} \xi L_R \\ \xi^\dagger L_L\end{pmatrix},
\end{align}
which no longer transform under $SU(2)_L$ or $U(1)_Y$, but have a
common transformation law as isospin doublets: $Q\to U_CQ$, $L\to U_CL$.
Similarly, $\xi$ factors make the resonance multiplets invariant under
$SU(2)_L\times U(1)_Y$, but transforming under $SU(2)_C$.

For a vector resonance $\rho$, the CCWZ formulation allows to
introduce it either as a matter field, or as the gauge field of local
$SU(2)_C$, with gauge couplings only.  In the Lagrangian above, we
have introduced the $\rho$ resonance as a matter field.  In
App.~\ref{app:rho}, we describe the alternative formulation with
$\rho$ as a gauge field and verify the equivalence of the
two approaches.

To summarize, while our formulation of the chiral Lagrangian coupled
to resonances is by no means unique, it is nevertheless equivalent to
any other formulation that correctly describes low-energy physics.  As
such, the chiral Lagrangian approach is model-independent.  We do use
model assumptions and truncations, however: no isospin violation
beyond hypercharge and fermion couplings, minimality in the number of
degrees of freedom (at most one resonance per channel), a minimal set
of couplings (no independent couplings to transversal gauge bosons, no
self-couplings of resonances), truncation of the low-energy expansion
(LO and NLO only), and minimality in the unitarization scheme (no
extra parameters).  As long as the new degrees of freedom are heavy,
these model assumptions are likely irrelevant for the experimental
precision that can be achieved at the LHC.  Extensions of our
approach, e.g., including secondary resonances, are easily possible,
but not worked out here to keep this paper compact.

In App.~\ref{app:models} we relate various specific models that are
frequently used in the analysis of weak-boson scattering to our
generic parameterization.

%%%%%%%%%%%%%%%%%%%%%%%%%%%%%%%%%%%%%%%%%%%%%%%%%%%%%%%%%%%%%%%%%%%%%%%%

\section{On-Shell Scattering Amplitudes}

\subsection{Low-energy effective theory}

Let us look first at the $W^+W^-\to ZZ$ weak-boson scattering
amplitude.  In the electroweak coupling expansion, the leading term is
of order $g^0$ and corresponds, at high energy, to the scattering of
longitudinally polarized particles.  This term rises with $s$, while
the scattering amplitudes of transversally polarized vector bosons come
with factors of $g$ and asymptotically do not rise with energy.  By
the GBET, the leading term is equal to the amplitude $A(s,t,u)$ for
$w^+w^-\to zz$ Goldstone scattering.  This amplitude is easily
computed using the Lagrangian~(\ref{chpt}).  At tree-level, but to NLO
in the $E/\Lambda$ expansion, it is
\begin{equation}
  A^{\text{tree}}(s,t,u) = \frac{s}{v^2}
  + 4\alpha_4\frac{t^2+u^2}{v^4}
  + 8\alpha_5\frac{s^2}{v^4}.
\end{equation}
The leading real part (order $g^0$) of the one-loop correction is
given by~\cite{Cheyette:1987jf}
\begin{equation}
  A^{\text{1-loop}}_C(s,t,u) =
  \frac{1}{16\pi^2}\left[
    \left(\frac12\ln\frac{\mu^2}{|s|}+8C_5\right)\frac{s^2}{v^4}
    + \left(\frac{t(s+2t)}{6v^4}\ln\frac{\mu^2}{|t|} 
            + 4C_4\frac{t^2}{v^4}\right)
    + (t\leftrightarrow u)
  \right],
\end{equation}
where $\mu$ is the renormalization scale, and $C_4$ and $C_5$ are
finite scheme-dependent matching coefficients.  For instance, in the
\MSbar\ scheme, $\mu$ is identified with the \MSbar\ scale, and
$C_4=C_5=0$.  By contrast, in the scheme where a fictitious (heavy)
Higgs boson is used as a regulator~\cite{Dawson:1989up}, we have
\begin{equation}
  \mu=M_H
  \qquad\text{and}\qquad
  C_4 = -\frac{1}{18}\approx -0.056,\qquad
  C_5 = \frac{9\pi}{16\sqrt3} - \frac{37}{36} \approx -0.0075.
\end{equation}
Note that these matching coefficients are numerically small, so the
difference between the two schemes may be neglected.  Other schemes
are possible, e.g., the QCD-inspired scheme used in
Ref.~\cite{Fabbrichesi:2007ad} is reproduced by $C_4=-13/72$,
$C_5=-5/72$.

\begin{figure}[hbt]
  \begin{center}
    \includegraphics[width=0.6\textwidth]{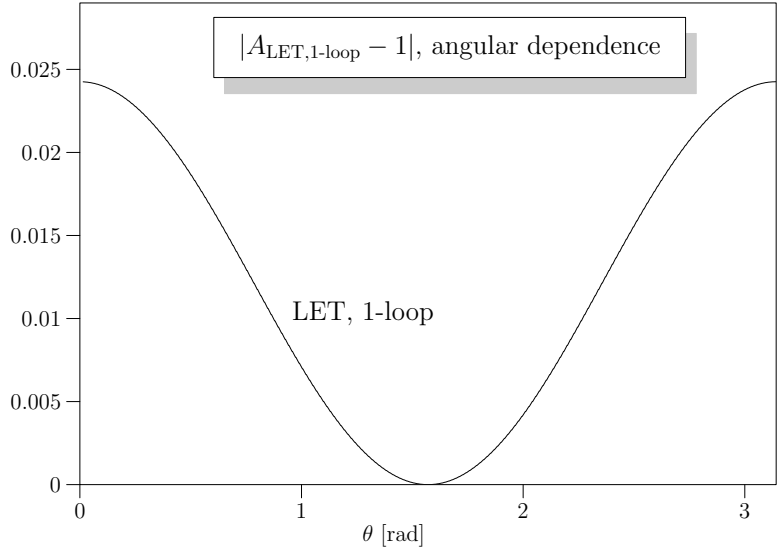}
    \caption{Dependence of the loop correction on the scattering
      angle $\theta$.  The parameters are $\sqrt{s}=\mu=1\;\TeV$.}
    \label{fig:loop-angle}
  \end{center}
\end{figure}

In Fig.~\ref{fig:loop-angle}, we plot the angular dependence of the
one-loop correction.  If the renormalization scale $\mu$ is chosen
equal to the energy $\sqrt{s}$, the loop correction, and thus the
angular dependence, is less than $2.5\,\%$.  Since the NLO correction
is proportional to $s^2$ (compared with the LO amplitude proportional
to $s$), it rapidly becomes important for $s>\mu^2$.  However, this
mainly indicates the breakdown of the low-energy expansion at high
energies.

We can transfer the scheme-dependent matching coefficients to the NLO
counterterms, so the above result is reproduced by maintaining only
the logarithmic terms in the amplitude,
\begin{align}
  A^{\text{1-loop}}(s,t,u) &=
  \frac{1}{16\pi^2}\left[
    \frac{s^2}{2v^4}\ln\frac{M^2}{|s|}
    + \frac{t(s+2t)}{6v^4}\ln\frac{M^2}{|t|}
    + (t\leftrightarrow u)
  \right],
%%   A^{(1)}_r(s,t,u) 
%%   &=
%%   \frac{1}{16\pi^2}\left[
%%     \frac{s^2+t^2+u^2}{3v^4}\ln\frac{M^2}{s}
%%     + \frac{t-u}{6v^4}\left(t\ln\frac{s}{-t}-u\ln\frac{s}{-u}\right)
%%   \right].
\end{align}
and adding one-loop matching contributions to $\alpha_4$ and
$\alpha_5$,
\begin{align}  \label{a45corr}
  \alpha_4^{(1)} &= \frac{1}{16\pi^2}C_4,
&
  \alpha_5^{(1)} &= \frac{1}{16\pi^2}C_5. 
\end{align}
The renormalization scale dependence of these coefficients is given by
\begin{align} \label{a45ren}
  \alpha_4(\mu) &= \alpha_4(\mu_0) 
    - \frac{1}{12}\,\frac{1}{16\pi^2}\ln\frac{\mu^2}{\mu_0^2},
&
  \alpha_5(\mu) &= \alpha_5(\mu_0)
    - \frac{1}{24}\,\frac{1}{16\pi^2}\ln\frac{\mu^2}{\mu_0^2}.
\end{align}
with some reference scale $\mu_0$.

Isospin symmetry determines all individual scattering amplitudes in
terms of the master amplitude $A(s,t,u)$:
\begin{subequations}
\begin{align}
  \label{A-LET-wwzz}
  A(w^+w^-\to zz)
  &=
  A(s,t,u)
\\
  A(w^+z\to w^+z)
  &=
  A(t,s,u)
\\
  A(w^+w^-\to w^+w^-)
  &=
  A(s,t,u) + A(t,s,u)
\\
  A(w^+w^+\to w^+w^+)
  &=
  A(t,s,u) + A(u,s,t)
\\
  \label{A-LET-zzzz}
  A(zz\to zz)
  &=
  A(s,t,u) + A(t,s,u) + A(u,s,t)
\end{align}
\end{subequations}
Expanding the amplitudes in powers of the energy, the order-$E^2$
term is known as the low-energy theorem (LET)~\cite{LET}:
\begin{subequations}
\begin{align}
  A^{(0)}(w^+w^-\to zz)
  &= s/v^2
\\
  A^{(0)}(w^+z\to w^+z)
  &= t/v^2
\\
  A^{(0)}(w^+w^-\to w^+w^-)
  &= -u/v^2
\\
  A^{(0)}(w^+w^+\to w^+w^+)
  &= -s/v^2
\\
  A^{(0)}(zz\to zz)
  &= 0
\end{align}
\end{subequations}
These expressions are model-independent and depend just on the
electroweak scale $v$. 

%%%%%

\subsection{Resonances}

In Sec.~\ref{sec:resonances}, we have introduced heavy resonances in
weak-boson scattering.  The interaction Lagrangians
(\ref{L-sigma}--\ref{L-a}) induce couplings to vector bosons and to
Goldstone bosons, which are related by electroweak gauge invariance,
maintaining the GBET.  Each resonance multiplet therefore contributes
additional terms to the Goldstone scattering amplitude $A(s,t,u)$,
which have poles at the appropriate locations.  We do not yet include
the resonance widths.  The new contributions are
\begin{subequations}
\begin{align}
  A^\sigma(s,t,u) &= - \frac{g_\sigma^2}{v^2}\frac{s^2}{s-M^2}
\\
  A^\phi(s,t,u) &= - \frac{g_\phi^2}{4 v^2}\left( \frac{t^2}{t-M^2} +
    \frac{u^2}{u-M^2} - \frac{2}{3} \frac{s^2}{s - M^2} \right)  
\\
  A^\rho(s,t,u) &= - g_\rho^2\left(\frac{s-u}{t-M^2} +
           \frac{s-t}{u-M^2} + 3 \frac{s}{M^2} \right)
\\
  A^f(s,t,u) &= 
  - \frac{g_f^2}{6 v^2} \frac{s^2}{s-M^2} P_2(s,t,u)
  + \frac{g_f^2}{12 v^2} \frac{s^2}{M^2}
\\
  A^a(s,t,u) &= 
  - \frac{g_a^2}{24 v^2} \left\{ \frac{t^2}{t-M^2}
  P_2(t,s,u) + \frac{u^2}{u-M^2}
  P_2(u,s,t) - \left( \frac23 
  \frac{s^2}{s-M^2} - \frac{s^2}{6M^2} \right) P_2(s,t,u) 
\right\}
\end{align}
\end{subequations}
where $P_2(s,t,u) = [3(t^2 + u^2) - 2 s^2]/s^2$.

Beyond the resonance location, for $g_\sigma=1$ the $\sigma$ exchange
amplitude cancels the rise of the LET amplitude. This is the SM case.
Otherwise, beyond the resonance all amplitudes rise with a power of
$s/M^2$.  This implies again unitarity violation, which has to be
cured by the unknown UV completion of the theory.

\subsection{Eigenamplitudes}
\label{sec:eigenamplitudes}

For the analysis of unitarity, we need the spin-isospin
eigenamplitudes, i.e., scattering amplitudes for superpositions of
states which scatter only into themselves.  We first list the isospin
eigenamplitudes
\begin{subequations}
\begin{align}
  A_0(s,t,u) 
  &=
  3A(s,t,u) + A(t,s,u) + A(u,s,t)
\\
  A_1(s,t,u)
  &=
  A(t,s,u) - A(u,s,t)
\\  
  A_2(s,t,u) 
  &= 
  A(t,s,u) + A(u,s,t)
\end{align}
\end{subequations}
which can be decomposed into partial waves using Legendre polynomials,
\begin{equation}
  A_I(s,t,u) = \sum_{J=0}^\infty A_{IJ}(s)\,(2J+1)\,P_J(s,t,u),
\end{equation}
where $A_{IJ}\neq 0$ only for $I-J$ even.  The coefficient functions
$A_{IJ}(s)$ are the spin-isospin eigenamplitudes.  They are obtained by
angular integration,
\begin{equation}
  \label{eq:projectout}
  A_{IJ}(s) = \int_{-s}^0 \frac{dt}{s}A_I(s,t,u)\,P_J(s,t,u).
\end{equation}
Below, we explicitly list the spin-isospin eigenamplitudes, treating
LO, NLO, and resonances separately:

(a) The eigenamplitudes for the LO Lagrangian:
\begin{align}
\label{eigenamp-LET}
  A_{00}^{(0)} &= 
    2\frac{s}{v^2} 
&
  A_{11}^{(0)} &=
    \frac{s}{3v^2}
&
  A_{20}^{(0)} &=
    -\frac{s}{v^2}
\end{align}
All other terms vanish at this order.

\begin{figure}[hbt]
  \begin{center}
    \includegraphics[width=0.6\textwidth]{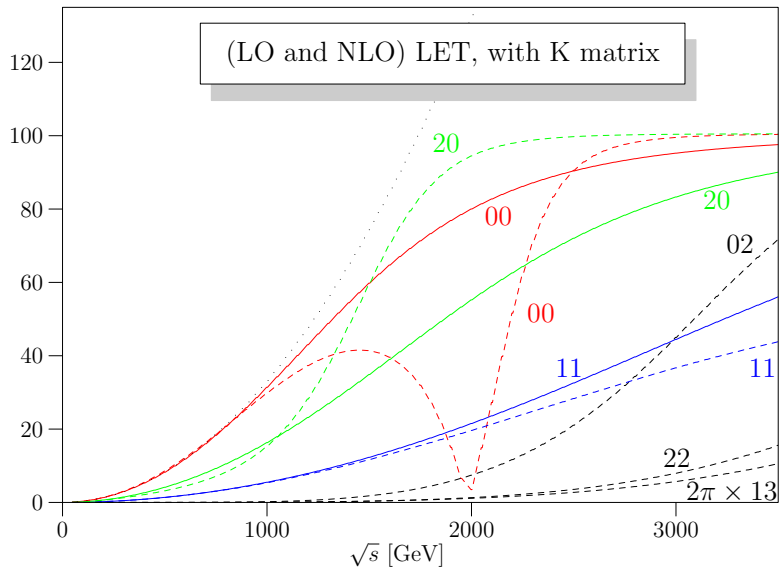}
    \caption{Spin-isospin eigenamplitudes as functions of the energy
      $\sqrt{s}$, unitarized according to the scheme described in
      Sec.~\ref{sec:unitarization}.  Solid curves: LO; dashed curves:
      NLO.  Dotted curve: $A_{00}(s)$ without unitarization.}
    \label{fig:amp-let}
  \end{center}
\end{figure}

(b) The one-loop correction with its logarithmic angular dependence
contains partial waves of arbitrary spin.  We extract the leading
logarithms $\ln(\mu^2/s)$, project out the partial waves and truncate
the series at spin~$3$, which numerically is an excellent
approximation.  Adding the tree-level NLO coefficients, which should
include their scheme-dependent and scale-dependent parts
(\ref{a45corr}, \ref{a45ren}), the real part of the result is
\begin{subequations}
\begin{align}
  A_{00}^{(1)} 
  &= 
  \left[
    \frac83\left(7\alpha_4(\mu) + 11\alpha_5(\mu)\right)
    + \frac{1}{16\pi^2}
    \left(\frac{25}{9}\ln\frac{\mu^2}{s} + \frac{11}{54}\right)
  \right]\frac{s^2}{v^4}
  \label{a00-1}
\\
  A_{02}^{(1)}
  &=
  \left[
    \frac{8}{15}\left(2\alpha_4(\mu) + \alpha_5(\mu)\right)
    + \frac{1}{16\pi^2}
    \left(\frac{1}{9}\ln\frac{\mu^2}{s} - \frac{7}{135}\right)
  \right]\frac{s^2}{v^4}
  \label{a02-1}
\\
  A_{11}^{(1)}
  &=
  \left[
    \frac43\left(\alpha_4(\mu)-2\alpha_5(\mu)\right)
    + \frac{1}{16\pi^2}
    \left( - \frac{1}{54} \right)
  \right]\frac{s^2}{v^4}
  \label{a11-1}
\\
  A_{13}^{(1)}
  &=
  \left[
    0
    + \frac{1}{16\pi^2}
    \left(\frac{7}{1080}\right)
  \right]\frac{s^2}{v^4}
  \label{a13-1}
\\  
  A_{20}^{(1)} 
  &= 
  \left[
    \frac{16}{3}\left(2\alpha_4(\mu)+\alpha_5(\mu)\right)
    + \frac{1}{16\pi^2}
    \left(\frac{10}{9}\ln\frac{\mu^2}{s} + \frac{25}{108}\right)
  \right]\frac{s^2}{v^4}
  \label{a20-1}
\\
  A_{22}^{(1)}
  &=
  \left[
    \frac{4}{15}\left(\alpha_4(\mu) + 2\alpha_5(\mu)\right)
    + \frac{1}{16\pi^2}
    \left(\frac{2}{45}\ln\frac{\mu^2}{s} - \frac{247}{5400}\right)
  \right]\frac{s^2}{v^4}
  \label{a22-1}
\end{align}
\end{subequations}
We note that the scale dependence of the $\alpha$ parameters cancels
the scale-dependence of the one-loop terms, as it should be the case.
The results are shown in Fig.~\ref{fig:amp-let}.  While the loop
correction is small below about $1\;\TeV$, for higher energies it
becomes important and, eventually, drastically changes the behavior.
For instance, in $A_{00}$ there is a cancellation between the LO and
NLO terms at $2\;\TeV$.  This clearly indicates the breakdown of the
low-energy expansion.

(c) For the decomposition of resonance corrections, we define the
following functions:
\begin{subequations}
\begin{align}
  \cS_J(s) &= \int_{-s}^0\frac{dt}{s} \frac{t^2}{t-M^2} P_0(t,s,u) P_J(s,t,u)
\\
  \cP_J(s) &= \int_{-s}^0 \frac{dt}{s} \frac{t}{t-M^2} P_1(t,s,u) P_J(s,t,u)
\\
  \cD_J(s) &= \int_{-s}^0 \frac{dt}{s} \frac{t^2}{t-M^2} P_2(t,s,u) P_J(s,t,u)
\end{align}
\end{subequations}
which we give explicitly in Appendix~\ref{app:integrals}.  We obtain
for the isosinglet scalar,
\begin{subequations}
\begin{align}
  A^\sigma_{00}(s) &=\;  - 3\frac{g_\sigma^2}{v^2} \frac{s^2}{s-M^2}
            - 2\frac{g^2}{v^2} \cS_0(s) 
            & \qquad  
  A^\sigma_{13}(s) &=\;
            - 2\frac{g_\sigma^2}{v^2} \cS_3(s) 
            \\
  A^\sigma_{02}(s) &=\;
            - 2\frac{g_\sigma^2}{v^2} \cS_2(s)
            & \qquad  
  A^\sigma_{20}(s) &=\; - 2 \frac{g_\sigma^2}{v^2} \cS_0(s) \\
  A^\sigma_{11}(s) &=\; - 2 \frac{g_\sigma^2}{v^2} \cS_1(s) 
            & \qquad  
  A^\sigma_{22}(s) &=\; - 2 \frac{g_\sigma^2}{v^2} \cS_2(s)
\end{align}
\end{subequations}
the isoquintet scalar,
\begin{subequations}
\begin{align}
  A^\phi_{00}(s) &=\; - \frac{5}{3}\frac{g_\phi^2}{v^2} \cS_0(s) 
            & \qquad  
  A^\phi_{13}(s) &=\;
            \frac{5}{6} \frac{g_\phi^2}{v^2} \cS_3(s) 
            \\
  A^\phi_{02}(s) &=\;
            - \frac{5}{3}\frac{g_\phi^2}{v^2} \cS_2(s)
            & \qquad  
  A^\phi_{20}(s) &=\; - \frac12 \frac{g_\phi^2}{v^2}
            \frac{s^2}{s-M^2} -
            \frac16 \frac{g_\phi^2}{v^2} \cS_0(s) \\
  A^\phi_{11}(s) &=\; \frac{5}{6} \frac{g_\phi^2}{v^2} \cS_1(s) 
            & \qquad  
  A^\phi_{22}(s) &=\; - \frac16 \frac{g_\phi^2}{v^2} \cS_2(s)
\end{align}
\end{subequations}
the isotriplet vector,
\begin{subequations}
\begin{align}
  A^\rho_{00}(s) &=\; 
    - 4 g^2_\rho \cP_0(s) - 3 g^2_\rho \frac{s}{M^2} & \qquad
  A^\rho_{13}(s) &=\; 
         - 2 g_\rho^2 \frac{2s+M^2}{M^4} \cS_3(s)
  \\
  A^\rho_{02}(s) &=\; - 4g^2_\rho 
    \; \frac{2s + M^2}{M^4} \cS_2(s)
    & \qquad 
  A^\rho_{20}(s) &=\;
    2 g^2_\rho \cP_0(s) + 3 g_\rho^2 \frac{s}{M^2}
    \\ 
  A^\rho_{11}(s) &=\;
             - \frac23 g^2_\rho \frac{s}{s-M^2} -
              g_\rho^2 \frac{s}{M^2} - 2 g_\rho^2 \cP_1(s) 
   & \qquad 
  A^\rho_{22}(s) &=\;
    2 g_\rho^2 \frac{2s+M^2}{M^4} \cS_2(s)
\end{align}
\end{subequations}
the isosinglet tensor,
\begin{subequations}
\begin{align}
  A^f_{00}(s) &=
    - \frac{g_f^2}{3 v^2} \cD_0(s) - \frac{11}{36} \frac{g_f^2}{v^2}
    \frac{s^2}{M^2}
    \\ 
  A^f_{02}(s) &= 
            - \frac{g_f^2}{10 v^2} \frac{s^2}{s-M^2} -
            \frac{g_f^2}{3 v^2} \left( 1  + 6 \frac{s}{M^2} + 6
            \frac{s^2}{M^4} \right) \cS_2(s)
            - \frac{1}{180} \frac{g_f^2}{v^2}
            \frac{s^2}{M^2}
            \\
  A^f_{11}(s) &= 
    - \frac{g^2_f}{3 v^2} \cD_1(s) + \frac{1}{36} \frac{g_f^2}{v^2}
            \frac{s^2}{M^2}
  \\ 
  A^f_{13}(s) &= 
            - \frac{g_f^2}{3 v^2} \left( 1  + 6 \frac{s}{M^2} + 6
            \frac{s^2}{M^4} \right) \cS_3(s)             
            \\
  A^f_{20}(s) &= 
    - \frac{g_f^2}{3 v^2} \cD_0(s) 
      - \frac{1}{18} \frac{g_f^2}{v^2}
            \frac{s^2}{M^2} 
            \\ 
  A^f_{22}(s) &= 
            - \frac{g_f^2}{3 v^2} \left( 1  + 6 \frac{s}{M^2} + 6
            \frac{s^2}{M^4} \right) \cS_2(s)
            - \frac{1}{180} \frac{g_f^2}{v^2}
            \frac{s^2}{M^2}
\end{align}
\end{subequations}
and the isoquintet tensor,
\begin{subequations}
\begin{align}
  A^a_{00}(s) &= 
                - \frac56 \frac{g_a^2}{3 v^2} \cD_0(s) 
                - \frac{5}{108} \frac{g_a^2}{v^2}
                \frac{s^2}{M^2}
                \\ %% \nonumber
  A^a_{02}(s) &= 
                - \frac56 \frac{g_a^2}{3 v^2} \left( 1 + 6
                \frac{s}{M^2} + 6 \frac{s^2}{M^4} \right) \cS_2(s)
                - \frac{1}{216} \frac{g_a^2}{v^2}
                \frac{s^2}{M^2}
                \\
  A^a_{11}(s) &= 
                \frac{5}{12} \frac{g_a^2}{3 v^2} 
                \cD_1(s)
                - \frac{5}{432} \frac{g_a^2}{v^2}
                \frac{s^2}{M^2}
                \\ %%\nonumber 
  A^a_{13}(s) &=	
                \frac{5}{12} \frac{g_a^2}{3 v^2} \left( 1 + 6
                \frac{s}{M^2} + 6 \frac{s^2}{M^4} \right)
                \cS_3(s)
                \\                
  A^a_{20}(s) &= 
                - \frac{1}{12} \frac{g_a^2}{3 v^2} 
                \cD_0(s) - \frac{5}{108} \frac{g_a^2}{v^2}
                \frac{s^2}{M^2}
                \\ %% \nonumber 
  A^a_{22}(s) &=
                - \frac{g_a^2}{60 v^2} 	\frac{s^2}{s-M^2} -  
                \frac{1}{12} \frac{g_a^2}{3 v^2} \left( 1 + 6
                \frac{s}{M^2} + 6 \frac{s^2}{M^4} \right)
                \cS_2(s)
                 - \frac{1}{2160} \frac{g_a^2}{v^2}
                \frac{s^2}{M^2}
\end{align}
\end{subequations}
The coefficient functions $A_{IJ}$ contain poles in $s-M^2$ as well as
finite parts.  The poles are confined to those $(I,J)$ combinations
which correspond to the $(I,J)$ assignments of the resonances.  Again,
we truncate the partial-wave expansion at $J=3$, so for each
spin-isospin combination we only keep the leading and one subleading
term.

%%%%%

\subsection{Unitarization scheme}

\begin{figure}[htb]
\begin{center}
\includegraphics{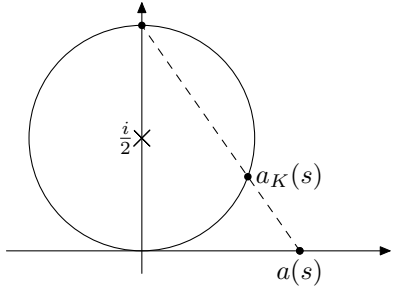}
\end{center}
\index{Argand circle}
\caption{$K$ matrix construction for projecting a real scattering
amplitude onto the Argand circle}
\label{fig:K-matrix}
\end{figure}

Elastic unitarity requires that the normalized eigenamplitudes
\begin{equation}
  a_{IJ}(s) = \frac{1}{32\pi}A_{IJ}(s),
\end{equation}
respect the Argand-circle condition
\begin{equation}
  |a_{IJ}(s) - i/2| = 1/2,
\end{equation}
which can also be stated as
\begin{equation}
  \Im\frac{1}{a_{IJ}(s)} = -1.
\end{equation}
Computed in finite-order perturbation theory, or deduced from some
model, the amplitude $a(s)$ will usually fail this requirement.
However, an arbitrary amplitude $a(s)$ can be transformed into a
unitary amplitude if we take the real part of $1/a(s)$ and add $-i$ as
the imaginary part, i.e.,
\begin{align}
  \hat a(s) &= \frac{1}{\Re(1/a(s)) - i}.
\\
  &= \frac{a(s)}{1 - ia(s)}
  \quad\text{if $a(s)$ is real.}
\end{align}
For the unnormalized eigenamplitudes $A_{IJ}(s)$, this can be
rephrased as
\begin{equation}
  \hat A_{IJ}(s) = A_{IJ}(s) + \Delta A_{IJ}(s),
  \quad\text{where}\quad
  \Delta A_{IJ}(s) 
  = 
  \frac{i}{32\pi}\,
  \frac{A_{IJ}(s)^2}{1 - \frac{i}{32\pi}A_{IJ}(s)}.
\end{equation}
This is the K-matrix unitarization scheme~\cite{kmatrix}, cf.\
Fig.~\ref{fig:K-matrix}. 

With this prescription, a LET amplitude $A(s)=s/v^2$ becomes
\begin{equation}
  \hat A(s) = \frac{s/v^2}{1 - \frac{i}{32\pi v^2} s}
  \stackrel{s\to\infty}{\longrightarrow}
  32\pi i,
\end{equation}
so instead of rising quadratically with energy, the absolute value of
$\hat A(s)$ asymptotically approaches saturation, formally a resonance at
infinity.

The K-matrix scheme transforms a simple-pole amplitude,
$A(s)=-c/(s-M^2)$, into Breit-Wigner form,
\begin{equation}
  \hat A(s) = \frac{-c}{s-M^2 + iM\Gamma}
  \quad\text{with}\quad
  \Gamma = \frac{c}{32\pi M},
\end{equation}
so it is an alternate implementation of Dyson resummation for
s-channel particle exchange.  If $c$ is not a constant but depends on
energy, we get a Breit-Wigner resonance with $s$-dependent width.  In
particular, the amplitude
\begin{equation}\label{Pade-pole}
  A(s) = -\frac{M^2}{v^2}\frac{s}{s-M^2}
\end{equation}
is transformed into
\begin{equation}\label{Pade}
  \hat A(s) = -\frac{M^2}{v^2}\,\frac{s}{s - M^2 + iM\Gamma\frac{s}{M^2}}
  \quad\text{with}\quad
  \Gamma = \frac{M^2}{32\pi v^2}M.
\end{equation}
Eq. (\ref{Pade-pole}) has the low-energy expansion
\begin{equation}\label{Pade-LE}
  A(s)\stackrel{s\to 0}{\longrightarrow}
  \frac{s}{v^2} + \frac{s^2}{M^2v^2}
  = A^{(0)}(s) + A^{(1)}(s)
\end{equation}
An expansion of this form can also be treated by the inverse-amplitude
method (IAM) for unitarization~\cite{Dobado:1996ps}.  The result is
\begin{equation} \label{IAM}
  \hat A(s) = \frac{A^{(0)}(s)^2}
                   {A^{(0)}(s) - A^{(1)}(s) - \frac{i}{32\pi}A^{(0)}(s)^2},
\end{equation}
which equals the $(1,1)$ Pad\'e approximant, and precisely coincides
with (\ref{Pade}).  We observe that, in the present context, the IAM
or Pad\'e unitarization scheme is a special case of the K-matrix
scheme, where the low-energy expansion of the amplitude is identified
with the low-energy tail of a single resonance.  In QCD, where the
$\rho$ meson dominates form factors at low energy, this turns out to
be a valid assumption which leads to accurate high-energy
extrapolations.  In the electroweak case, physics may be different,
and the actual (unitary) weak-boson scattering amplitudes need not
follow the extrapolation of the K-matrix/IAM/Pad\'e or any other given
unitarization scheme.  

In QCD, low-energy parameters can be computed, to good accuracy, by
integrating out the $\rho$ resonance.  This may also be the case for
the leading resonances in electroweak interactions (we list the
necessary formulas in Sec.~\ref{sec:low-energy}), but there may well
be extra contributions that can be assigned to further resonances, or
to other physical effects.  For this reason, we keep $\alpha_4$ and
$\alpha_5$ as independent parameters in our implementation.

The detailed shape of resonances in weak-boson scattering may also
differ from the (running-width) Breit-Wigner that our parameterization
provides.  However, the experimental resolution of weak-boson pair
invariant masses at the LHC will be limited, so there is little hope
for precise resonance scans.  A parameterization in terms of mass and
width, augmented by extra $\alpha_{4,5}$ parameters which describe
deviations in the low-energy tail, is sufficient.

Beyond a resonance peak, our expressions suggest a definite
prediction, such as a new rise of the amplitude with a definite power
of $s$.  We should emphasize that this is misleading: the behavior in
this region is arbitrary and can only be modeled, introducing further
parameters.  However, any precise measurements of the high-energy tail
of a heavy resonance will be challenging, if not impossible at the
LHC.  The only property of unitarized amplitudes that we really make
use of is: that they do not exceed the unitarity limits.

%%%%%

\subsection{Unitarized Amplitudes}
\label{sec:unitarization}

In this section, we apply the unitarization scheme defined above to
the generic parameterization of scattering amplitudes.  Collecting
everything, each eigenamplitude consists of a LO (LET) part, a NLO
correction which includes the one-loop part and finite extra
contributions to the $\alpha$ parameters, and resonance terms:
\begin{equation}
  A_{IJ}(s) = A^{(0)}_{IJ}(s) + A^{(1)}_{IJ}(s) 
  + \sum_{R=\sigma,\phi,\rho,f,a} A^{R}_{IJ}(s)
\end{equation}
which we write in the form
\begin{equation}
  A_{IJ}(s) = A^{(0)}_{IJ}(s) + F_{IJ}(s) + \frac{G_{IJ}(s)}{s-M^2},
\end{equation}
where $F_{IJ}(s)$ is finite, and $G_{IJ}(s)$ is proportional to $s$
(vector), or $s^2$ (scalar, tensor).  According to the prescription in
the previous section, the unitarized amplitude becomes
\begin{equation}
\label{pole decomposition}
  \hat A_{IJ}(s) = \frac{A_{IJ}(s)}{1 - \frac{i}{32\pi}A_{IJ}(s)}
  = A^{(0)}_{IJ}(s) + \Delta A_{IJ}(s),
\end{equation}
where the correction to the LET amplitude is given by
\begin{equation}
\label{unitarization}
  \Delta A_{IJ}(s) = 32\pi i
  \left(1 + \frac{i}{32\pi}A^{(0)}(s)
        + \frac{s - M^2}
               {\frac{i}{32\pi} G_{IJ}(s)
                - (s - M^2)\left[1 - \frac{i}{32\pi}
                                 (A^{(0)}(s) + F_{IJ}(s))\right]
                }
  \right)
\end{equation}

\begin{figure}[p]
  \begin{center}
    \includegraphics[width=0.45\textwidth]{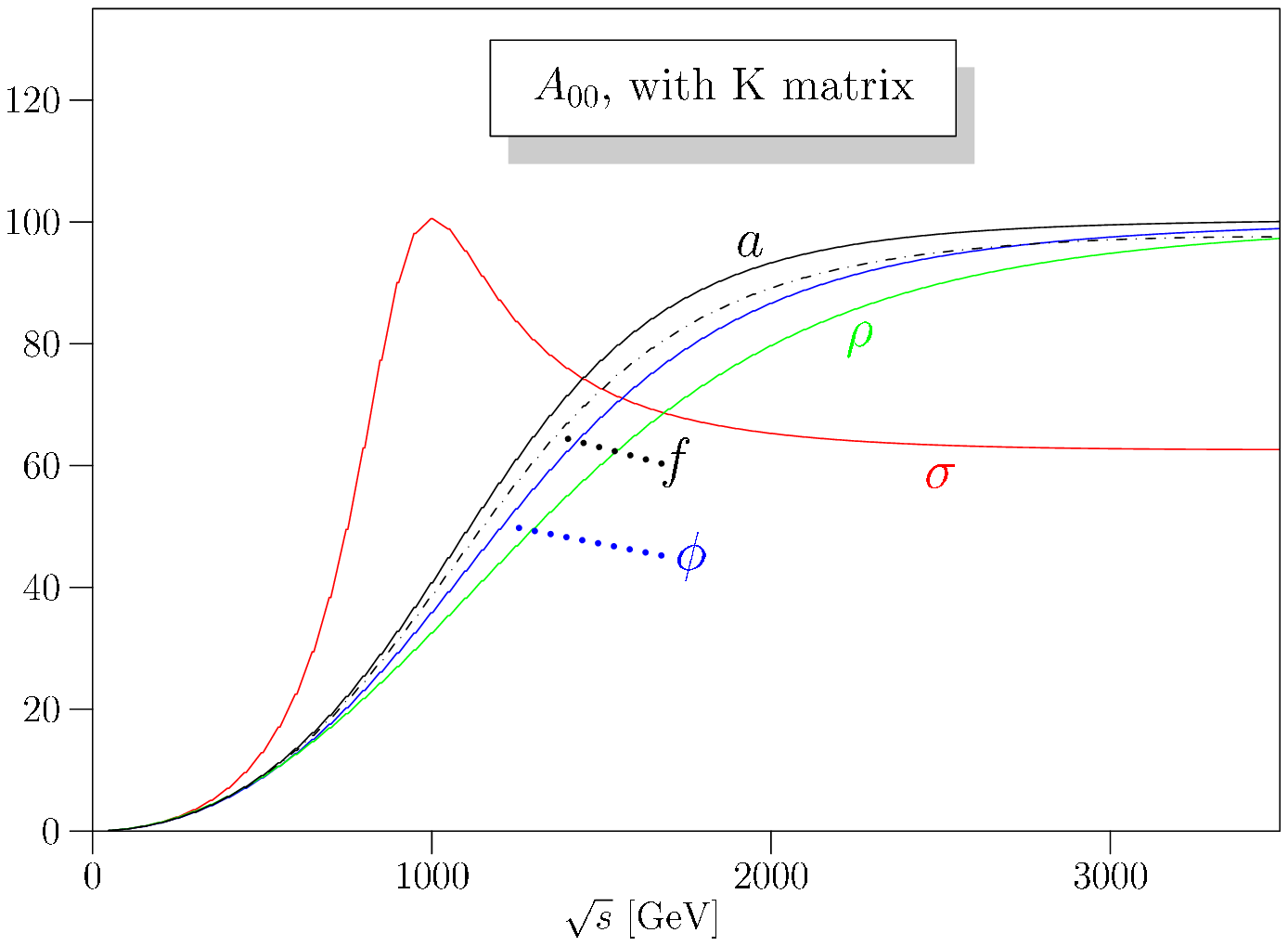}
    \hspace{3mm}
    \includegraphics[width=0.45\textwidth]{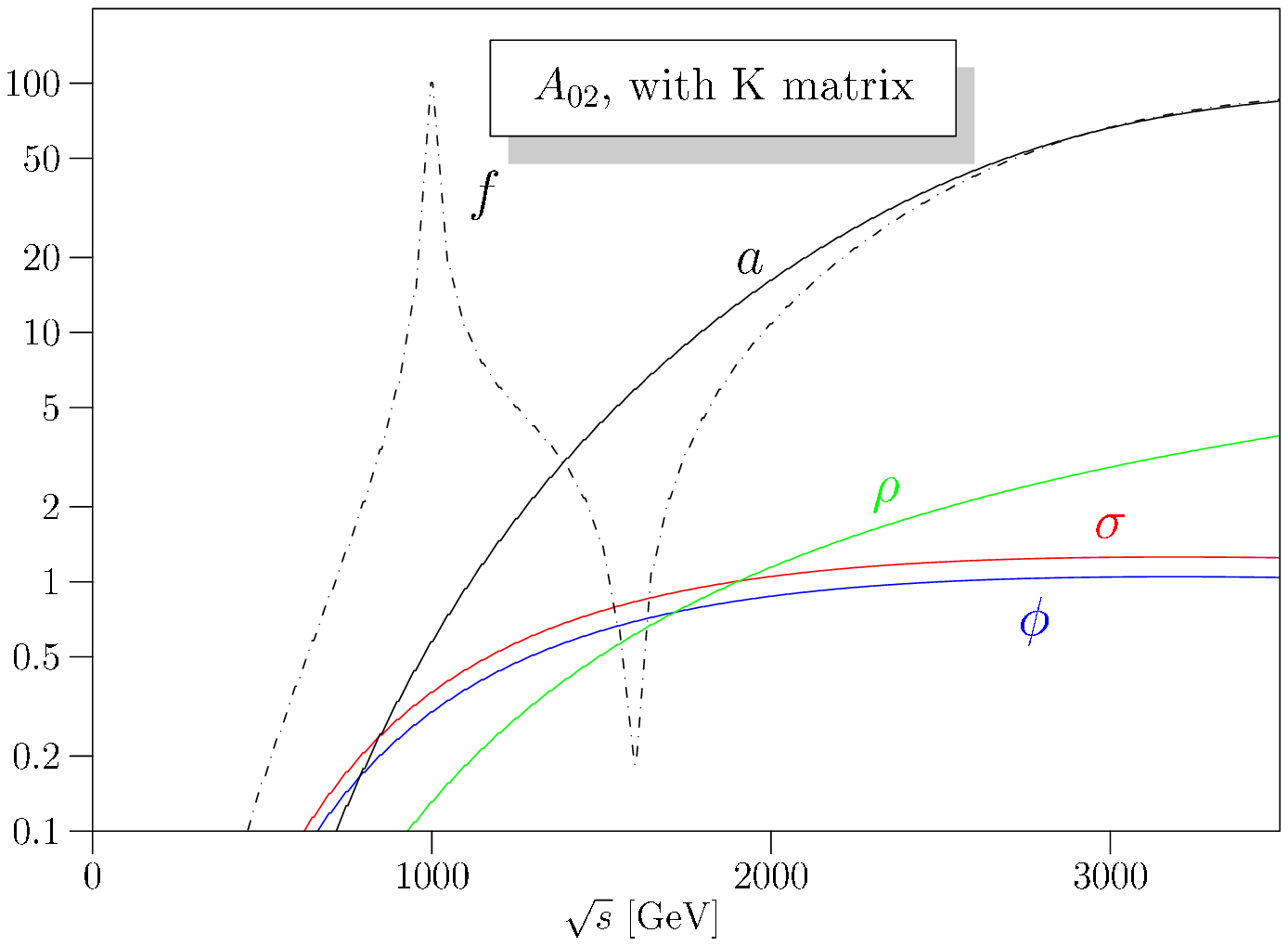}
    \\[3mm]
    \includegraphics[width=0.45\textwidth]{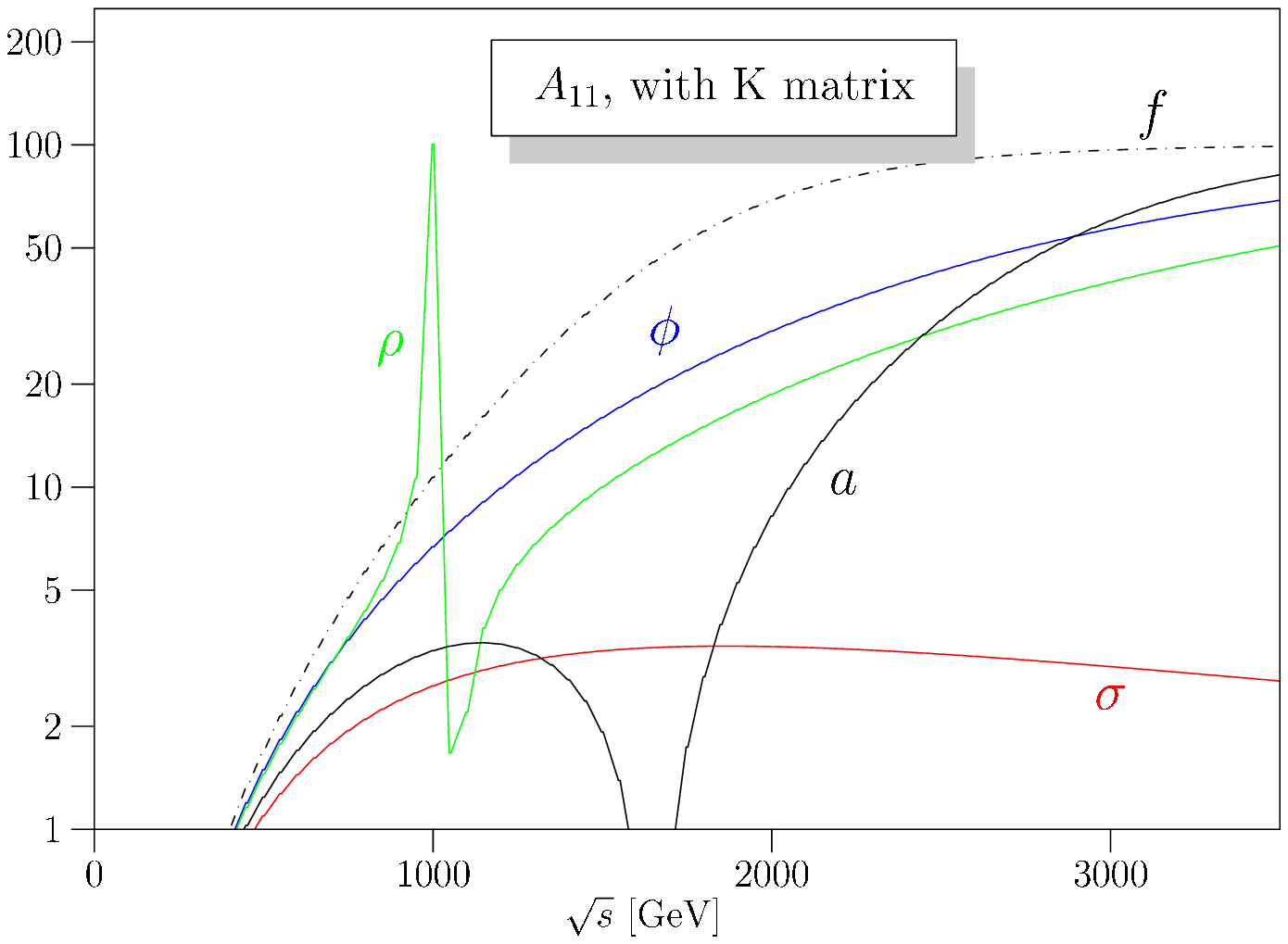}
    \hspace{3mm}
    \includegraphics[width=0.45\textwidth]{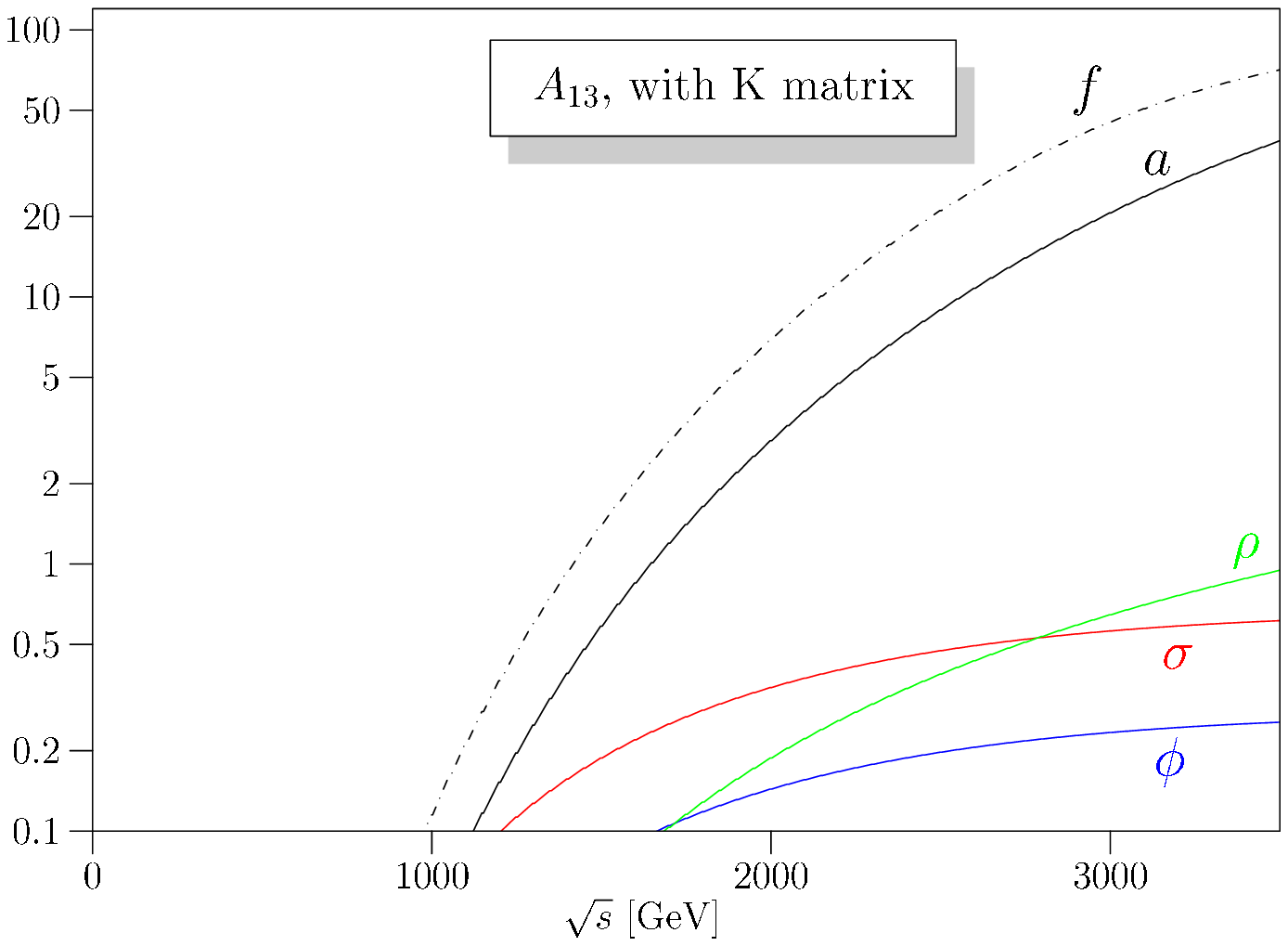}
    \\[3mm]
    \includegraphics[width=0.45\textwidth]{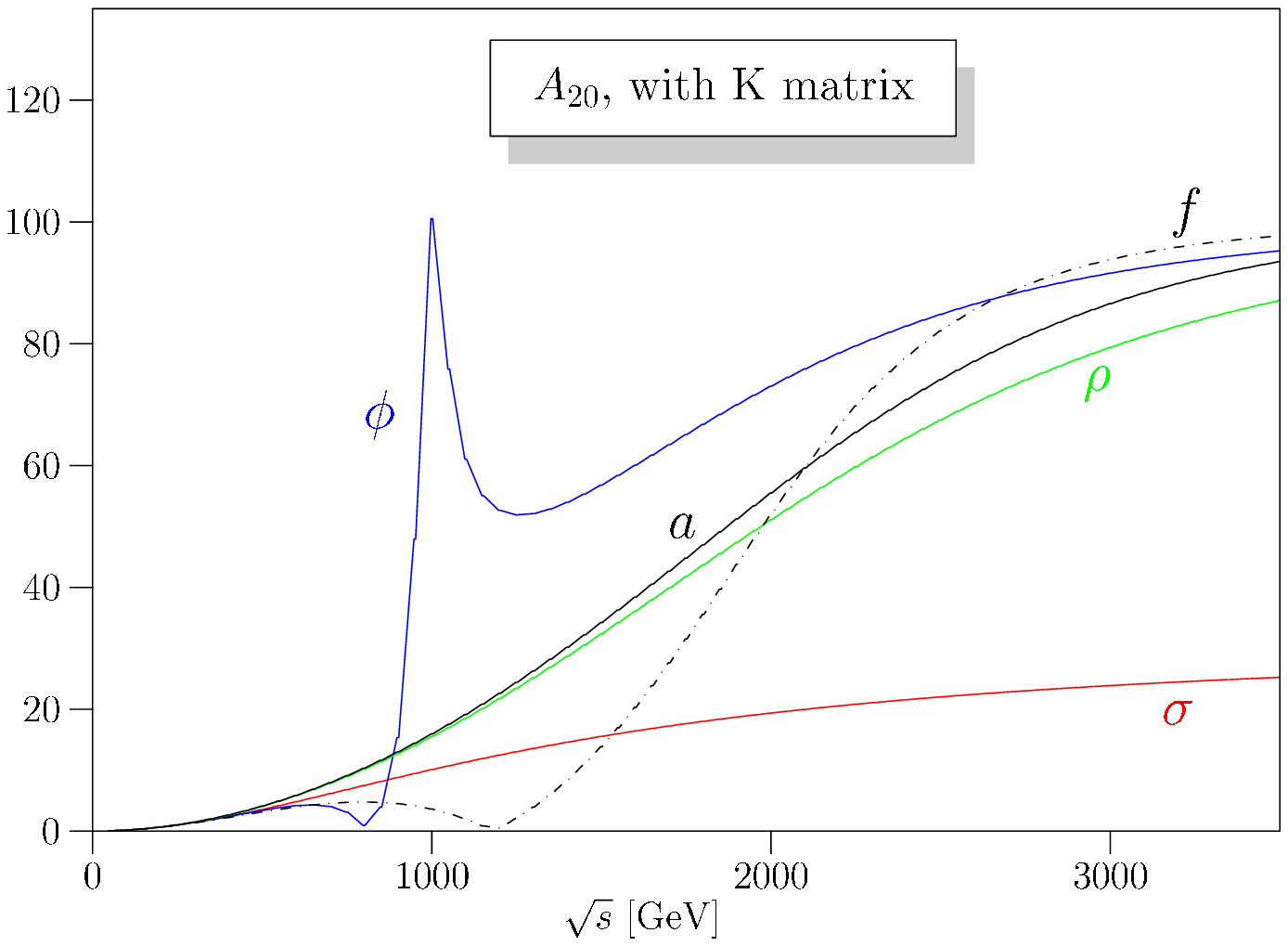}
    \hspace{3mm}
    \includegraphics[width=0.45\textwidth]{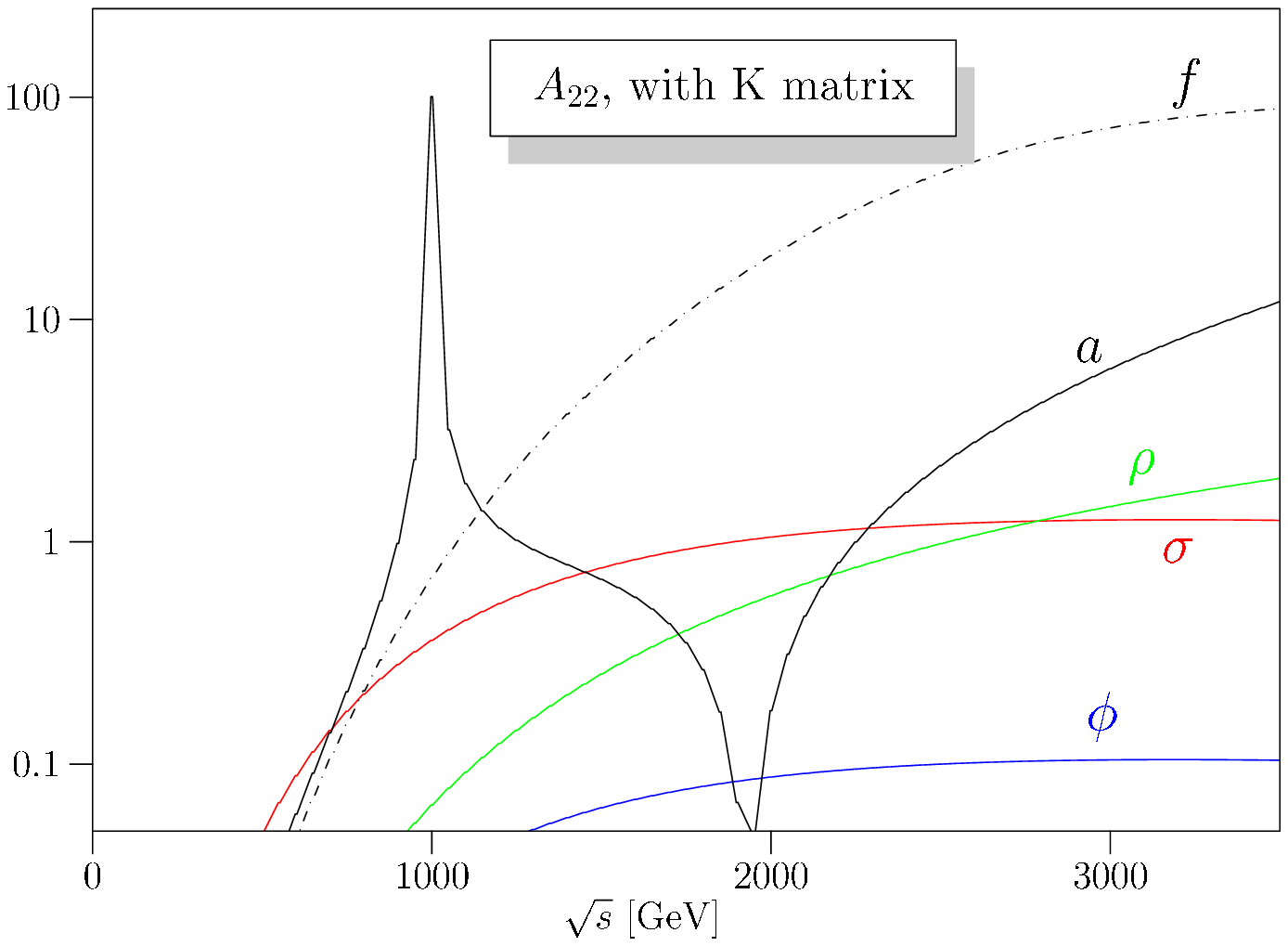}
  \end{center}
  \caption{Unitarized spin-isospin-eigenamplitudes for Goldstone-boson
    scattering.  In each plot, we display the eigenamplitudes for a
    definite spin and isospin value, one curve for each of the five
    possible resonances $\sigma,\phi,\rho,f,a$.  The resonance masses
    are fixed at $1\;\TeV$, and their couplings to Goldstone bosons
    have been set to unity.}
  \label{fig:eigenamp}
\end{figure}

In Fig.~\ref{fig:eigenamp} we draw the absolute values of the
resulting unitarized eigenamplitudes, including the LET part
$A_{IJ}^{(0)}$~(\ref{eigenamp-LET}).  Since the resonances have
definite spin and isospin quantum number assignments, each plot
contains exactly one curve with a resonance, while the other curves
are non-resonant.  The resonance masses $M_R$
($R=\sigma,\phi,\rho,f,a$) have been set to $1\;\TeV$, and the
couplings $g_R$ to unity.  The unitarization prescription smoothly
cuts off the amplitudes, so their absolute values do not exceed the
limit $32\pi\approx 100$.  Some of the amplitudes (e.g.,
$A_{00}^\rho$) contain terms rising like a power with the energy and
eventually saturate this bound, while others (e.g., $A_{13}^\rho$)
rise logarithmically at most, so at accessible energies they stay much
below this limit.

For the scalar isosinglet $\sigma$, the choice $g_\sigma=1$
corresponds to the SM with a heavy Higgs.  In this case, unitarity is
restored already by the scalar resonance exchange. Hence, as long as
$M_\sigma$ is below about $1.2\;\TeV$, the asymptotic values of all
$A_{IJ}^\sigma$ stay below the limit of $32 \pi$.  At tree level, they
are constants that depend on the ratio $M_\sigma^2/v^2$.  This is
slightly modified by loop corrections and by the unitarization
prescription.  For $g_\sigma\neq 1$, the cancellations are incomplete,
and the amplitudes $A_{IJ}^\sigma$ behave in the same way as the
other amplitudes.

Several of the curves exhibit a zero, which in the logarithmic plots
manifests itself as a sharp down-pointing spike.  In fact, in our
parameterization this happens for all resonant amplitudes, with the
exception of the SM Higgs case.  The reason is negative interference
between the resonant propagator and the contact term; the latter is
necessary for satisfying the LET and rises with a higher power of the
energy.  For vector resonances, cancellation typically occurs at very
high energies (above $10\;\TeV$), while for tensor resonances the
effect is visible in the energy range that we have chosen for our
plots.  However, if such a zero occurs beyond the resonance mass, it
should not be taken seriously, because in this range the amplitude
contains further, undetermined contributions, and the energy behavior
of the contact term as given by our formulae is not a prediction.
Only if this zero appears below the resonance a dip should actually be
expected.  This is the case for $A_{20}$ in the presence of a scalar
isoquintet.

\begin{figure}%[hbt]
  \begin{center}
    \includegraphics[width=0.6\textwidth]{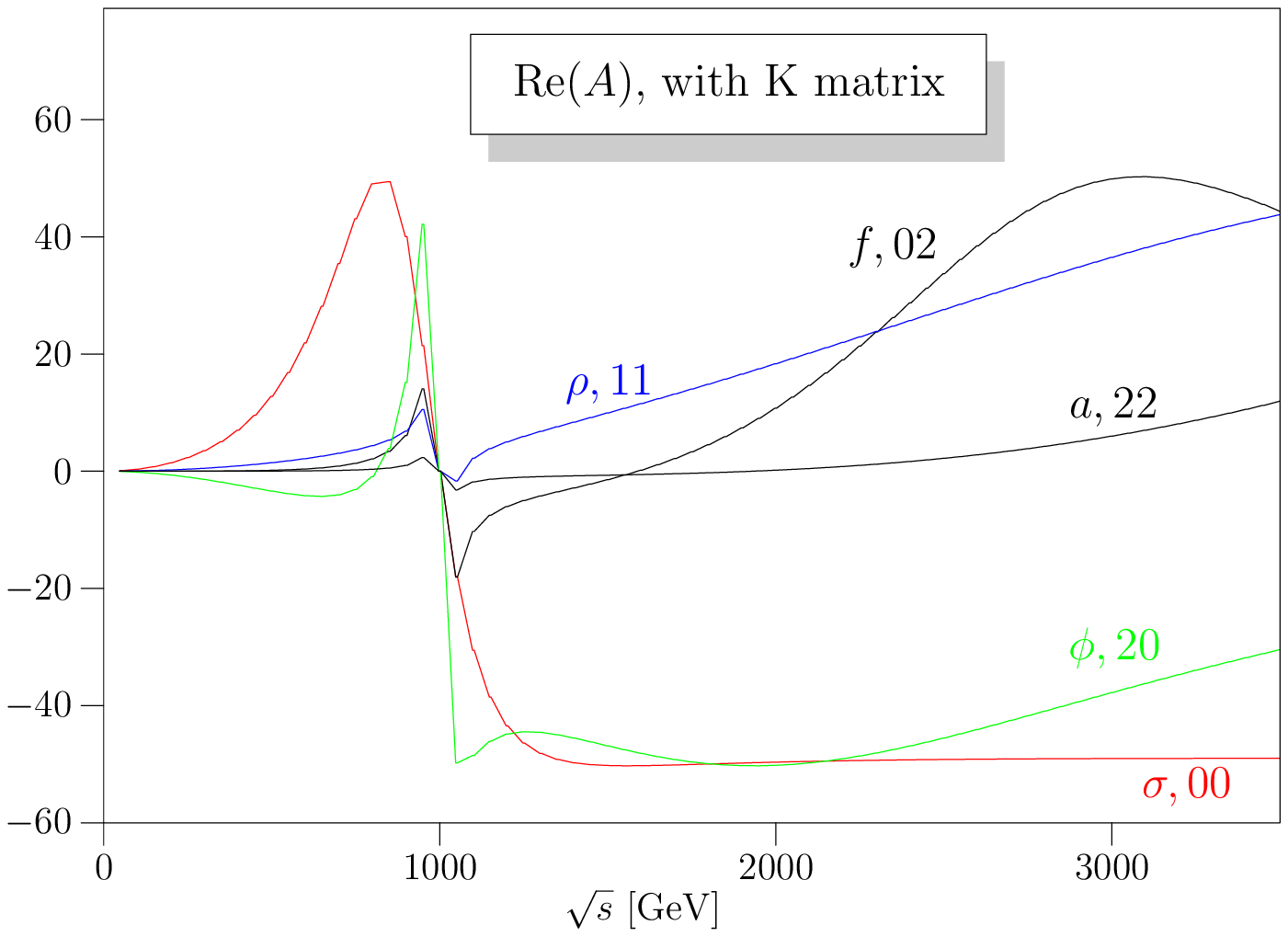}
  \end{center}
  \caption{Real part of the eigenamplitudes $|A_{IJ}(s)|$, each with
    the corresponding resonance(s) switched on; $M_R=1\;\TeV$.}
  \label{fig:real}
\end{figure}

The analytic behavior of the amplitudes is transparent if we plot the
real part, which vanishes on a resonance.  This is illustrated in
Fig.~\ref{fig:real}.  All curves cross zero at $1\;\TeV$, the
resonance mass.  Beyond this, they rise and asymptotically approach
zero again.  This is the resonance at infinity generated by the
unitarization procedure.  The exception is the $\sigma$ resonance
which approaches a constant, since in this model (the SM) there is no
unitarity problem.

\begin{figure}%[hbt]
  \begin{center}
    \includegraphics[width=0.6\textwidth]{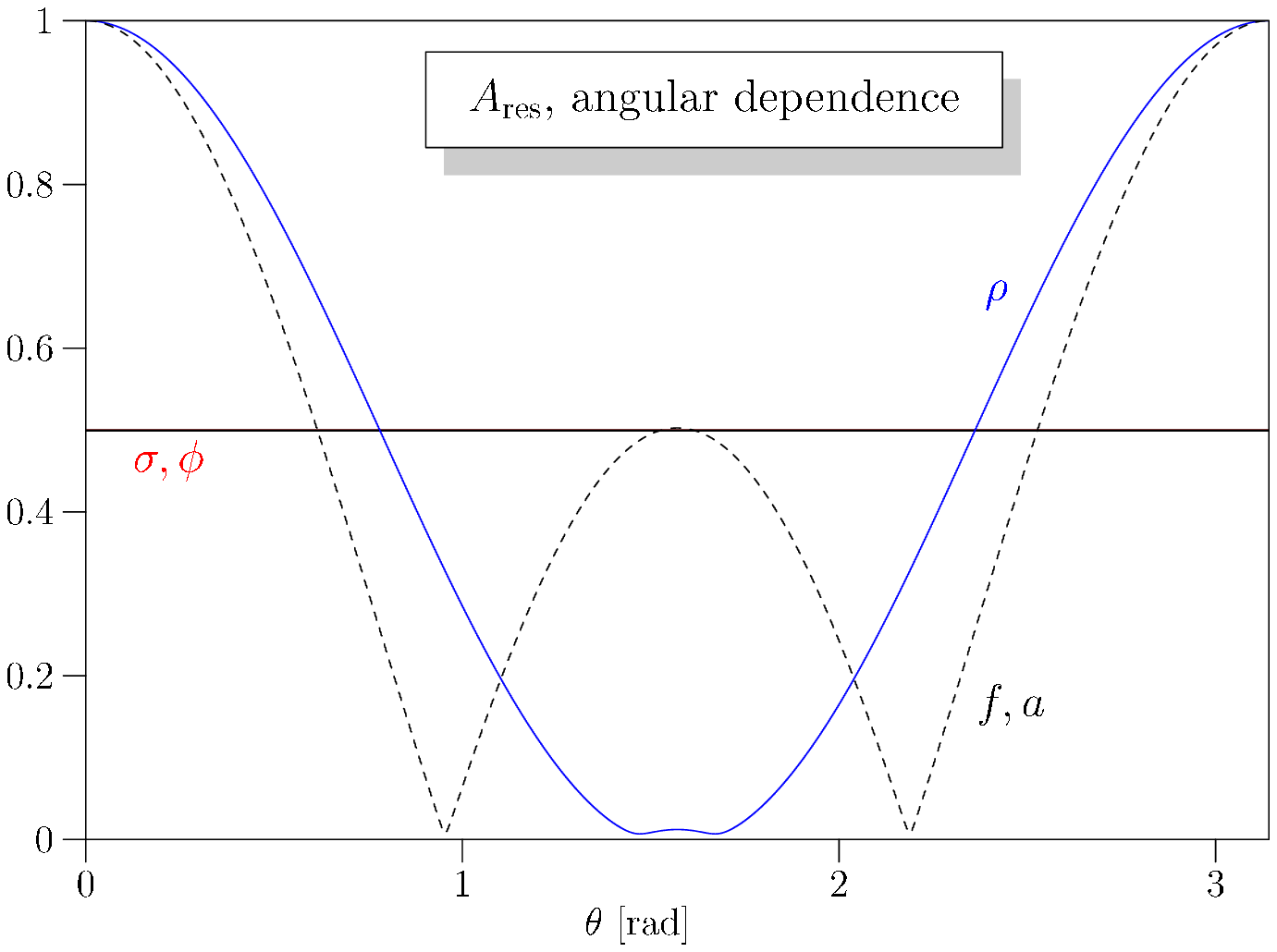}
  \end{center}
  \caption{Angular dependence of the amplitudes $|A_I(s,t,u)|$ for
    $I=0,1,2$, each with the corresponding resonance(s) switched on and
    evaluated at $\sqrt{s}$ equal to the resonance mass.}
  \label{fig:angle}
\end{figure}

For a concrete Monte-Carlo implementation, we need the unitarized
amplitudes for physical states, e.g., $w^+w^-$, $zz$, etc.  Therefore,
we first translate the spin-isospin eigenamplitudes back into
corrections to the isospin eigenamplitudes as functions of $s,t,u$,
\begin{subequations}
\begin{align}
  \Delta A_0(s,t,u) &= 
  \Delta A_{00}(s)\,P_0(s,t,u) + \Delta A_{02}(s)\,5P_2(s,t,u),
\\
  \Delta A_1(s,t,u) &= 
  \Delta A_{11}(s)\,3P_1(s,t,u) + \Delta A_{13}(s)\,7P_3(s,t,u),
\\
  \Delta A_2(s,t,u) &= 
  \Delta A_{20}(s)\,P_0(s,t,u) + \Delta A_{22}(s)\,5P_2(s,t,u),
\end{align}
\end{subequations}
The result is shown in Fig.~\ref{fig:angle}.  The plot clearly
exhibits the characteristic angular dependence of the resonances with
$J=0,1,2$, respectively, while the continuum background that we have
included is negligible for $s=M^2$.  The nonresonant part is
important, however, to describe the off-peak amplitude behavior.

This, in turn, is translated into corrections to the individual
scattering amplitudes,
\begin{subequations}
\begin{align}
  \label{Delta-A-wwzz}
  \Delta A(w^+w^-\to zz)
  &=
  \frac13\Delta A_0(s,t,u)-\frac13\Delta A_2(s,t,u)
\\
  \label{Delta-A-wzwz}
  \Delta A(w^+z\to w^+z)
  &=
  \frac12\Delta A_1(s,t,u)+\frac12\Delta A_2(s,t,u)
\\
  \label{Delta-A-wwww}
  \Delta A(w^+w^-\to w^+w^-)
  &=
  \frac13\Delta A_0(s,t,u) 
  + \frac12\Delta A_1(s,t,u)
  + \frac16\Delta A_2(s,t,u)
\\
  \label{Delta-A-wwww1}
  \Delta A(w^+w^+\to w^+w^+)
  &=
  \Delta A_2(s,t,u)
\\
  \label{Delta-A-zzzz}
  \Delta A(zz\to zz)
  &=
  \frac13\Delta A_0(s,t,u)+\frac23\Delta A_2(s,t,u)
\end{align}
\end{subequations}
Unitarization breaks crossing symmetry, since it is applied only in
the s-channel.  Explicitly, we obtain
\begin{subequations}
\begin{align}
  \Delta A(w^+w^-\to zz)
  &= 
%%   \frac{s}{v^2}
%%   \nonumber\\ &\quad
%%   +
  8\left[
    \alpha_5 
    + \frac{v^4}{24s^2}\left(\Delta A_{00}(s) - \Delta A_{20}(s) \right) 
    - \frac{5v^4}{12s^2}\left(\Delta A_{02}(s) - \Delta A_{22}(s)\right)
  \right]
  \frac{s^2}{v^4}
  \nonumber\\ &\quad
  +
  4\left[
    \alpha_4 
    + \frac{5v^4}{4s^2}\left(\Delta A_{02} - \Delta A_{22}
    \right)
  \right]
  \frac{t^2+u^2}{v^4}
  \label{a-wwzz}
\\
  \Delta A(w^+z\to w^+ z)
  &=
%%   \frac{t}{v^2}
%%   \nonumber\\ &\quad
%%   +
  4\left[
    \alpha_4 
    + \frac{v^4}{8s^2}\Delta A_{20}(s)
    - \frac{5v^4}{4s^2}\Delta A_{22}(s)
  \right]
  \frac{s^2}{v^4}
  \nonumber\\ &\quad
  +
  8\left[
    \alpha_5 
    - \frac{3v^4}{16s^2}\Delta A_{11}(s)
    + \frac{15v^4}{16s^2}\Delta A_{22}(s)
  \right]
  \frac{t^2}{v^4}
  \nonumber\\ &\quad
  +
  4\left[
    \alpha_4 
    + \frac{3v^4}{8s^2}\Delta A_{11}(s)
    + \frac{15v^4}{8s^2}\Delta A_{22}(s)
  \right]
  \frac{u^2}{v^4}
\\
  \Delta A(w^+w^-\to w^+w^-)
  &=
%%   -\frac{u}{v^2}
%%   \nonumber\\ &\quad
%%   + 
  4\left[
    \alpha_4 + 2\alpha_5
    + \frac{v^4}{24s^2}\left(2\Delta A_{00}(s) + \Delta A_{20}(s)\right)
  \right.\nonumber\\ &\qquad\qquad\qquad\qquad\left.
    - \frac{5v^4}{12s^2}\left(2\Delta A_{02}(s) + \Delta A_{22}(s)\right)
  \right]
  \frac{s^2}{v^4}
  \nonumber\\ &\quad
  + 
  4\left[
    \alpha_4 + 2\alpha_5
    + \frac{v^4}{8s^2}\left(10\Delta A_{02}(s) - 3 \Delta A_{11}(s)
      + 5\Delta A_{22}(s)\right)
  \right]
  \frac{t^2}{v^4}
  \nonumber\\ &\quad
  + 
  8\left[
    \alpha_4
    + \frac{v^4}{16s^2}\left(10\Delta A_{02}(s) + 3\Delta A_{11}(s)
      + 5\Delta A_{22}(s)\right)
  \right]
  \frac{u^2}{v^4}
\\
  \Delta A(w^+w^+\to w^+w^+)
  &=
%%   -\frac{s}{v^2}
%%   \nonumber\\ &\quad
%%   + 
  8\left[
    \alpha_4
    + \frac{v^4}{8s^2}\left(\Delta A_{20}(s) - 10\Delta A_{22}(s)\right)
  \right]
  \frac{s^2}{v^4}
  \nonumber\\ &\quad
  +
  4\left[
    \alpha_4 + 2\alpha_5
    + \frac{15v^4}{4s^2}\Delta A_{22}(s)
  \right]
  \frac{t^2+u^2}{v^4}
\\
  \Delta A(zz\to zz)
  &=
  8\left[
    \alpha_4 + \alpha_5
    + \frac{v^4}{24s^2}\left(\Delta A_{00}(s) + 2\Delta A_{20}(s)\right)
    \right.\nonumber\\ &\qquad\qquad\qquad\left.
    - \frac{5v^4}{12s^2}\left(\Delta A_{02}(s) + 2\Delta A_{22}(s)\right)
  \right]
  \frac{s^2}{v^4}
  \nonumber\\ &\quad
  + 8\left[
    \alpha_4 + \alpha_5
    + \frac{5v^4}{8s^2}\left(\Delta A_{02}(s) + 2\Delta A_{22}(s)\right)
  \right]
  \frac{t^2 + u^2}{v^4}
  \label{a-zzzz}
\end{align}
\end{subequations}
Here, the coefficients functions $\Delta A_{IJ}(s)$ are determined by
decomposing the results from Sec.~\ref{sec:eigenamplitudes} according
to (\ref{pole decomposition}) and inserting this into the
unitarization formula~(\ref{unitarization}).

%%%%%%%%%%

Adding the above correction terms to the LET scattering amplitudes
(\ref{A-LET-wwzz}--\ref{A-LET-zzzz}), we have a complete and unitary
parameterization of on-shell Goldstone scattering.  The
parameterization depends on $\alpha_4$ and $\alpha_5$, on a
renormalization scale $\mu$, and on the mass and width parameters of
the five possible resonances.

%%%%%

\subsection{Off-shell Implementation}

For realistic calculations, we want to transform the unitarized
Goldstone scattering amplitudes into matrix elements for off-shell
weak-boson scattering.  We first note that the complete SM without a
Higgs and without anomalous couplings, already yields the LET result
for weak-boson scattering.  In order to avoid double-counting, we
therefore have to remove the LET part from any extra contributions
that we add to the theory.  This is achieved by considering only the
correction terms (\ref{Delta-A-wwzz}--\ref{Delta-A-zzzz}) instead of
the complete unitarized amplitudes.

The chiral Lagrangian with NLO parameters (i.e., $\alpha_4$ and
$\alpha_5$) provides an off-shell formulation for the low-energy
effective theory.  We can determine Feynman rules and compute complete
matrix elements of $2\to 6$ fermion processes which include weak-boson
interactions with anomalous couplings.  The Feynman rules of
four-boson couplings depend on $\alpha_4$ and $\alpha_5$.  In
unitarity gauge, they are derived from the quartic gauge interactions
\begin{align}
\label{QGC}
  \LL_{QGC} &=
  e^2\left[ g_1^{\gamma\gamma} A^\mu A^\nu W^-_\mu W^+_\nu
           -g_2^{\gamma\gamma} A^\mu A_\mu W^{-\nu} W^+_\nu\right]
\nonumber\\ &\quad
  + e^2\frac{\cw}{\sw}\left[ g_1^{\gamma Z} A^\mu Z^\nu
                               \left(W^-_\mu W^+_\nu + W^+_\mu W^-_\nu\right)
              -2g_2^{\gamma Z} A^\mu Z_\mu W^{-\nu} W^+_\nu \right]
\nonumber\\ &\quad
  + e^2\frac{\cw^2}{\sw^2}\left[ g_1^{ZZ} Z^\mu Z^\nu W^-_\mu W^+_\nu
                  -g_2^{ZZ} Z^\mu Z_\mu W^{-\nu} W^+_\nu\right]
\nonumber\\ &\quad
  + \frac{e^2}{2\sw^2}\left[ g_1^{WW} W^{-\mu} W^{+\nu} W^-_\mu W^+_\nu
                       -g_2^{WW}\left(W^{-\mu} W^+_\mu\right)^2\right]
%\nonumber\\ &\quad
  + \frac{e^2}{4\sw^2\cw^4} h^{ZZ} (Z^\mu Z_\mu)^2,
\end{align}
where the SM values of the couplings\footnote{In these expressions,
the numerical values of the electroweak gauge couplings and the weak
mixing angle depend on the precise definition of the electroweak
renormalization scheme.}  are given by
\begin{equation}
  g_1^{VV'}=g_2^{VV'}=1
  \quad\text{($VV'=\gamma\gamma,\gamma Z, ZZ, WW$),}\qquad\qquad
  h^{ZZ} = 0.
\end{equation}
If we include the dependence on all five isospin-symmetric NLO chiral
parameters $\alpha_i$ (\ref{L1}--\ref{L5}), the deviations from the SM
values are
\begin{subequations}
\begin{align}
  \Delta g_1^{\gamma\gamma} &= \Delta g_2^{\gamma\gamma} = 0
&
  \Delta g_1^{\gamma Z} &= \Delta g_2^{\gamma Z}
                         = \frac{g^\pp}{\cw^2-\sw^2} \alpha_1
                           + \frac{g^2}{\cw^2}\alpha_3
\\
  \Delta g_1^{ZZ} &= 2\Delta g_1^{\gamma Z}
                     + \frac{g^2}{\cw^4}\alpha_4
&
  \Delta g_2^{ZZ} &= 2\Delta g_1^{\gamma Z}
                     - \frac{g^2}{\cw^4}\alpha_5
\\
  \Delta g_1^{WW} &= 2\cw^2\Delta g_1^{\gamma Z}
                     + g^2\alpha_4
&
  \Delta g_2^{WW} &= 2\cw^2\Delta g_1^{\gamma Z}
                     - g^2\left(\alpha_4 + 2\alpha_5\right)
\vphantom{\frac{g^2}{\cw^2}}
\\
  h^{ZZ} &= g^2\left(\alpha_4 + \alpha_5\right).
\end{align}
\end{subequations}

We can now construct a generic off-shell parameterization of
weak-boson scattering that corresponds to the unitary on-shell
Goldstone scattering
amplitudes~(\ref{Delta-A-wwzz}--\ref{Delta-A-zzzz}).  When the Feynman
rule for a given quartic gauge vertex is inserted in a physical
process, we replace the dependence on the constant parameters
$\alpha_4$ and $\alpha_5$ by form factors which depend on $s$.  For
$W^+W^-\to ZZ$ and $W^+W^+\to W^+W^+$, this is straightforward: the
two expressions $s^2/v^4$ and $(t^2+u^2)/v^4$ correspond to the two
different interaction terms for $WWZZ$ ($WWWW$) in the
Lagrangian~(\ref{QGC}), so we simply can replace the constants
$\alpha_4$ and $\alpha_5$ ($\alpha_4$ and $\alpha_4+2\alpha_5$) by the
full expressions (\ref{Delta-A-wwzz}, \ref{Delta-A-wwww1}),
respectively.  For the other processes, this assignment cannot be done
in the interaction Lagrangian, but it is obvious in the Feynman rule,
where each term $s^2/v^4$, $t^2/v^4$, and $u^2/v^4$ corresponds to a
definite combination of $g_{\mu\nu}$ Lorentz factors.

As a first result, we can compute on-shell scattering amplitudes for
physical $W$ and $Z$ bosons.  These combine the features of the chosen
resonance model with SM effects such as photon and $W/Z$ exchange.
Since on-shell initial vector bosons cannot be prepared in practice,
we defer this discussion to Appendix~\ref{app:onshell-scattering}.

Such an algorithm breaks crossing symmetry, but this is natural since
the unitarization scheme already breaks crossing symmetry.  In a
practical implementation, for a given vertex we implement all possible
orientations of the time arrow as alternatives, and determine the
orientation that is actually realized when we insert the vertex into a
physical process.  This is straightforward to do for an automatic
matrix-element generator.

Two sources for ambiguities appear in this construction.
(i) The GBET relates Goldstone scattering amplitudes to weak-boson
scattering amplitudes only in the high-energy limit, and only for
longitudinal polarization.  We do not specify couplings to transversal
gauge bosons, which are not directly related to EWSB and formally
subleading in the physics of strongly interacting weak bosons.
Corrections to the GBET therefore can be computed only up to further
free parameters.  Keeping this in mind, we translate the
Goldstone amplitudes to weak-boson amplitudes using the leading-order
GBET.\footnote{If we adopt the EWA, part of the difference w.r.t.\
complete amplitudes is formally of the same order as the GBET
corrections.  However, the EWA strongly affects kinematics and ignores
a large set of irreducible background diagrams, so the numerical
impact of this approximation for LHC analyses is much more important
than model-dependent ambiguities in corrections to the GBET.} (ii)
Strictly speaking, the Mandelstam variables $s,t,u$ in form factors
are defined for on-shell scattering of massless particles.  $t$, $u$
can be replaced by Lorentz factors which are unambiguous, but in the
off-shell continuation, the subenergy squared $s$ is evaluated for
massive off-shell $W/Z$ bosons.  This affects the unitarization corrections,
but these are scheme-dependent anyway.  Their main property -- to
cancel any unphysical rise of subamplitudes -- is preserved off-shell.
It also affects the location of resonance poles.  However, as
discussed in Sec.~\ref{sec:resonances}, off-shell effects in the
latter are accounted for by higher-dimensional operators and translate
into corrections to $\alpha_{4,5}$.  Finally, we recall that the
off-shell continuation of $W/Z$ propagators is controlled by
electroweak gauge invariance.  We keep $SU(2)_L\times U(1)_Y$ symmetry
manifest in the gauge and fermion sectors by using covariant
derivatives, so this is not an issue.

As a cross-check, we can compute $2\to 6$ fermion processes for the
ordinary SM with a Higgs boson.  In our parameterization, this is the
chiral Lagrangian with $\alpha_4=\alpha_5=0$ and a $\sigma$ resonance
with $g_\sigma=1$.  The form factors for the $WWZZ$, $WWWW$, and
$ZZZZ$ vertices contain exactly the Higgs propagator factors that we
would have obtained with the Higgs boson as an ordinary particle.  In
the $s$-channel, the propagator pole turns out to be regularized by a
running width $\Gamma\theta(s)\times s/M^2$, which is a sensible
treatment of the width of a heavy Higgs boson in SM scattering
amplitudes~\cite{Kilian:1998bh}.  So, despite the fact that we have
used the leading-order GBET, our off-shell formulation exactly reproduces the
tree-level SM result, both on-shell and off-shell.  The only missing
parts are double-Higgs and Higgs-fermion couplings (see
e.g.~\cite{silh}), but those couplings do not contribute to the
processes we are interested in.

%%%%%%%%%%%%%%%%%%%%%%%%%%%%%%%%%%%%%%%%%%%%%%%%%%%%%%%%%%%%%%%%%%%%%%%%

\section{LHC Processes}

\subsection{Monte-Carlo simulation}

We have implemented our parameterization of vector-boson scattering in
the multi-particle event generator \whizard~\cite{whizard,Omega}. The
program generates matrix elements for partonic processes via optimized
helicity amplitudes while avoiding the redundancies inherent in a
Feynman diagram expansion. These optimized matrix elements together
with a highly efficient phase-space setup enable the simulation of six
and eight-particle final states. \whizard\ contains the infra\-structure
for simulations of complex collider environments like structured
beams, parton shower, and interfaces to fragmentation and
hadronization. 

As the starting point for the implementation in \whizard, we have
chose the SM extension with anomalous three-boson and four-boson
couplings which has been used for the simulation of anomalous triple
and quartic gauge
operators~\cite{Beyer:2006hx,menges,anomaloussim}. The algorithm 
for the symbolic generation of the matrix elements in \whizard, which
is especially suited for the inclusion of beyond the Standard Model
(BSM) physics~\cite{omwhiz_bsm}, allows for the insertion of operators
in specific time directions necessary by the crossing-symmetry
breaking effects of the K-matrix unitarization prescription.

%%%%%

\subsection{Comparison with the Effective $W$ approximation (EWA)}

In $2\to 6$ fermion processes that contain weak-boson scattering
(Fig.~\ref{fig:signal}) the $W/Z$ bosons that initiate the interaction
are represented by their propagators with a spacelike momentum.  The
main contribution comes from the region with small virtuality, and we
are interested in the region of large c.m.\ energy of the vector boson
pair.  In this region, the virtualities and the masses of the vector
bosons induce only small corrections to the amplitude, so the initial
vector bosons can be treated as approximately on-shell.

We can thus approximate the dominant Feynman graphs by a convolution
of massless splitting (of the initial quark into a quark and a vector
boson) with the vector-boson interaction, which is called {\em
  effective} $W$ {\em approximation (EWA)}~\cite{EWA}:
\begin{equation}
  \label{sigma-convolution}
  \sigma(q_1q_2\to q_1'q_2'V_1'V_2') \approx
  \sum_{\lambda_1,\lambda_2}\int dx_1\,dx_2\,
  F^{\lambda_1}_{q_1\to q_1'V_1}(x_1)\,
  F^{\lambda_2}_{q_2\to q_2'V_2}(x_2)\,
  \sigma^{\lambda_1\lambda_2}_{V_1V_2\to V_1'V_2'}(x_1 x_2 s)
\end{equation}
This has to be convoluted with the quark structure functions to yield
the cross section for the $pp$ initial state. 

Eq.~(\ref{sigma-convolution}) contains integrations over $x_{1,2}$,
the energy fractions of the vector bosons that are radiated from the
initial quarks, and a sum over vector-boson helicities.  In contrast to
the analogous Weizs\"acker-Williams approximation for photons, there
is a longitudinal polarization direction in addition to the two
transversal polarization directions.  Explicitly, the structure
functions are
\begin{subequations}
\begin{align}
  \label{F+}
  F^{+}_{q\to q'V}(x) &= 
  \frac{1}{16\pi^2}\,\frac{(v_V-a_V)^2 + (v_V+a_V)^2\bar x^2}{x}
  \left[
  \ln\left(\frac{\ptmax^2 + \bar x m_V^2}{\bar x m_V^2}\right)
  -
  \frac{\ptmax^2}{\ptmax^2 + \bar x m_V^2}
  \right]
  \\
  \label{F-}
  F^{-}_{q\to q'V}(x) &= 
  \frac{1}{16\pi^2}\,\frac{(v_V+a_V)^2 + (v_V-a_V)^2\bar x^2}{x}
  \left[
  \ln\left(\frac{\ptmax^2 + \bar x m_V^2}{\bar x m_V^2}\right)
  -
  \frac{\ptmax^2}{\ptmax^2 + \bar x m_V^2}
  \right]
  \\
  \label{F0}
  F^0_{q\to q'V}(x) &= 
  \frac{v_V^2+a_V^2}{8\pi^2}\,\frac{2\bar x}{x}\,
  \frac{\ptmax^2}{\ptmax^2 + \bar x m_V^2}
\end{align}
\end{subequations}
with $\bar x\equiv 1-x$.  The vector and axial couplings for a fermion
branching into a $W$ are
\begin{align}
  v_W &= \frac{g}{2\sqrt 2},
& a_W &= \frac{g}{2\sqrt 2}.
\end{align}
For $Z$ emission, this is replaced by
\begin{align}
  v_Z &= \frac{g}{2\cos\theta_w}\left(t_3 - 2q\sin^2\theta_w\right),
& a_Z &= \frac{g}{2\cos\theta_w}t_3,
\end{align}
where $t_3=\pm\frac12$ is the fermion isospin, and $q$ its charge.

These structure functions depend on a transverse-momentum cutoff
$\ptmax$.  The kinematical limit for the cutoff is
\begin{equation}\label{ptmax}
  \ptmax \leq \bar x \sqrt{s}/2.
\end{equation}
In the derivation of (\ref{F+}--\ref{F0}), one integrates over
$p_\perp$ under the assumption that it is small compared to the
subprocess energy, so the subprocess cross section does not depend on
it.  For the longitudinal structure function that we are most
interested in, this can be justified because the limit
$\ptmax\to\infty$ is finite.  This structure function is concentrated
near $p_\perp=\bar xm_V$.  The transverse structure functions have a
logarithmic divergence in $\ptmax$, so the cutoff is needed there.
This already suggests that the EWA is more reliable for longitudinal
than for transversal vector bosons.

\begin{figure}[hbt]
  \begin{center}
    \includegraphics{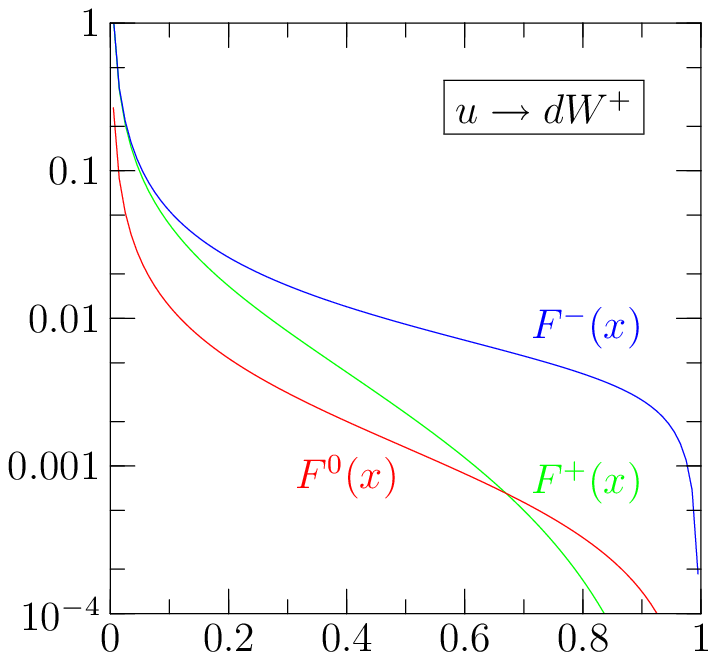}
    \includegraphics{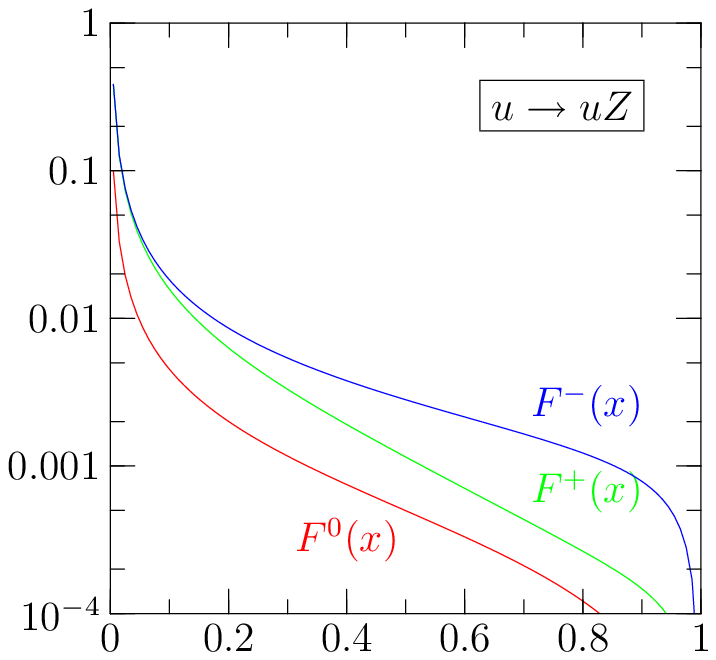}
  \end{center}
  \caption{EWA structure functions for $W$ (left) and $Z$ (right)
    emission from an up-type quark, for $\sqrt{s}=2\;\TeV$ and
    $\ptmax$ given by~(\ref{ptmax}).}
  \label{fig:strfun}
\end{figure}

In Fig.~\ref{fig:strfun}, we display the structure functions of $W$
and $Z$ bosons, separately for positive, longitudinal, and negative
helicity.  The emitting quark has been chosen to be an up-type quark;
for down-type quarks or electrons the $Z$ curves have to be
renormalized according to the respective charges.  For antiquarks or
positrons, the transverse polarizations have to be interchanged.  The
plots illustrate the fact that emission of a $W$ or $Z$, in particular
at high energies, is more likely for a transversally polarized vector
boson.  In effect, the production of longitudinally polarized $VV$
pairs which couple to the symmetry-breaking sector is suppressed
compared to this irreducible background.

%%%%%%

\begin{figure}[p]
  \begin{center}
    \includegraphics[width=0.45\textwidth]{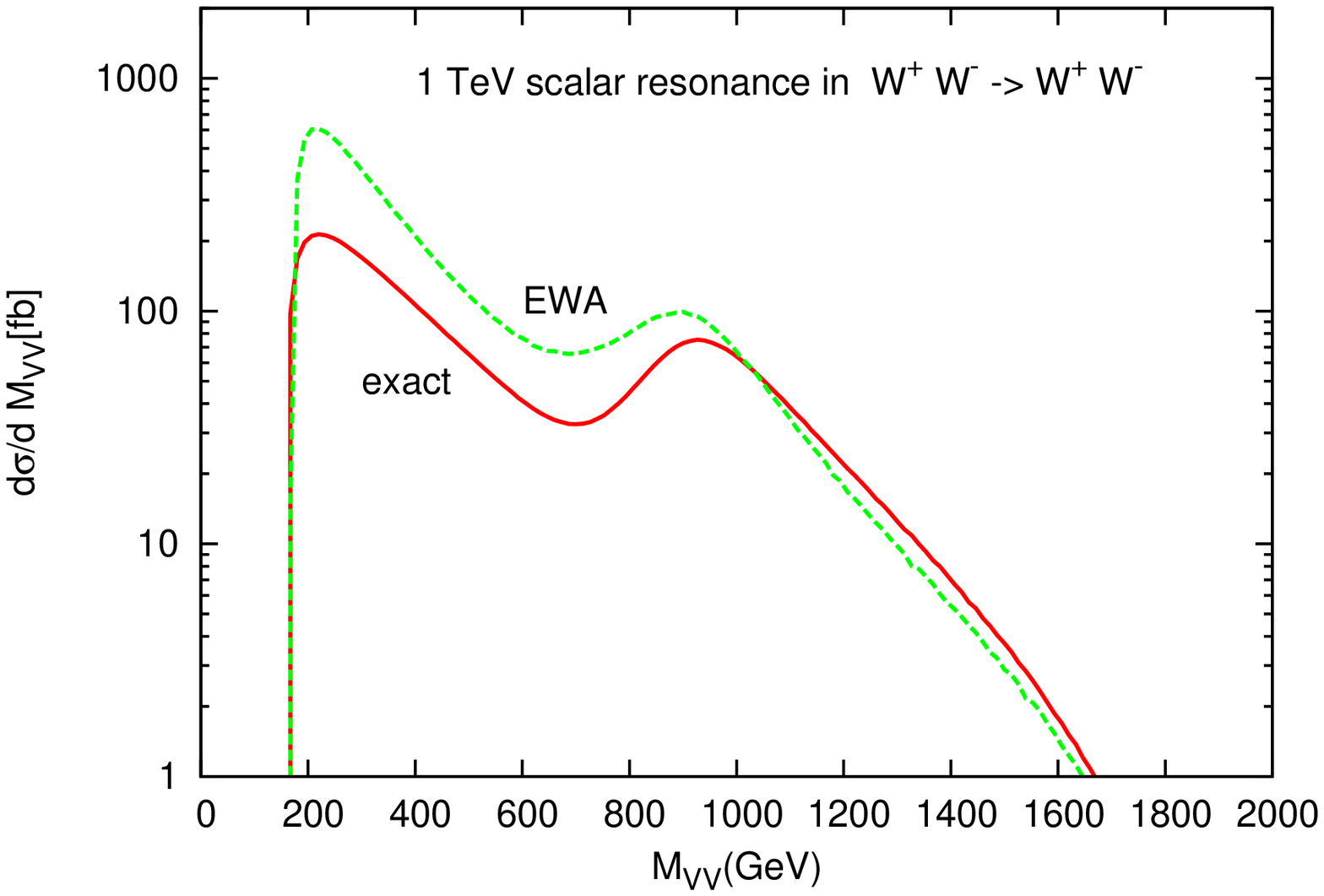}
    \includegraphics[width=0.45\textwidth]{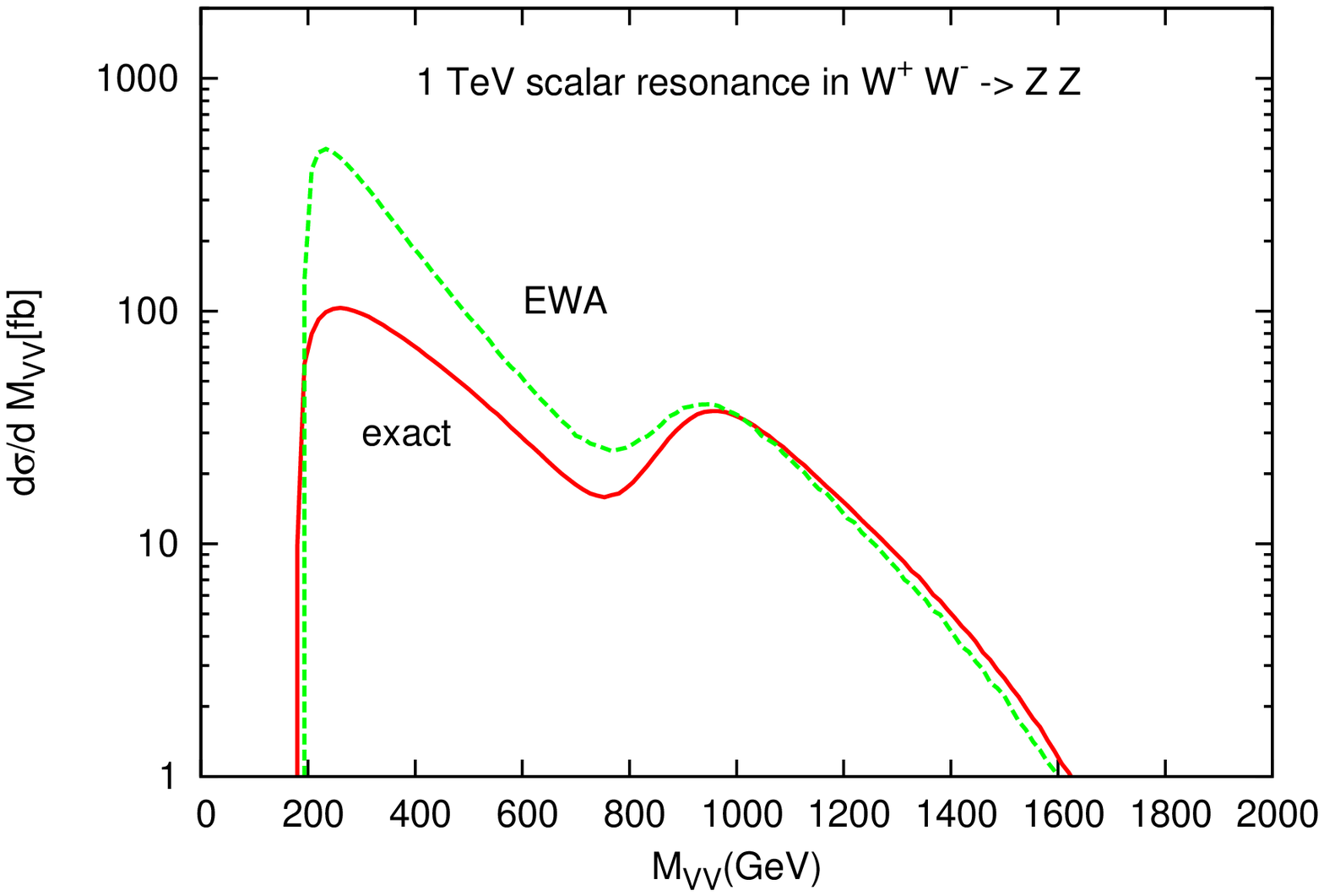}
    \\
    \includegraphics[width=0.45\textwidth]{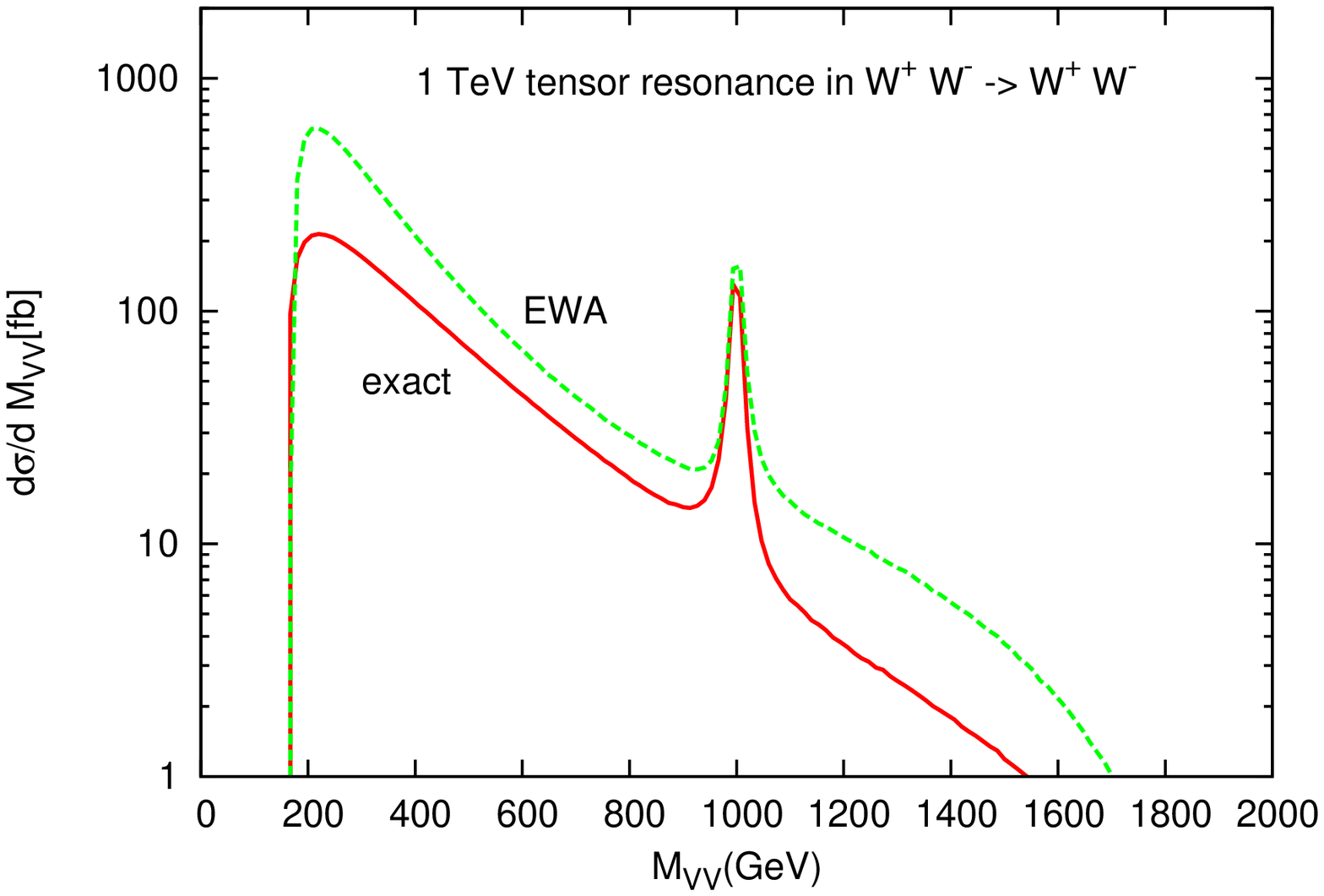}
    \includegraphics[width=0.45\textwidth]{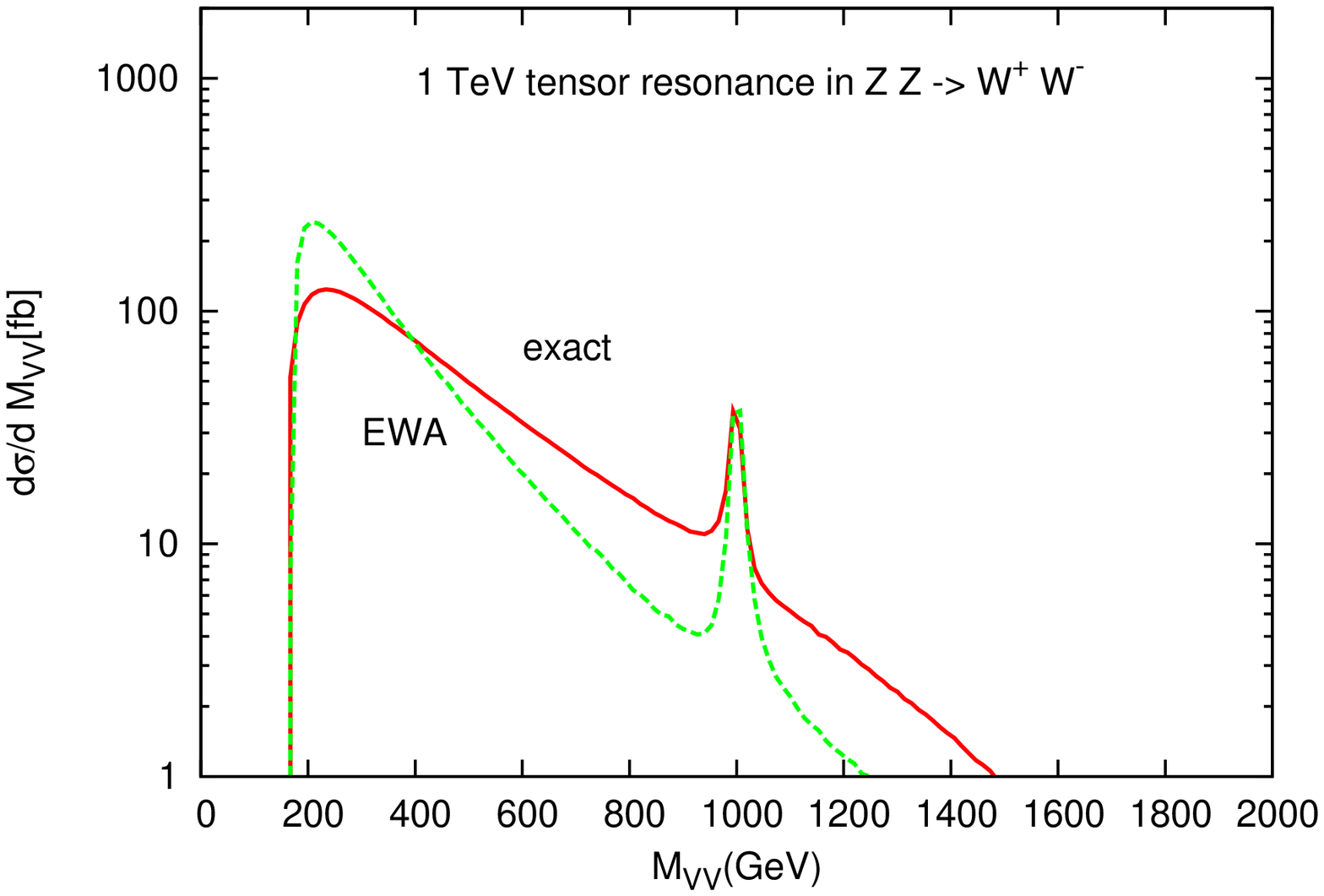}
  \end{center}
  \caption{Comparison of the exact (red) and EWA (green, dashed)
    results for weak-boson scattering for processes of the type
    $q_1q_2\to q_1'q_2'VV$ for $\sqrt{s}_{q_1q_2}=2\;\TeV$.  Upper
    line: scalar isosinglet resonance, lower line: tensor isosinglet
    resonance.  The resonance masses and couplings are $M_R=1\;\TeV$
    and $g_R=1$, respectively, the amplitudes are unitarized by
    the K-matrix scheme of Sec.~\ref{sec:unitarization}, and a $p_T$
    cut of $30\;\GeV$ has been applied to the vector bosons.}
    \label{fig:ewa_vs_exact}
  \medskip
  \begin{center}
    \includegraphics{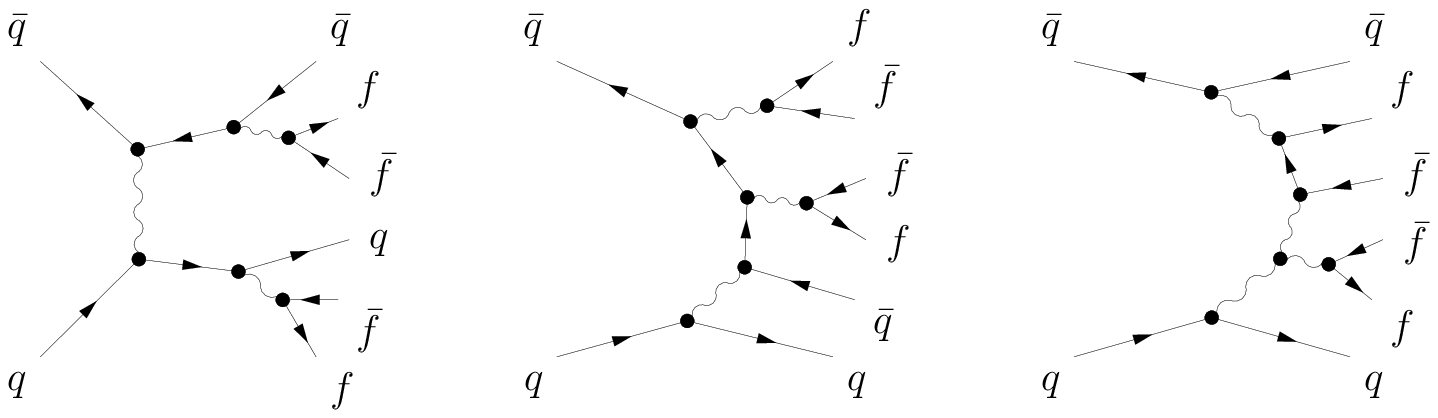}
  \end{center}
  \caption{Feynman graphs that contribute an irreducible background to
    weak-boson scattering in $2\to 6$ fermion
    processes. E.g. double final state and
    double initial state radiation, as well as $t$-channel like diagrams.}
  \label{fig:bg}
\end{figure}

Fig.~\ref{fig:ewa_vs_exact} exemplifies the differences between the
exact result for $qq\to qq+VV$ processes which contain resonant
weak-boson scattering.  To make a meaningful comparison, we first
recall that in the EWA the initial vector bosons are on-shell, while
in the exact process they are off shell.  The on-shell amplitudes have
a Coulomb singularity due to photon and $Z,W$ exchange.  In
particular, an on-shell cross section with photon exchange is
infinite, while $Z/W$ exchange yields a Coulomb peak proportional to
$\hat s^2/M_V^4$.  Here, $\hat s$ is the c.m.\ energy of the
vector-boson subsystem, equal to the invariant mass squared $M_{VV}^2$
of the outgoing vector bosons.  To reduce this effect which in the
exact result is regulated by the vector-boson virtuality, we cut the
$p_T$ of the outgoing vector bosons at $30\;\GeV$.

A particular choice of this cut allows us to approximate the
high-energy end of the $M_{VV}$ distribution for the SM with a heavy
Higgs (Fig.~\ref{fig:ewa_vs_exact}, top) quite well~\cite{EWA}.  This
is misleading, however: with the same cut, the prediction of the
tensor resonance case (Fig.~\ref{fig:ewa_vs_exact}, lower left) with
its unitarity saturation beyond the peak is considerably worse.  If we
are looking at $ZZ\to WW$ instead of $WW\to ZZ$, the EWA background
undershoots the exact value (Fig.~\ref{fig:ewa_vs_exact}, lower
right).  More importantly, while the peak can be approximated up to
better than a factor~$2$, the background is predicted with less accuracy.
Since $M_{VV}$ cannot be reconstructed experimentally (apart from $ZZ$
final states), so sideband subtraction is not possible, this
significantly affects the analysis.

Part of the deviation is due to the kinematical simplifications
inherent in the derivation, which can be improved in
principle~\cite{Johnson:1987tj,Boos:1997gw}.  Unfortunately, this only
marginally improves the EWA, since the main error comes from the
existence of irreducible background diagrams for on-shell vector boson
pair + jets production, and additional irreducible background for the
complete six-fermion process, cf.\ Fig.~\ref{fig:bg}, which cannot be
accounted for in this way.  Off-shell, those background diagrams are
connected to the signal diagrams by gauge invariance and cannot be
neglected: simply omitting them would disrupt detailed cancellations,
similar to the familiar s/t-channel cancellation in $W$ pair
production~\cite{Accomando:2006mc}.

%%%%%

\subsection{Complete Simulation}

The implementation of the off-shell continued amplitudes in the
Monte-Carlo generator \whizard\ allows us to get rid of the EWA and to
simulate event samples for the complete process $pp\to q q'+4f$, where
the four additional fermions are the decay products of the vector
bosons, or come from the irreducible background.  Using, e.g.,
\pythia\ for parton showering and fragmentation, this results in
physical LHC events that can be analyzed by detector simulation and
eventually compared to real data.  

\begin{figure}[p]
  \begin{center}
    \includegraphics[width=0.8\textwidth]{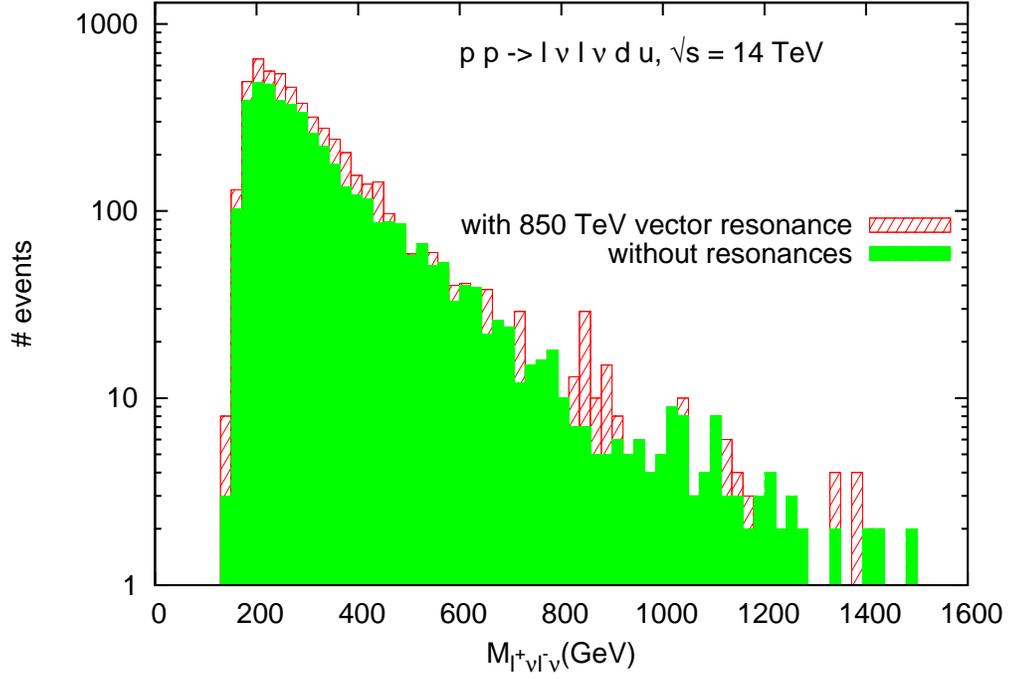}
    \\
    \includegraphics[width=0.8\textwidth]{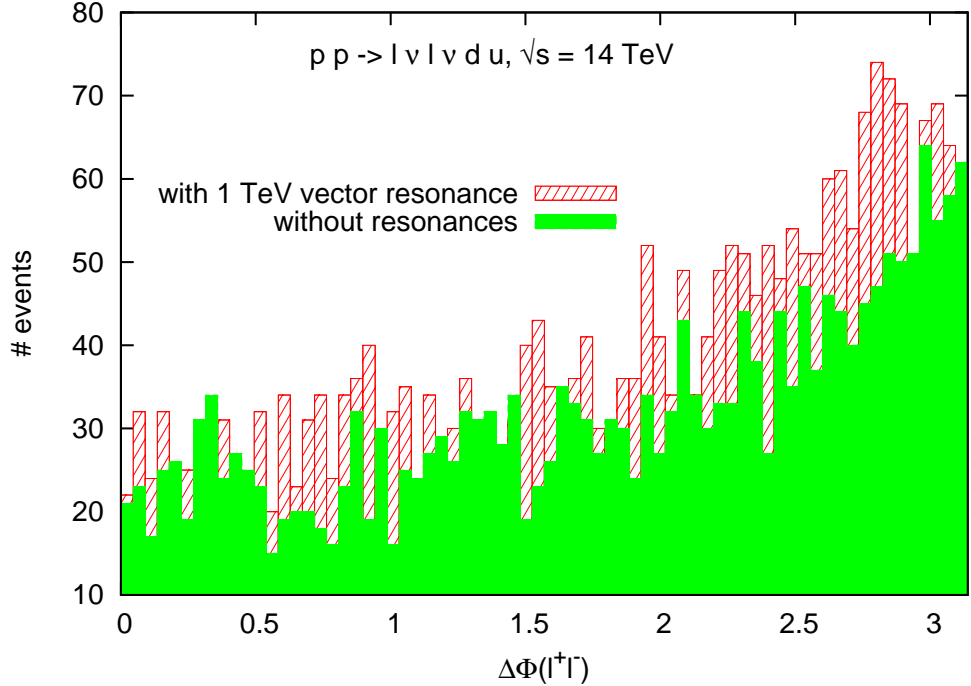}
  \end{center}
  \caption{Unweighted event samples for $pp\to ud + e^+\nu_e
    e^-\bar\nu_e$.  Upper: Invariant mass of the $e^+\nu_e
    e^-\bar\nu_e$ system.  Lower: Azimuthal distance of the charged
    leptons.  The red histogram (hatched) corresponds to a model with
    a vector resonance ($M_\rho=850\;\GeV$ and $g_\rho=1$).  The green
    histogram (filled) is the LET result.  Both models have been
    unitarized by the K-matrix scheme.  Cuts:
    $p_\perp(\ell\nu)>30\;\GeV$, $|\eta(\ell\nu)|<1.5$,
    $\theta(u/d)>0.5^\circ$.  The integrated luminosity is $4\times
    225\;\fb^{-1}$ (the factor $4$ accounts for the sum over
    $e,\mu$).}
  \label{fig:pp-sample}
\end{figure}

For illustration, in Fig.~\ref{fig:pp-sample} we present the result of
a parton-level simulation of $WW/ZZ$ scattering, using complete
six-fermion matrix elements.  In these plots, we compare the effect of
a $850\;\GeV$ vector resonance with the nonresonant (unitarized) LET
model, which serves as a reference model for the higgsless case.  In
the four-lepton invariant mass, the resonance is clearly visible.
However, this quantity is not an observable.  The azimuthal distance
of the two decay leptons is observable; there, vector-resonance
exchange in $s$- and $t$-channel leads to a significant excess.

A realistic study would be based on a sum over all possible final
states with parton shower and hadronization, using cuts and
distributions in observable quantities.  Furthermore, it would include
a complete account of background and detector effects.  A cut-based
analysis strategy was proposed in
Refs.~\cite{Bagger:1993zf,Bagger:1995mk}.  An ATLAS study that makes
use of the parameterization of the present paper is currently under
way~\cite{atlas-study}.

%%%%%%%%%%%%%%%%%%%%%%%%%%%%%%%%%%%%%%%%%%%%%%%%%%%%%%%%%%%%%%%%%%%%%%%%

\section{Summary and Conclusions}

We have described a generic approach to extrapolating vector-boson
scattering into the energy range where no perturbative predictions
exist.  Nontrivial features of the amplitudes are possible, which will
likely appear as resonances.  In addition to the classical alternative
of a heavy scalar-isoscalar (Higgs) or a vector-isovector (technirho
or $W'$) resonance, we account for scalar-isotensor resonances which
are present in extended models, and for tensor resonances that could,
for instance, be associated with gravity in extra dimensions.
Furthermore, we connect the model-dependent part to the
model-independent low-energy effective theory and keep this relation
transparent in the implementation.  Unitarization of the on-shell
amplitudes avoids the problem of unphysical behavior at the highest
energies that plagues a naive tree-level approach.

Our approach is economical in the number of free parameters, but
intended as a sufficiently general description of those energy regions
where the LHC will have sensitivity.  If necessary, refinements of the
models, such as recurring resonances or more exotic behavior of the
amplitudes, are straightforward to add.  The resulting amplitudes are
translated into effective form-factors for vector-boson vertices in
unitarity gauge.  This allows for an implementation in universal
Monte-Carlo event generators, which we have realized for the case of
the \whizard\ event generator.

While the leading electroweak loop corrections for vector-boson
scattering are included, QCD corrections are not yet implemented.
These have been considered in Ref.~\cite{Jager:2006zc} and should be
combined with the effects modeled by our approach.  

With the event generator at hand, model-independent studies and
analyses of vector-boson scattering, both in SM extensions and in
Higgsless models, become feasible.  No approximations beyond those
inherent in the modeling are involved, as it is essential for
unbiased data analysis.  A particular feature of our implementation is
the smooth transition to the SM case (with a Higgs boson) or,
alternatively, to a featureless LET model of strong $WW$ scattering
without resonances, respecting unitarity.  In data analysis, the
signal can be defined as the deviation with respect to either one of
those reference models.

%%%%%%%%%%%%%%%%%%%%%%%%%%%%%%%%%%%%%%%%%%%%%%%%%%%%%%%%%%%%%%%%%%%%%%%%

\subsection*{Acknowledgments}

We thank Michael Kobel, Wolfgang Mader, Markus Schumacher and
especially Jan Schumacher for their fruitful discussions and helpful
comments during the development of this work. JR was partially
supported by the Bundesministerium f\"ur Bildung und Forschung,
Germany, under Grant No. 05HA6VFB. During its early stage this
research was supported by the Helmholtz-Gemeinschaft under Grant
No.\ VH-NG-005.

%%%%%%%%%%%%%%%%%%%%%%%%%%%%%%%%%%%%%%%%%%%%%%%%%%%%%%%%%%%%%%%%%%%%%%%%

\clearpage
\appendix

\section{Conventions and algebra}
\label{app:notation}

\subsection{$SU(2)$ algebra}
\label{app:algebra}

%% We'll need some well-known formulae for $SU(2)$ algebra and traces.
%% The Pauli matrices are $\tau^1,\tau^2,\tau^3$.  The transformation to
%% the charge eigenbasis uses
%% \begin{align}
%%   \tau^1 &= \tau^+ + \tau^-       & \tau^+ &= \tfrac12(\tau^1 + \ii\tau^2)
%% \nonumber\\
%%   \tau^2 &= -\ii(\tau^+ - \tau^-) & \tau^- &= \tfrac12(\tau^1 - \ii\tau^2)
%% \end{align}
%% Commutators:
%% \begin{align}
%%   [\tau^+,\tau^-] &= \tau^3
%% &
%%   [\tau^3,\tau^+] &= 2\tau^+
%% &
%%   [\tau^3,\tau^-] &= -2\tau^-
%% \end{align}
%% Special traces:
%% \begin{align}
%%   \tr{\tau^+\tau^-} &= 1      & \tr{\tau^3\tau^3} &= 2
%% \end{align}
%% Generic traces ($a,b,c,d$ are linear combinations of Pauli matrices, i.e.,
%% $a = a^k\tau^k$ etc.):
%% \begin{align}
%%   \tr 1 &= 2
%% \\
%%   \tr{ab} &= 2(a\cdot b)
%% \\
%%   \tr{abcd} &= \tfrac12
%%                \left(\tr{ab}\tr{cd} - \tr{ac}\tr{bd} + \tr{ad}\tr{bc}\right)
%% \end{align}

Throughout this paper, we use boldface notation for objects that are
defined in the adjoint of $SU(2)$, e.g.,
\begin{equation}
  \vW_\mu = W_\mu^a\frac{\tau^a}{2}
\end{equation}
with the Pauli matrices $\tau^a$ ($a=1,2,3$), and summation over $a$
understood.

For describing isospin quintet resonances, we introduce tensor
products of Pauli matrices:
\begin{align}
  \tau^{++} &= \tau^+\otimes\tau^+ \\
  \tau^{+} &= \tfrac12(\tau^+\otimes\tau^3 + \tau^3\otimes\tau^+) \\
  \tau^0 &= \tfrac1{\sqrt{6}}(\tau^3\otimes\tau^3
            - \tau^+\otimes\tau^- - \tau^-\otimes\tau^+) \\
  \tau^{-} &= \tfrac12(\tau^-\otimes\tau^3 + \tau^3\otimes\tau^-) \\
  \tau^{--} &= \tau^-\otimes\tau^-
\end{align}
These are normalized:
\begin{equation}
  \tr{\tau^{++}\tau^{--}} =
  \tr{\tau^+\tau^-} =
  \tr{\tau^0\tau^0} = 1
\end{equation}
Isospin singlet:
\begin{align}
  \tau^{aa} &\equiv \tau^a\otimes\tau^a 
  = \tau^3\otimes\tau^3 + 2\tau^+\otimes\tau^- + 2\tau^-\otimes\tau^+
\end{align}
Tracing this with something else gives
\begin{equation}
  \tr{(A\otimes B)\tau^{aa}} = 2\tr{AB},
\end{equation}
in particular
\begin{gather}
  \tr{\tau^{++}\tau^{aa}} =
  \tr{\tau^{+}\tau^{aa}} =
  \tr{\tau^{0}\tau^{aa}} = 0,
\qquad
  \tr{\tau^{33}\tau^{aa}} = 4,
\qquad
  \tr{\tau^{aa}\tau^{bb}} = 12.
\end{gather}
Furthermore we need:
\begin{equation}
  \tr{\tau^0 (\tau^3 \otimes \tau^3)} = \frac{4}{\sqrt{6}} 
    \qquad\qquad
    \tr{\tau^0 (\tau^+ \otimes \tau^- + \tau^- \otimes \tau^+)} = -
    \frac{2}{\sqrt{6}} 
\end{equation}

%%%%%%%%%%

\subsection{Goldstone bosons and Gauge fields}
\label{app:gauge}

We define the Goldstone scalar triplet $w^{1,2,3}$ or, alternatively,
$w^+,w^-,z$ such that
\begin{align}
  w^1 &= \tfrac1{\sqrt2}(w^+ + w^-)    & w^+ &= \tfrac1{\sqrt2}(w^1 - \ii w^2)
\nonumber\\
  w^2 &= \tfrac\ii{\sqrt2}(w^+ - w^-)  & w^- &= \tfrac1{\sqrt2}(w^1 + \ii w^2)
\end{align}
and $w^3 = z$.
Contractions:
\begin{align}
  \vw \equiv w^a\tau^a &= \sqrt2(w^+\tau^+ + w^-\tau^-) + z\tau^3
\\
  (\vw)^2 &= w^+w^- + w^-w^+ + zz
\end{align}
The Higgs-field matrix is given by
\begin{equation}
  \Sigma = \exp\left(-\frac{\ii}{v}\vw\right)
\end{equation}
The covariant derivative of the Higgs field is
\begin{equation}
  D\Sigma = \pd\Sigma + \ii g\vW\Sigma 
                        - \ii g'\Sigma \left(B\frac{\tau^3}{2}\right),
\end{equation}
Unitary gauge would mean $\vw\equiv 0$, i.e., $\Sigma\equiv
1$. Herewith, we define the vector field
\begin{equation}
  \vV = \Sigma(D\Sigma)^\dagger = -(D\Sigma)\Sigma^\dagger \; ,
\end{equation}
which is in the adjoint representation of $SU(2)_L$, and is
a linear combination of Pauli matrices. Hence, $\tr\vV = 0$.
Note that $\vV$ is antihermitian, $\vV^\dagger = -\vV$.

Gauge fields for the electroweak and strong interactions are defined
such that they transform under $SU(2)_L\times U(1)_Y$ as $\vW\to
U_L\vW U_L^\dagger$,
\begin{align}
  \vW_{\mu\nu} &= \pd_\mu\vW_\nu - \pd_\nu\vW_\mu + ig[\vW_\mu, \vW_\nu], \\
  \vB_{\mu\nu} &= \Sigma\left(\pd_\mu B_\nu - \pd_\nu B_\mu\right)
                  \frac{\tau^3}{2}\Sigma^\dagger;
\end{align}
furthermore there is the QCD gauge field
$\vG_{\mu\nu}=G_{\mu\nu}^a\frac{\lambda^a}{2}$.

In the gaugeless limit, the expansion in terms of Goldstone fields is
$\vV \Rightarrow \frac{\ii}{v} \pd w^k \tau^k + O(v^{-2})$. Expressing
this in terms of charge eigenstates, we derive
\begin{align}
  \vV =&\; \phantom{-} \frac{\ii}{v} \left[
    \sqrt2 \pd w^+ \tau^+ + \sqrt{2} \pd w^- \tau^- 
    + \pd z \tau^3 \right] \notag \\
  &\; - \frac{1}{v^2} \left[ \sqrt{2} w^+ \lrpd z \tau^+ - \sqrt{2}
    w^- \lrpd z \tau^- - w^+ \lrpd w^- \tau^3 \right] + O(v^{-3})
\intertext{and thus}
  \tr{\vV_\mu\vV_\nu}
  &= -\frac{2}{v^2}\left(\pd_\mu w^+ \pd_\nu w^- + \pd_\mu w^- \pd_\nu w^+
                         + \pd_\mu z \pd_\nu z\right)
     + O(v^{-3})
\end{align}
Hence,
\begin{align}
  - \frac{v^2}{4} \tr{\vV\cdot\vV} =&\quad \pd w^+ \pd w^- + \frac12 \pd z
    \pd z 
\end{align}
In the notation used for couplings to isospin quintets, we have
\begin{align}
  \frac12 \vV_{\left\{\mu\right.} \otimes \vV_{\left.\nu\right\}} =&\; 
  - \frac{1}{v^2} \biggl\{ 2 \pd_\mu w^+ \pd_\nu w^+ \tau^{++} + 2
  \pd_\mu w^- \pd_\nu w^- \tau^{--} \notag 
  \\ & \; \phantom{ - \frac{1}{v^2} \biggl\{ }
  + \sqrt{2} \left( \pd_\mu w^+ \pd_\nu z + \pd_\nu w^+ \pd_\mu z
  \right) \tau^+ + \sqrt{2} \left( \pd_\mu w^- \pd_\nu z 
  + \pd_\nu w^- \pd_\mu z \right) \tau^-  \notag 
  \\ & \; \phantom{ - \frac{1}{v^2} \biggl\{ }
  + \pd_\mu z \pd_\nu z \tau^3 \otimes \tau^3 
  + \left( \pd_\mu w^+ \pd_\nu w^- + \pd_\mu w^- \pd_\nu w^+ \right)
  (\tau^+ \otimes \tau^- + \tau^- \otimes \tau^+) 
  \biggr\}
\end{align}
And,
\begin{align}
  \vV_\mu \otimes \vV^\mu =&\; 
  - \frac{1}{v^2} \biggl\{ 2 \pd w^+ \cdot \pd w^+ \tau^{++} + 2
  \pd w^- \cdot \pd w^- \tau^{--} \notag 
  %% \\ & \; \phantom{ - \frac{1}{v^2} \biggl\{ }
  + 2 \sqrt{2} \pd w^+ \cdot \pd z \tau^+ 
  + 2 \sqrt{2} \pd w^- \cdot \pd z \tau^-  \notag 
  \\ & \; \phantom{ - \frac{1}{v^2} \biggl\{ }
  + \pd z \cdot \pd z \tau^{33} 
  + 2 \pd w^+ \cdot \pd w^- \left( \tau^{+-} + \tau^{-+} \right) 
  \biggr\}
\end{align}

%%%%%%%%%%%%%%%%%%%%
\subsection{Tensor Fields}
\label{app:tensor}

A massive tensor field $f^{\mu\nu}$ is subject to the conditions
\begin{align}
  f^{\mu\nu} &= f^{\nu\mu},
&
  f^\mu{}_\mu &= 0
&
  \pd_\mu f^{\mu\nu} = \pd_\nu f^{\mu\nu} = 0.
\end{align}
Its spin sum is given by
\begin{equation}\label{tensor-spinsum}
  \sum_\lambda\epsilon^*_\lambda{}^{\mu\nu}\epsilon_\lambda^{\rho\sigma}
%  = T^{\mu\nu,\rho\sigma}
  = \frac12\left(P^{\mu\rho}P^{\nu\sigma} +  P^{\mu\sigma}P^{\nu\rho}\right)
  -\frac13\left(P^{\mu\nu}P^{\rho\sigma}\right),
\end{equation}
where
\begin{equation}
  P^{\mu\nu}(k) = g^{\mu\nu} - \frac{k^\mu k^\nu}{M^2}.
\end{equation}
The free Lagrangian is
\begin{equation}
  \LL_f = \LL_{\rm kin} - \frac{M^2}{2}f_{\mu\nu} f^{\mu\nu}
\end{equation}
where the kinetic part corresponds to the spin
sum~(\ref{tensor-spinsum}).

%%%%%%%%%%%%%%%%%%%%

\subsection{Integrals in spin-isospin eigenamplitudes}
\label{app:integrals}

To get compact expressions for the spin-isospin eigenamplitudes, we
define the following integrals:
\begin{subequations}
\begin{align}
  \cS_J(s) &= \int_{-s}^0\frac{dt}{s} \frac{t^2}{t-M^2} P_0(t,s,u) P_J(s,t,u)
\\
  \cP_J(s) &= \int_{-s}^0 \frac{dt}{s} \frac{t}{t-M^2} P_1(t,s,u) P_J(s,t,u)
\\
  \cD_J(s) &= \int_{-s}^0 \frac{dt}{s} \frac{t^2}{t-M^2} P_2(t,s,u) P_J(s,t,u)
\end{align}
\end{subequations}
The integrals over $u^2/(u-M^2)$ are $(-1)^J$ times those over $t^2/(t-M^2)$.

Explicit expressions for these integrals are:
\begin{subequations}
  \begin{align}
    \mathcal{S}_0 (s) &=\;
    M^2 - \frac{s}{2} + \frac{M^4}{s} \logms  
    \\
    \mathcal{S}_1 (s) &=\;
    2 \frac{M^4}{s} + \frac{s}{6} + \frac{M^4}{s^2} (2 M^2 + s) \logms    
    \\
    \mathcal{S}_2(s) &=\; 
    \frac{M^4}{s^2} \left( 6 M^2 + 3s \right) + 
    \frac{M^4}{s^3} \left( 6 M^4 + 6 M^2 s + s^2 \right) \logms
    \\
    \mathcal{S}_3(s) &=\; 
    \frac{M^4}{3 s^3} \left( 60 M^4 + 60 M^2 s + 11s^2 \right) + 
    \frac{M^4}{s^4} (2 M^2 + s) 
    \left( 10 M^4 + 10 M^2 s + s^2 \right) \logms 
    \\
    \mathcal{P}_0(s) &=\;
    1 + \frac{2 s + M^2}{s} \logms 
    \\
    \mathcal{P}_1(s) &=\;   
    \frac{M^2 + 2s}{s^2} \left( 2s + (2M^2 + s) \logms \right)
    \\
    \mathcal{D}_0(s) &=\; 
    \frac12 (2M^2+11s) + \frac{1}{s} (M^4 + 6M^2s + 6 s^2) \logms
    \\
    \mathcal{D}_1(s) &=\;
    \frac{1}{6s^2}\left\{ s (12 M^4 + 72 M^2 s + 73 s^2) + 6 (2 M^2
    +s) 
    (M^4 + 6M^2s + 6 s^2) \logms \right\} 
  \end{align}
\end{subequations}

%%%%%%%%%%%%%%%%%%%%
\section{Feynman rules for scalar and tensor resonances}
\label{app:feynrules}

We briefly summarize the Feynman rules for scalar and tensor
resonances that derive from the interaction
Lagrangians~(\ref{L-sigma})-(\ref{L-a}). The $k$s in this section are
the momenta of the Goldstone bosons. 

Scalar isoscalar:
\begin{equation}
  \label{eq:feyn_sigma}
  \sigma w^+ w^-: \; - \frac{2 i g}{v} (k_+ \cdot k_-) \qquad\qquad
  \sigma z z: \; - \frac{2 i g}{v} (k_1 \cdot k_2) \qquad\qquad
\end{equation}

Scalar isotensor:
\begin{align}
  \label{eq:feyn_phi}
  \phi^{\pm\pm} w^\mp w^\mp: &\; - \frac{\sqrt{2} i g}{v} (k_1
  \cdot k_2) \qquad\qquad 
  & \phi^\pm w^\mp z: &\; - \frac{i g}{v} (k_\mp \cdot k_z)
   \\
  \phi^0 z z: &\; - \frac{2 i g}{\sqrt{3} v} (k_1 \cdot k_2)
  \qquad\qquad
  & \phi^0 w^+ w^-: &\; + \frac{i g}{\sqrt{3} v} (k_+ \cdot k_-)
\end{align}

For the Feynman rules of the tensor resonances we use the symbol
$C_{\mu\nu,\rho\sigma} := g_{\mu\rho} g_{\nu\sigma} +
g_{\mu\sigma}g_{\nu\rho} - \frac12 g_{\mu\nu} g_{\rho\sigma}$ to get
(momenta incoming).

Tensor isoscalar:
\begin{equation}
  \label{eq:feyn_f}
  f^{\mu\nu} w^+ w^- : - \frac{i g_f}{v} C_{\mu\nu,\rho\sigma} k_+^\rho
  k_-^\sigma 
  \qquad\qquad
  f^{\mu\nu} z z: - \frac{i g_f}{v} C_{\mu\nu,\rho\sigma} k_1^\rho
  k_2^\sigma 
\end{equation}

Tensor isotensor:
\begin{align}
  \label{eq:feyn_a}
  a^{\pm\pm}_{\mu\nu} w^\mp w^\mp: &\; - \frac{i g_a}{\sqrt{2} v} 
  C_{\mu\nu,\rho\sigma} k_1^\rho k_2^\sigma
  %%% (k_{1,\mu} k_{2,\nu} + k_{1,\nu} k_{2,\mu}) 
  \qquad\quad 
  & a^\pm_{\mu\nu} w^\mp z: &\; - \frac{i g_a}{2 v} 
  C_{\mu\nu,\rho\sigma} k_\mp^\rho k_z^\sigma
  %%% (k_{\mp,\mu} k_{z,\nu} + k_{z,\mu} k_{\mp,\nu})
   \\
  a^0_{\mu\nu} z z: &\; - \frac{i g_a}{\sqrt{3} v} 
  C_{\mu\nu,\rho\sigma} k_1^\rho k_2^\sigma
  %%% (k_{1,\mu} k_{2,\nu} + k_{1,\nu} k_{2,\mu})
  \qquad\quad
  & a^0_{\mu\nu} w^+ w^-: &\; + \frac{i g_a}{2 \sqrt{3} v} 
  C_{\mu\nu,\rho\sigma} k_+^\rho k_-^\sigma
  %%% (k_{+,\mu} k_{-,\nu} + k_{+,\nu} k_{-,\mu})
\end{align}

Note that taking the conditions on the tracelessness as well as the
transversality not necessarily demands the coupling of the tensor
resonance to a conserved current (like the energy-momentum tensor)
which leads to the same Feynman rules as in~\cite{Han:1998sg}. The
constraint of the LET on the other hand results in an (off-shell
continued) amplitude that is identical to the one of a massive
graviton resonance.

%%%%%%%%%%%%%%%%%%%%

\section{Vector Resonance Exchange}
\label{app:rho}

Heavy vector resonances have been studied many times in the
literature, and various different formalisms describe their
interactions with the SM particles.  In this section, we demonstrate
the equivalence of some popular approaches.  In particular, we look at
the correction to the amplitude $A(s,t,u)$ for Goldstone-Goldstone
scattering which via the GBET and spin/isospin symmetry yields the
leading term for all channels of quasi-elastic $WW$ scattering, $w^+
w^- \to z z$.  Since we maintain manifest $SU(2)_L\times U(1)_Y$ gauge
invariance by using covariant derivatives, the GBET holds in any
formalism that we describe.  If desired, this can be verified by
switching to unitarity gauge and computing the $W/Z$ scattering
amplitudes directly.

We ignore couplings to fermion currents, which generically will be
present and sizable.  While such couplings get shifted by some of the
transformations described below, in our model-independent approach
they are considered as independent parameters which are determined by
independent measurements.  The shifts of fermion couplings induced by
reparameterizations merely change a set of undetermined parameters
into another set of undetermined parameters, so for our purposes there
is no need to calculate them.  However, in the context of a specific
model, one should always treat fermionic and bosonic sectors together
when applying reparameterizations~\cite{Nyffeler:1999ap}. A specific
example for this in the context of Little Higgs models can be found
in~\cite{Kilian:2003xt}.

\begin{enumerate}
\item
  We use the representation with the Goldstone field
  $\Sigma=\exp(-i\vw/v)$ and introduce the $\rho$ resonance as a
  vector field in the iso-triplet representation.  The Goldstone
  kinetic term is
  \begin{equation}
    \LL_{\text{kin}} = \frac{v^2}{4}\tr{(D_\mu\Sigma)^\dagger(D^\mu\Sigma)}
  \end{equation}
  and can be expanded up to second order as
  \begin{align}
    \LL_{\text{kin}}^{(0)} 
    &= 
    \pd w^+\pd w^- + \frac12\pd z\,\pd z,
    \\
    \LL_{\text{kin}}^{(2)}
    &=
    - \frac{1}{2v^2}  \left(w^+\pd w^- + w^-\pd w^+ + z\,\pd
    z\right)^2 .
  \end{align}
  The interaction Lagrangian is
  \begin{equation}
    \LL_{\text{int}} = \frac{ig_\rho v^2}{2}\tr{\vrho^\mu\vV_\mu}
  \end{equation}
  Ignoring gauge fields, we carry out the expansion up to third order:
  \begin{align}
    \LL_{\text{int}}^{(1)}
    &=
    -g_\rho v\left(\rho^+\pd w^- + \rho^-\pd w^+ + \rho^0\pd z\right),
    \\
    \LL_{\text{int}}^{(2)}
    &=
    ig_\rho\left[\rho^+\left(w^-\pd z - z\,\pd w^-\right)
               - \rho^-\left(w^+\pd z - z\,\pd w^+\right)
               + \rho^0\left(w^+\pd w^- - w^-\pd w^+\right)\right],
    \\
    \LL_{\text{int}}^{(3)}
    &=
    - \frac{2g_\rho}{3v}\left[
      \rho^+\left(\left(w^+w^- + z^2\right)\pd w^- 
                  - w^{-2}\pd w^+ - w^- z\,\pd z\right)
    \right. \nonumber\\
    &\quad
    \left.\qquad{}
    + \rho^-\left(\left(w^+w^- + z^2\right)\pd w^+
                  - w^{+2}\pd w^- - w^+ z\,\pd z\right)
    \right. \nonumber\\
    &\quad
    \left.\qquad{}
    + \rho^0\left(-w^- z\,\pd w^+ - w^+ z\pd w^- + 2w^+w^-\pd
    z\right)\right] \; ,
  \end{align}
  leading to the corresponding Feynman rules 
  \begin{align}
    \label{eq:feynvector}
    \rho^\pm_\mu w^\mp : &\; i g v k_{\mp,\mu} \qquad\qquad
    & \rho^0_\mu w^+ w^-: &\; i g 
    v k_{z,\mu} \\
    \rho^\pm_\mu w^\mp z: &\; \pm i g (k_z - k_\mp)_\mu \qquad\qquad
    & \rho^0_\mu w^+ w^-: &\; i g 
    (k_- - k_+)_\mu \\
    \rho^\pm_\mu w^\mp zz: &\; - i \frac{g}{3 v} (2 k_\mp - k_1 -
    k_2)_\mu \qquad\qquad 
    & \rho^0_\mu w^+ w^- z: &\; - i \frac{g}{3 v} (2 k_z - k_+ -
    k_-)_\mu
  \end{align}  
  %%%%%
  \begin{figure}
    \begin{center}
      \includegraphics{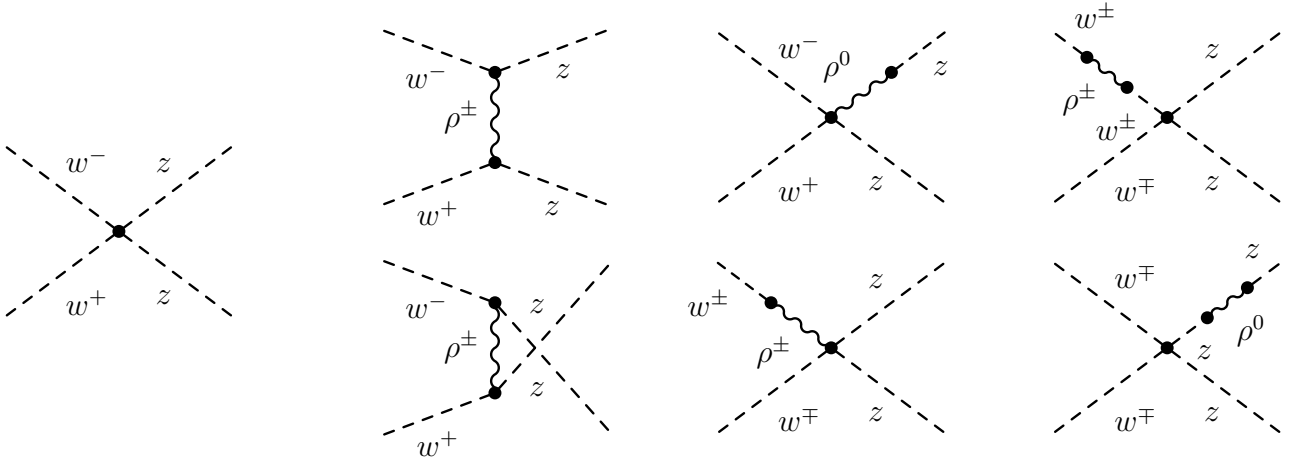}
    \end{center}
    \caption{
      \label{fig:rho_diag}
      Feynman graphs for  $w^+w^-\to zz$: contact term from the
      non-linear Lagrangian leading to the LET on the left, resonance
      $t$- and $u$-channel exchange in the middle, and contact terms
      from $\rho/w/z$ mixing on the right.}
  \end{figure}
  %%%%%
  The resulting Feynman graphs for $w^+w^-\to zz$ are shown in
  Fig.~\ref{fig:rho_diag}.  There is a Goldstone contact interaction from the
  Goldstone kinetic term; this yields the low-energy theorem (LET)
  $A(s,t,u)=s/v^2$.  Resonance exchange adds $t$- and $u$-channel
  contributions,
  \begin{equation}
    A_\rho'(s,t,u) = 
    - g_\rho^2\left(\frac{s-u}{t-M^2} + \frac{s-t}{u-M^2}\right),
  \end{equation}
  which in the limit $s\to 0$ become $A_\rho'\to -3g_\rho^2 s/M^2$.
  Furthermore, there are contributions where the resonance mixes with
  the external Goldstone scalar, either as an external wave-function
  correction, or with the $\rho$ coupling to three scalars.  In both
  graph types the resonance pole cancels, so they are additional
  contact terms proportional to $s/M^2$.  In unitarity gauge, this
  translates into a correction to $W/Z$ masses and interactions.

  The appearance of a contact term looks like a violation of the LET.
  However, the measured value of $v$ (e.g., as extracted from $W/Z$
  pole data) includes those additional contributions, so they merely
  renormalize a fictitious bare $v$ value.  This renormalization can
  be made explicit by adding a counterterm to the $\rho$ interaction
  Lagrangian, which by power counting and symmetries must be of the
  form $a\frac{gv^2}{M^2}\LL_{\text{kin}}$ with an appropriate
  prefactor $a$.  In effect, expressed in terms of the \emph{observed}
  scale $v$, the LET holds, and the vector-exchange amplitude is given
  by
  \begin{equation}
    \label{amp-rho-renorm}
    A_\rho(s,t,u) = 
    - g_\rho^2\left(\frac{s-u}{t-M^2} + \frac{s-t}{u-M^2} 
                   + 3\frac{s}{M^2}\right),
  \end{equation}
  which vanishes as $s^2$ as $s\to 0$.

\item
  In the previous paragraph, the vector resonance was coupled to $W/Z$
  bosons by a mass mixing term, $\tr{\vV\vrho}$.  Alternatively, we
  could couple it by a kinetic mixing term,
  \begin{equation}
    \LL_{\text{int}} = -g_\rho\tr{\vW_{\mu\nu}\vrho^{\mu\nu}}
  \end{equation}
  where the resonance ``field strength'' is $\vrho^{\mu\nu} =
  D^\mu\vrho^\nu - D^\nu\vrho^\mu$ with the covariant derivative in
  the adjoint representation.  Partial integration gives
  \begin{equation}
    \LL_{\text{int}} = 2g_\rho\tr{(D^\mu\vW_{\mu\nu})\vrho^\nu}.
  \end{equation}
  Here, we can apply the $W$ field equation
  \begin{equation}
    D^\mu\vW_{\mu\nu} = i\frac{v^2}{4}\vV_\nu
  \end{equation}
  to obtain
  \begin{equation}
    \LL_{\text{int}} = \frac{ig_\rho v^2}{2}\tr{\vV_\mu\vrho^\mu}
  \end{equation}
  as before, so we get the same scattering amplitude.  Using the
  equations of motion is precisely an application of the UET.

\item
  In the CCWZ formalism~\cite{CCWZ}, the elementary building block is
  $\xi$ with $\xi\xi=\Sigma$.  From $\xi$, we can construct a vector
  and an axial vector field,
  \begin{equation}
    \cV_\mu = 
    \frac{i}{2}\left(\xi^\dagger D_\mu\xi + \xi D_\mu\xi^\dagger\right)
    \qquad\text{and}\qquad
    \cA_\mu =
    \frac{i}{2}\left(\xi^\dagger D_\mu\xi - \xi D_\mu\xi^\dagger\right).
  \end{equation}
  Under $SU(2)_C$, these transform like a gauge field and a matter
  field, respectively,
  \begin{equation}
    \cV \to U_C\cV U_C^\dagger - (D_\mu U_C)U_C^\dagger
    \qquad\text{and}\qquad
    \cA \to U_C\cA U_C^\dagger.
  \end{equation}
  $\cA$ is related to the vector current that we have used in our
  previous formulation: $\xi^\dagger V_\mu\xi = 2i\cA_\mu$.  We just
  have to redefine $\vrho_\mu \to \xi^\dagger\vrho_\mu\xi$ to obtain
  \begin{equation}
    \LL_{\text{kin}} = -2v^2\tr{\cA_\mu\cA^\mu}
    \qquad\text{and}\qquad
    \LL_{\text{int}} = -g_\rho v^2\tr{\vrho_\mu\cA^\mu},
  \end{equation}
  so a matter field $\vrho$ coupled to the axial vector $\cA$ yields
  the same scattering amplitude again.
  
\item
  Alternatively, we can couple $\vrho$ to the vector field $\cV$ by
  assigning to it a gauge-field transformation law under $SU(2)_C$,
  \begin{equation}
    \vrho \to U_C\vrho U_C^\dagger 
    - i\frac{2g_\rho v}{M}(D_\mu U_C)U_C^\dagger,
  \end{equation}
  so the leading invariant term containing $\vrho$ is
  \begin{align}
    \LL_{\text{int}} 
    &= 
    - g_\rho^2v^2\tr{\left(\cV + i\frac{M}{2g_\rho v}\vrho\right)^2}
%    \nonumber\\ &=
    =
    - g_\rho^2v^2\tr{\cV\cV}
    - i g_\rho v M \tr{\cV\vrho}
    %% + i \frac{g_\rho v M}{2}\tr{\cV\vrho}
    + \frac{M^2}{4}\tr{\vrho\vrho}
  \end{align}
  The expansion of $\cV$ is even in
  the number of Goldstone fields.  Therefore, in this expression, the
  last term is the $\rho$ mass, the second term yields the $\rho^0
  w^+w^-$ and $\rho^\pm w^\mp z$ couplings, and the first term is a
  contact term.  Note that the gauge coupling is proportional to
  $1/g_\rho$.  The resulting $w^+w^-\to zz$ amplitude is again
  (\ref{amp-rho-renorm}), this time without the need for
  renormalizing~$v$.

\item
  The BESS model~\cite{BESS} has a similar setup.  Instead of gauging
  just $SU(2)_C$, we can extend the local symmetry by an extra local,
  nonlinearly realized $SU(2)_L\times SU(2)_R$.  This results in two
  vector isotriplets, which can be combined to a vector and an axial
  vector isotriplet, respectively.  Only the vector couples to
  longitudinal $W/Z$ pairs, and the amplitude~(\ref{amp-rho-renorm})
  can be derived as before.

\end{enumerate}
The different formalisms for coupling vector resonances all result in
the same scattering amplitude.  This is not surprising since this
amplitude is completely determined by spin and isospin conservation
together with the LET.  In order to give the CCWZ interpretation of the
vector resonance as a gauge field (in contrast to a generic matter
field) a physical meaning, we would have to measure triple $\rho$
couplings, analogous to the LEP2 measurements of triple gauge
couplings.  Unfortunately, such measurements are beyond the reach of
LHC.

%%%%%%%%%%%%%%%%%%%%

\section{Specific Models}
\label{app:models}

In the literature, a variety of ``benchmark'' models has been
formulated that test weak-boson scattering.  In this section, we
relate some of them to our parameterization:
\begin{enumerate}
\item \textbf{The SM.}
  As discussed in the main text, for $g_\sigma=1$ the scalar resonance
  model precisely reproduces SM Higgs exchange.  Alternatively, one
  can switch to the default SM implementation (in \whizard) without
  extra resonances.
\item \textbf{Scalar resonances.}
  Refs.~\cite{Bagger:1993zf,Bagger:1995mk} introduce a collection of
  models, among them two with a scalar resonance (``$O(2N)$'' and
  ``chirally coupled scalar'').  The latter model is identical to our
  scalar resonance parameterization.  The $O(2N)$ model is essentially
  a special case of this with fixed mass and width; the only
  distinction is a logarithmic cutoff-dependent modification, which
  manifests itself beyond the resonance.  This detail is unlikely to
  be detectable at the LHC.
\item \textbf{Vector resonances.}  
  The chirally-coupled vector resonance model
  of~\cite{Bagger:1993zf,Bagger:1995mk} is identical to ours (see the
  discussion of the CCWZ formalism in App.~\ref{app:rho}), where we
  identify $a = (2g_\rho v/M_\rho)^2$ and $g = M_\rho^2/(2v^2
  g_\rho)$.  An analogous identification holds for the BESS
  model~\cite{Casalbuoni:1997bv}, with $a$ replaced by $\beta$ in
  their notation.
\item \textbf{Pad\'e/IAM unitarization model.}

  As discussed in Sec.~\ref{sec:unitarization}, this scheme is a
  special case of the K-matrix scheme as defined in the present
  paper. For a given combination $(\alpha_4,\alpha_5)$ we use
  Eqs.~(\ref{a00-1}, \ref{a11-1}, \ref{a20-1}) to determine the NLO
  correction $A_{IJ}^{(1)}$ to the three amplitudes $A_{00}$,
  $A_{11}$, and $A_{20}$ which without correction would violate
  unitarity.  Then, we can use (\ref{IAM}) to identify scalar, vector,
  and tensor resonance masses and widths.  If we neglect the loop
  corrections in (\ref{a00-1}--\ref{a20-1}), we obtain
  \begin{subequations}
  \begin{align}
    M_\sigma^2 &= 
      \frac{3v^2}{4(7\alpha_4(\mu) + 11\alpha_5(\mu)}
    &
    \Gamma_\sigma &= \frac{M_\sigma}{16\pi}
    \\
    M_\rho^2 &= \frac{v^2}{4(\alpha_4(\mu) - 2\alpha_5(\mu))}
    &
    \Gamma_\rho &= \frac{M_\rho}{96\pi}
    \\
    M_f^2 &= -\frac{3v^2}{16(2\alpha_4(\mu) + \alpha_5(\mu))}
    &
    \Gamma_f &= -\frac{M_f}{32\pi}
  \end{align}
  \end{subequations}
  where we have to define a renormalization scheme and fix the scale
  $\mu$.  Note that the tensor-resonance parameters are unphysical.
  This is due to the negative sign of $A_{20}^{(0)}$ in
  Eq.~(\ref{eigenamp-LET}).  This model ignores the possibility of
  isotensor resonances $\phi$ or $a$.
\end{enumerate}

%%%%%%%%%%%%%%%%%%%%

%\clearpage
\section{On-shell vector boson scattering}
\label{app:onshell-scattering}

\begin{figure}[p]
\begin{center}    
  \includegraphics[width=.48\textwidth]{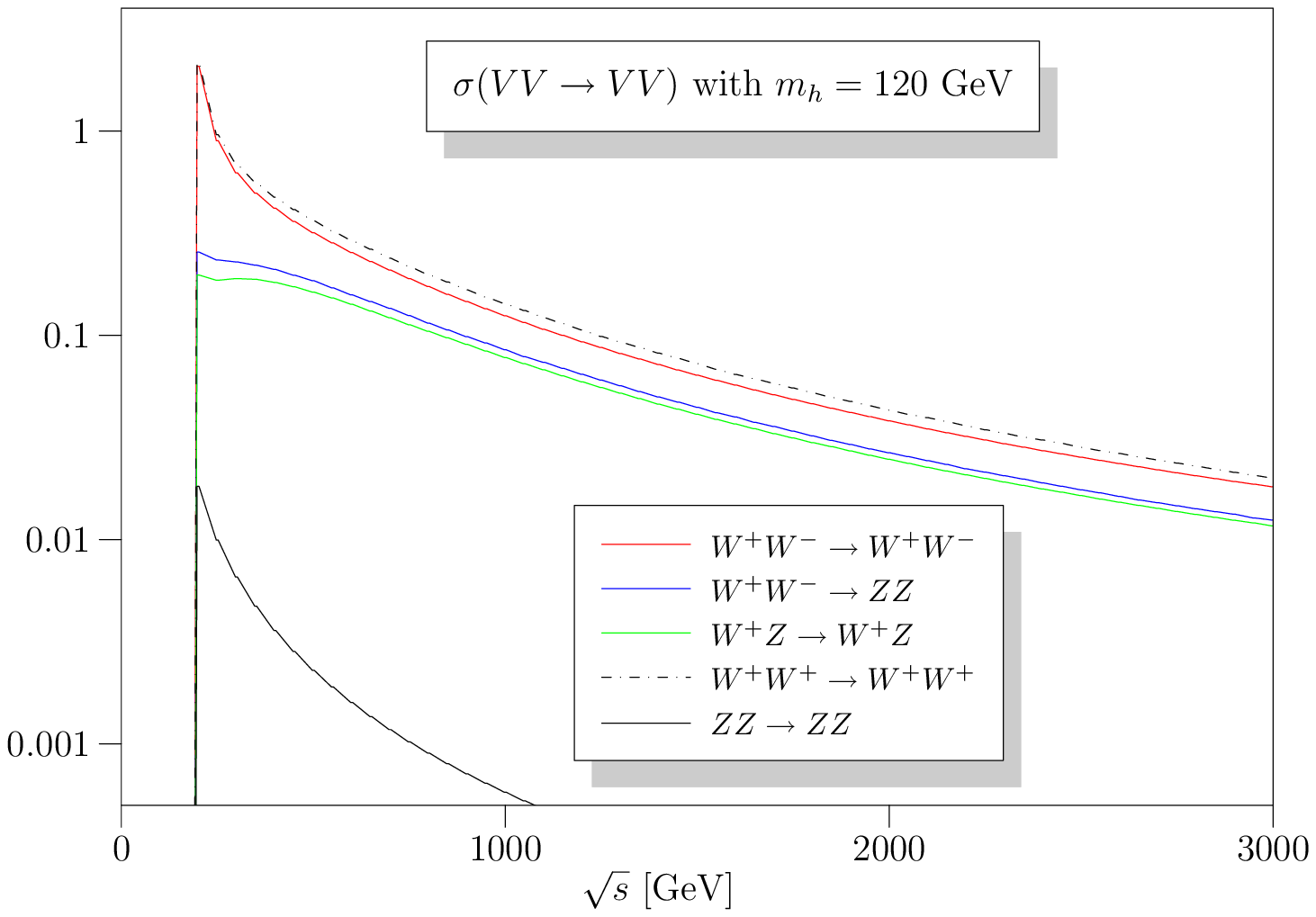}  \quad
  \includegraphics[width=.48\textwidth]{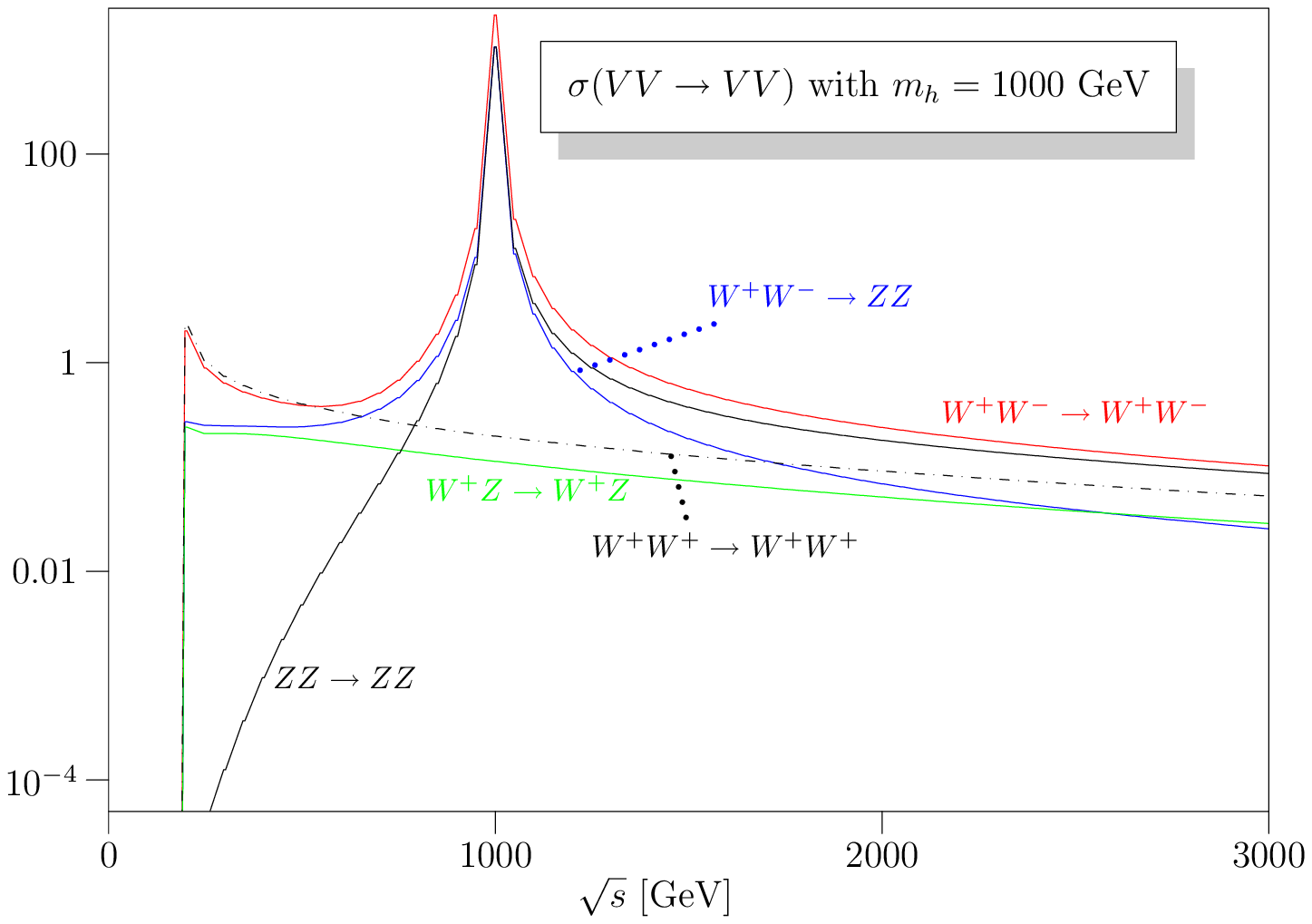}
  \vspace{2mm}
  \newline
  \vspace{2mm}
  \includegraphics[width=.48\textwidth]{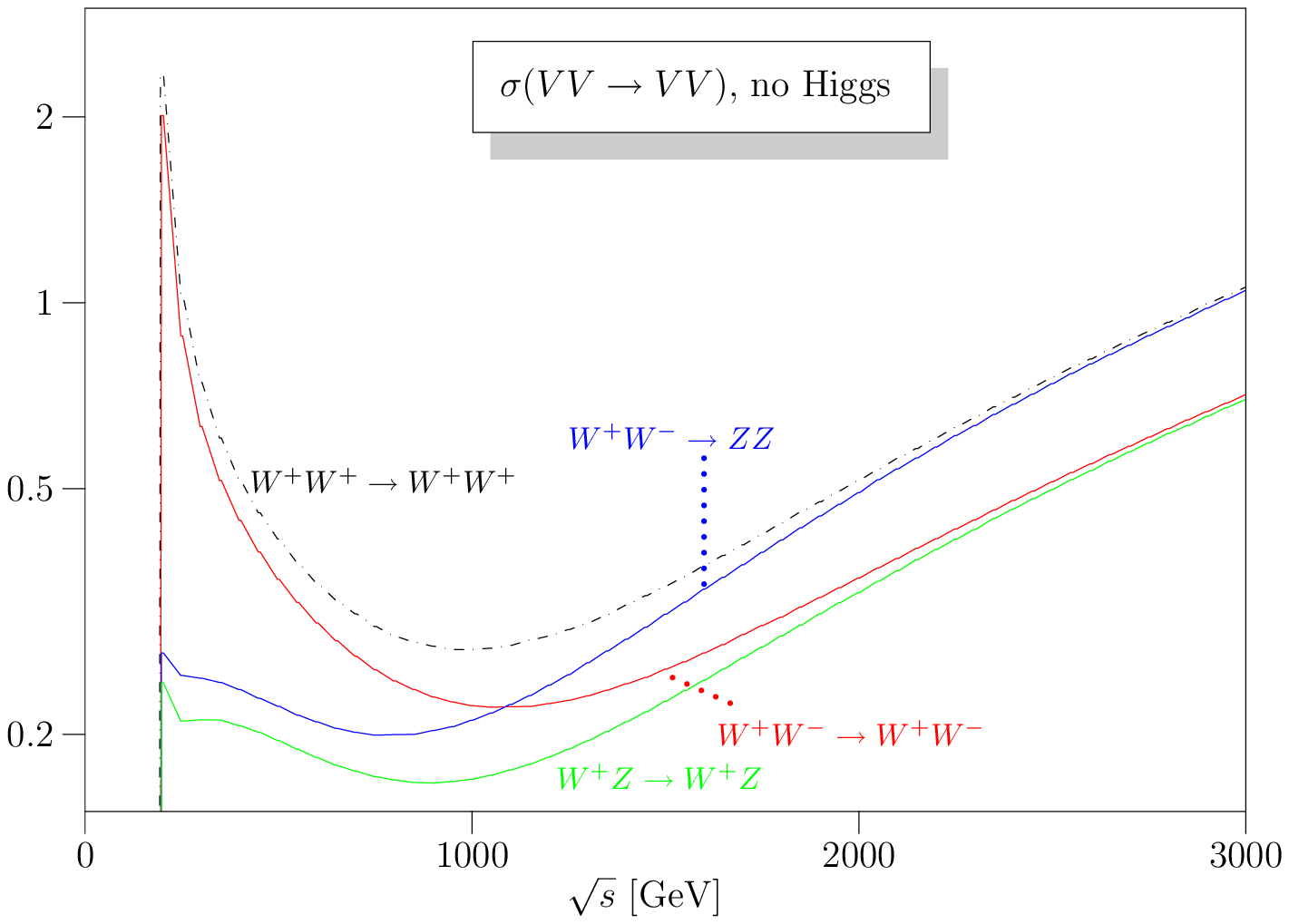} \quad
  \includegraphics[width=.48\textwidth]{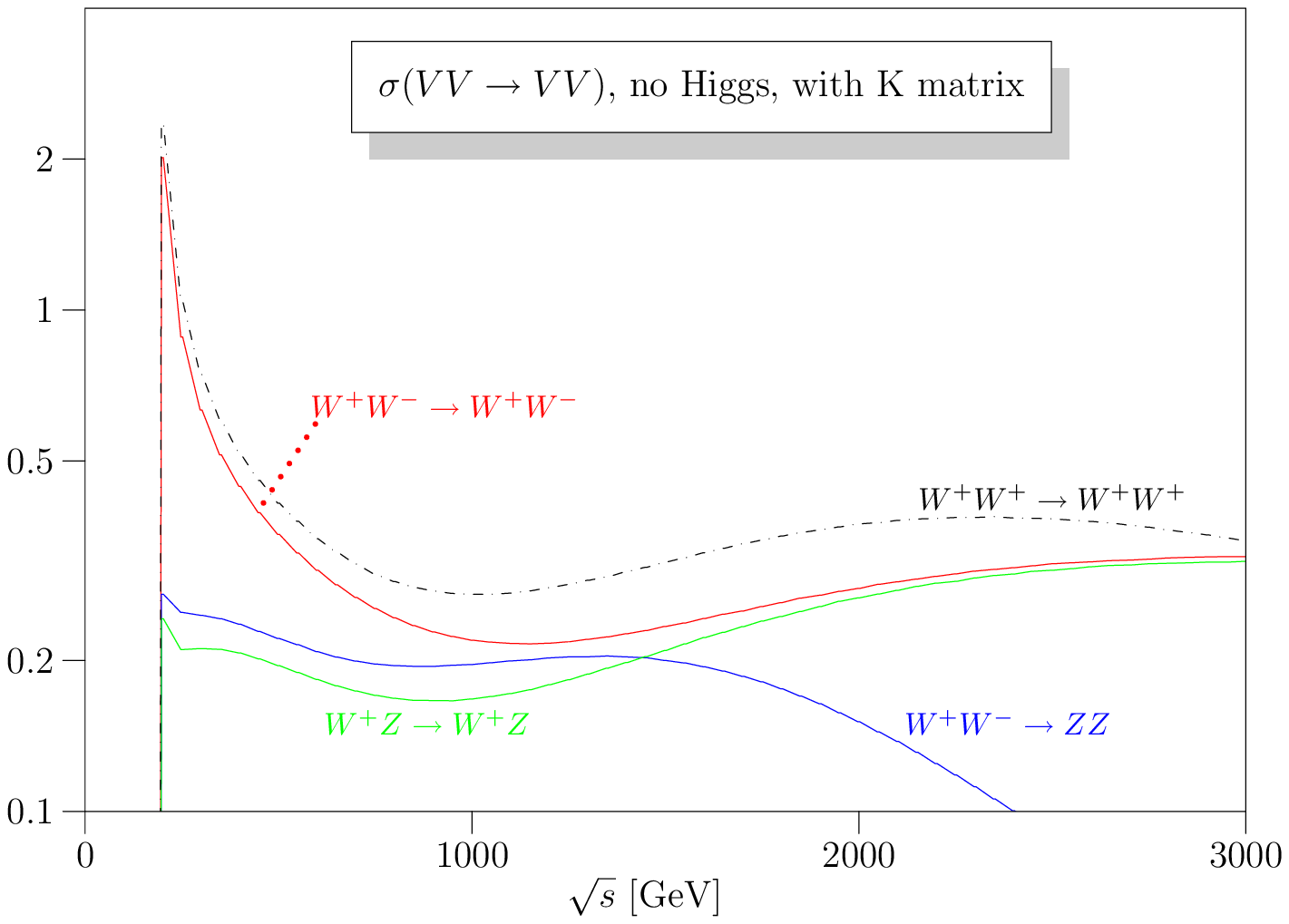} 
  \vspace{2mm}
  \newline
  \vspace{2mm}
  \includegraphics[width=.48\textwidth]{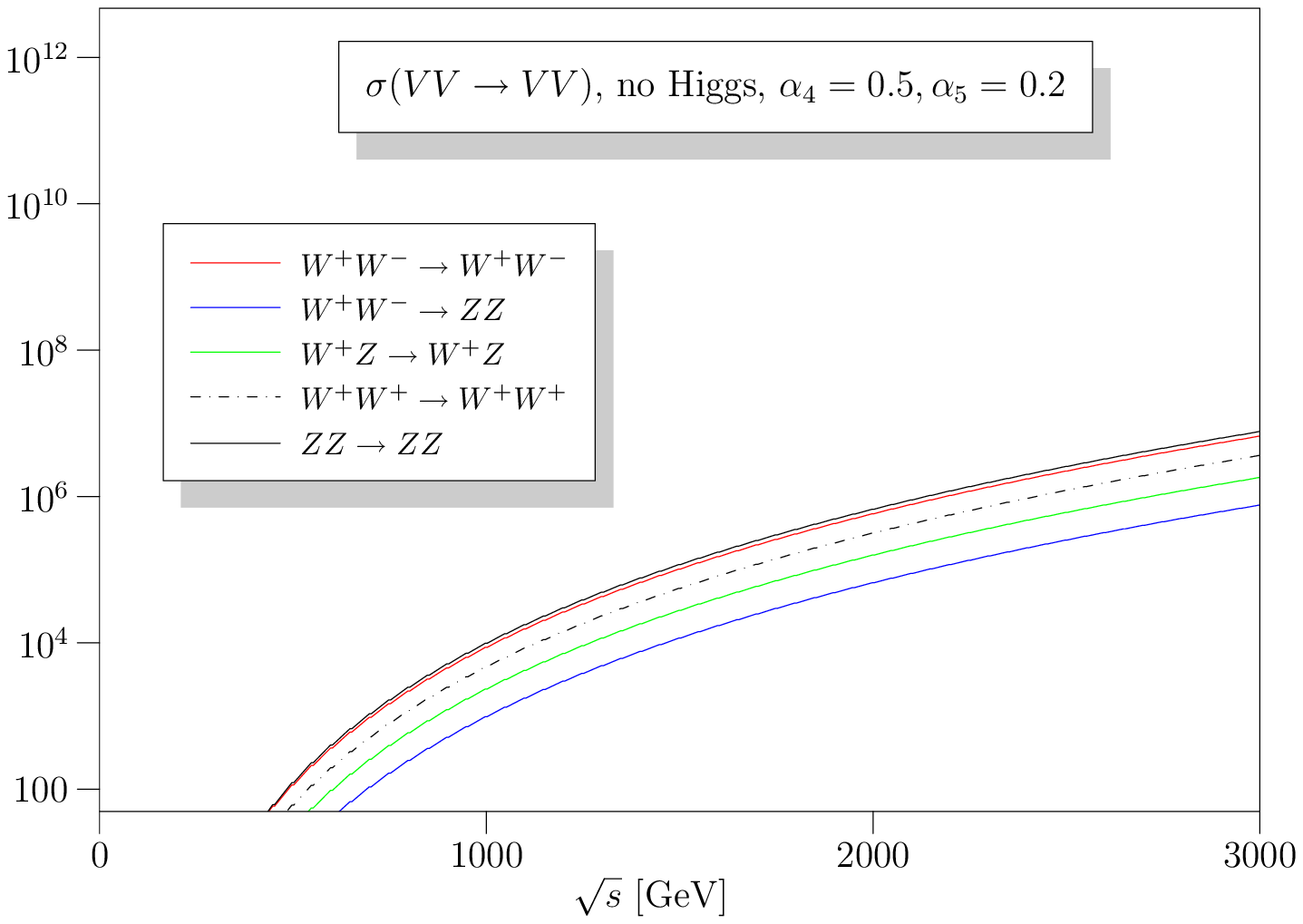} \quad
  \includegraphics[width=.48\textwidth]{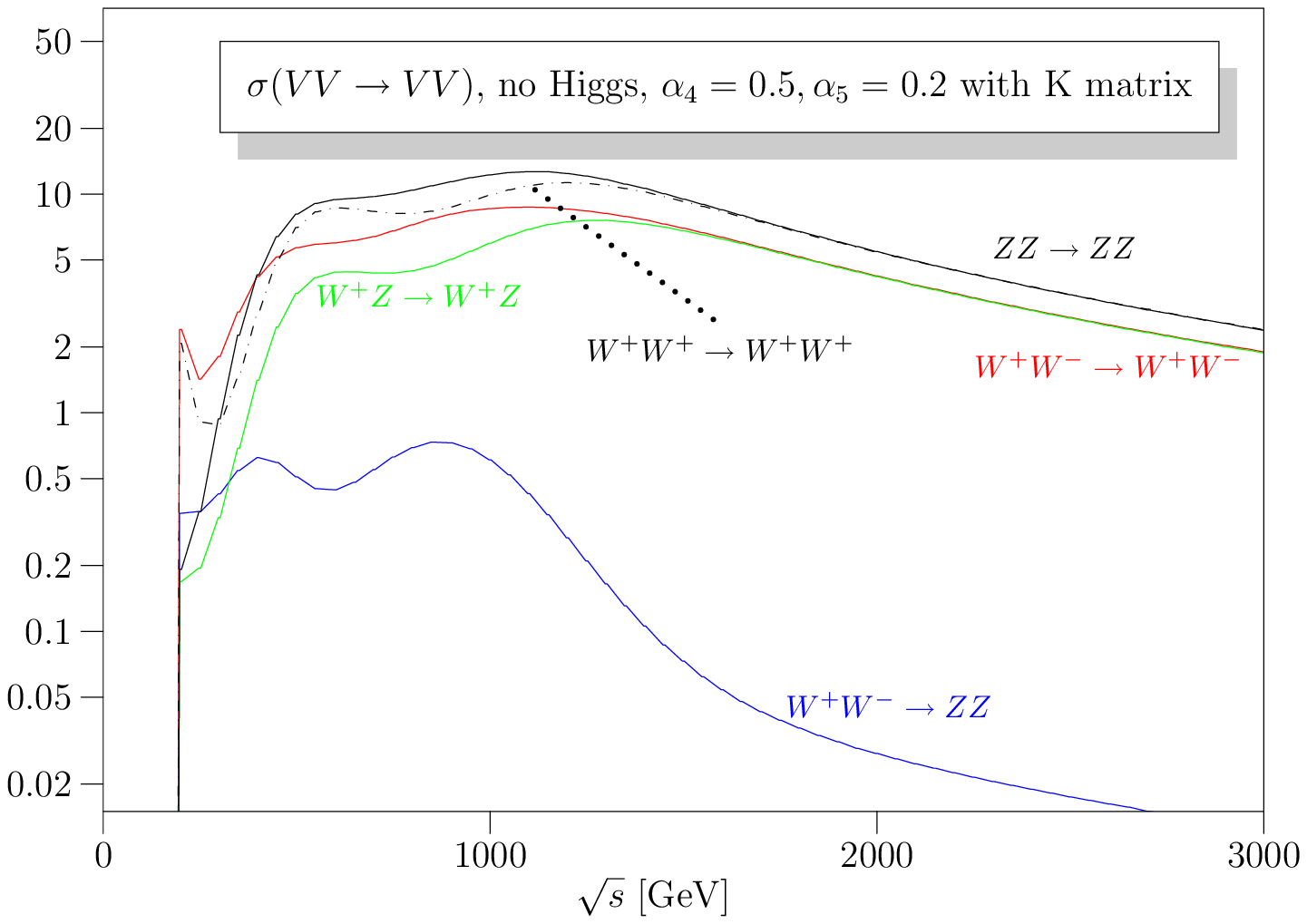} 
  \caption{
    Cross sections (in nanobarns) for the five different scattering
    processes of longitudinal weak gauge bosons: SM with a 120 GeV and
    a 1 TeV Higgs in the upper line, in the middle: SM without a Higgs
    without and with K-matrix unitarization, respectively. In the
    lower line, the case of $\alpha_{4,5}$ switched on are shown, on
    the left without, on the right with K matrix unitarization. The
    contribution from the forward region is cut out by a 15 degree cut
    around the beam axis.
    \label{fig:sm_cases}}    
\end{center}
  
\end{figure}

In this section we summarize the plots for ``partonic'' scattering of
spin-averaged and summed vector bosons.  In all these pictures, the EW
gauge bosons are treated on-shell, hence the cross sections start when
the physical $WW$ or $ZZ$ threshold is reached. Since we did not
switch off the electromagnetic coupling in those plots, we applied a
cut of 15 degrees around the beam axis to cut out the Coulomb
scattering part. Fig.~\ref{fig:sm_cases} shows in the 
upper line the SM with a 120 GeV Higgs on the left and a heavy 1 TeV
on the right. Unitarity is preserved in those cases because of the
($s$-channel) Higgs exchange. Besides the dominant resonance for a
heavy Higgs, the amplitudes show a saturation for the high-energy
tails which starts again violating partial-wave unitarity for 1.2,
3.5, and 1.7 TeV for the $I=0,1,2$ isospin channels,
respectively. Completely removing the Higgs as in the middle line
of~\ref{fig:sm_cases} leads to a rise of the amplitudes (and hence the
cross sections) with $s$ (the $zz \to zz$ process is absent in that
case). Switching on the $K$-matrix unitarization damps the amplitudes
back to the Argand circle, thereby restoring unitarity. This happens
for the above mentioned values for the corresponding isospin
eigenamplitudes. In the lower line of Fig.~\ref{fig:sm_cases}, the
case of the LET extended by nonzero values for the parameters
$\alpha_4$ and $\alpha_5$ are shown, on the left the badly diverging
case without unitarization, and the K-matrix unitarized case on the
right.

\begin{figure}[p]
\begin{center}
  \includegraphics[width=.48\textwidth]{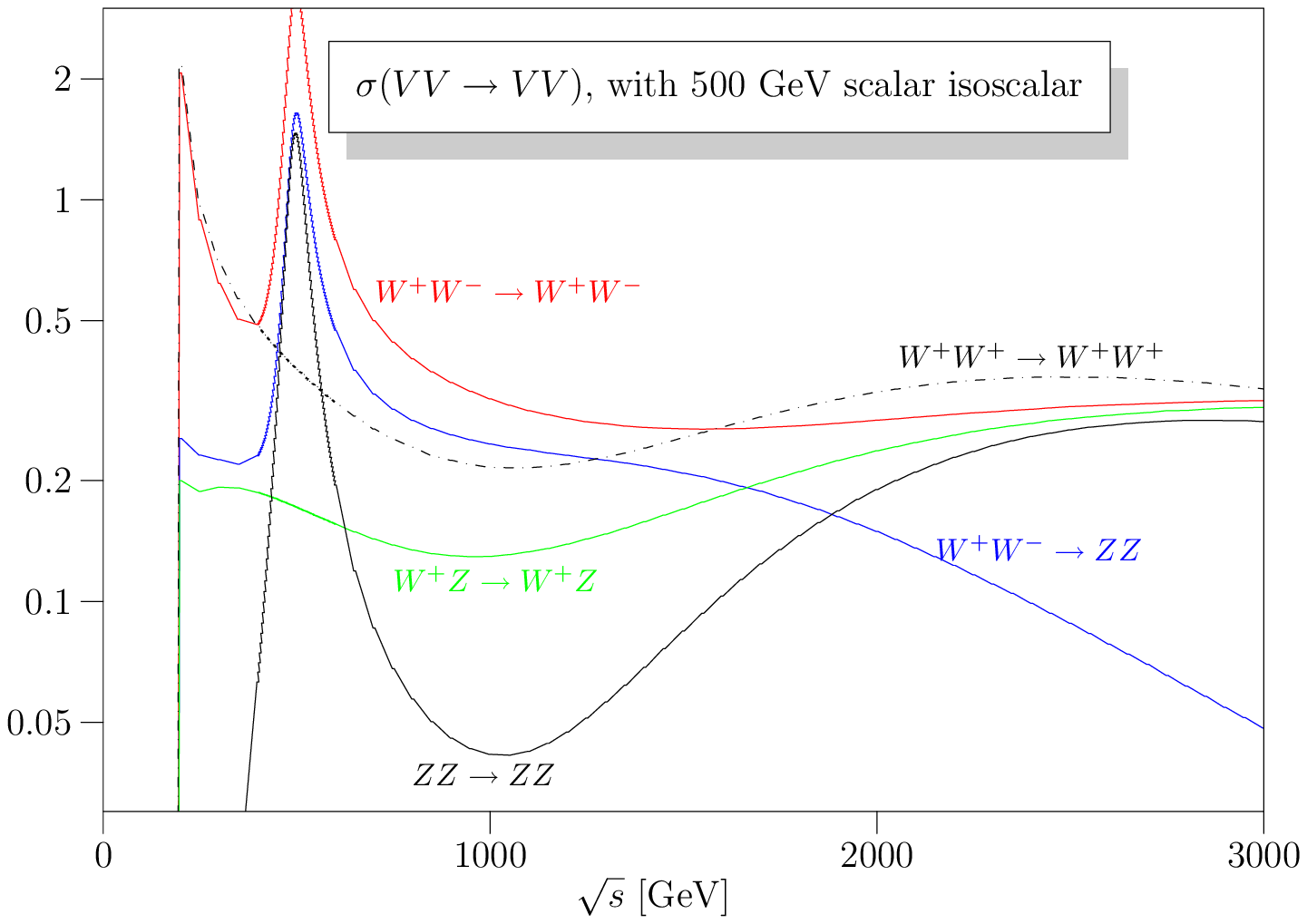} \quad
  \includegraphics[width=.48\textwidth]{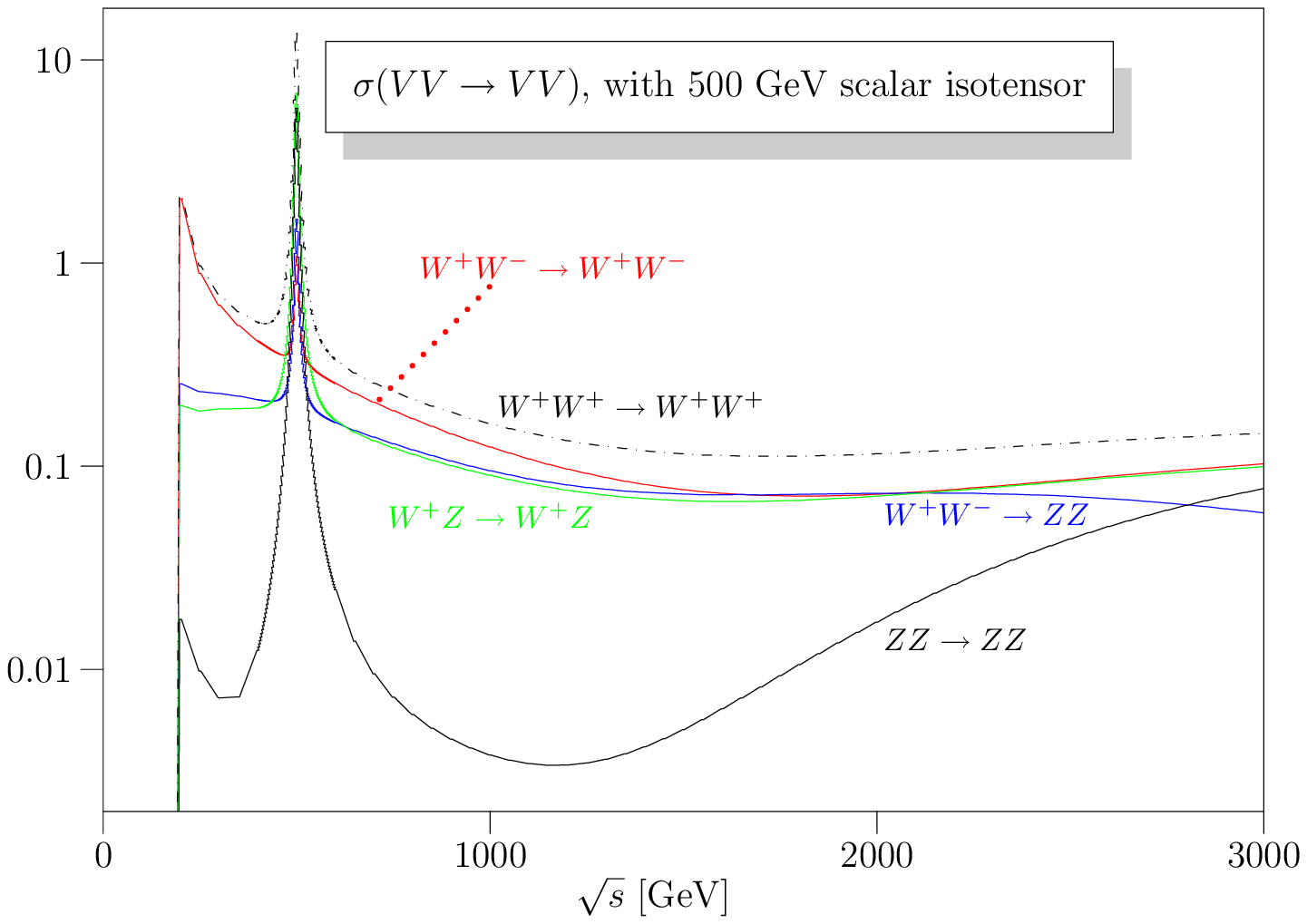} 
  \vspace{2mm}
  \newline
  \vspace{2mm}
  \includegraphics[width=.48\textwidth]{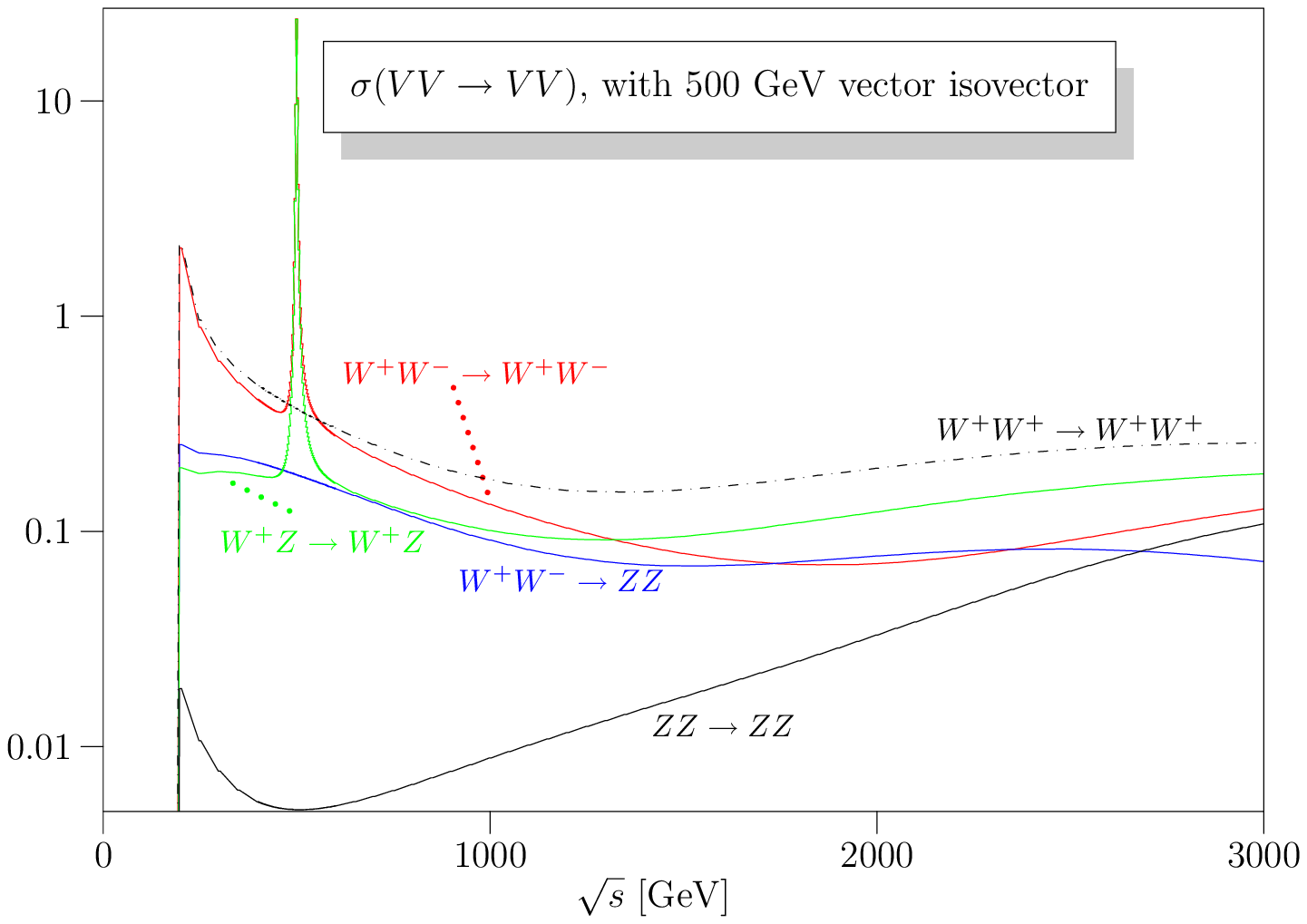}
  \vspace{2mm}
  \newline
  \vspace{2mm}
  \includegraphics[width=.48\textwidth]{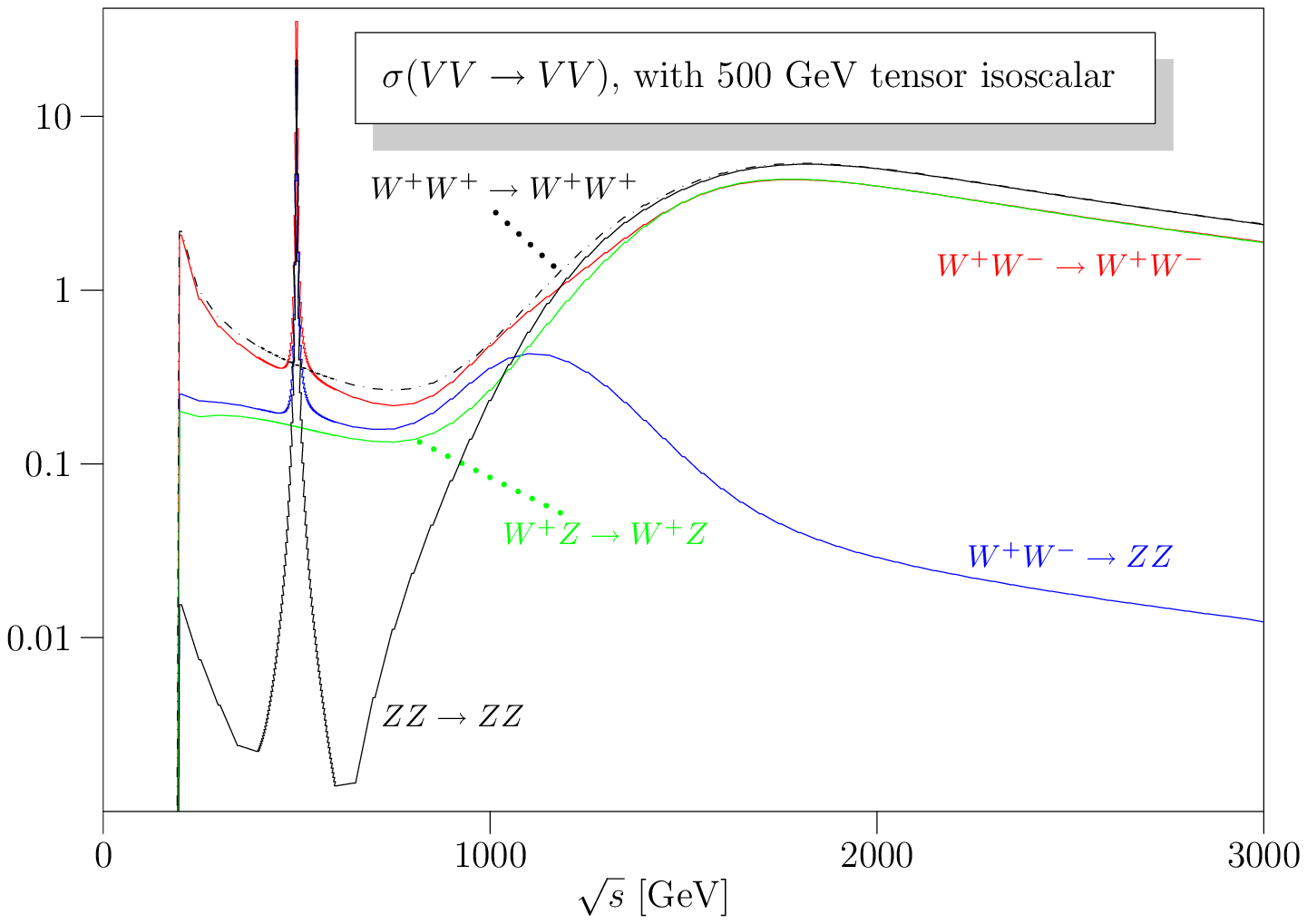} \quad
  \includegraphics[width=.48\textwidth]{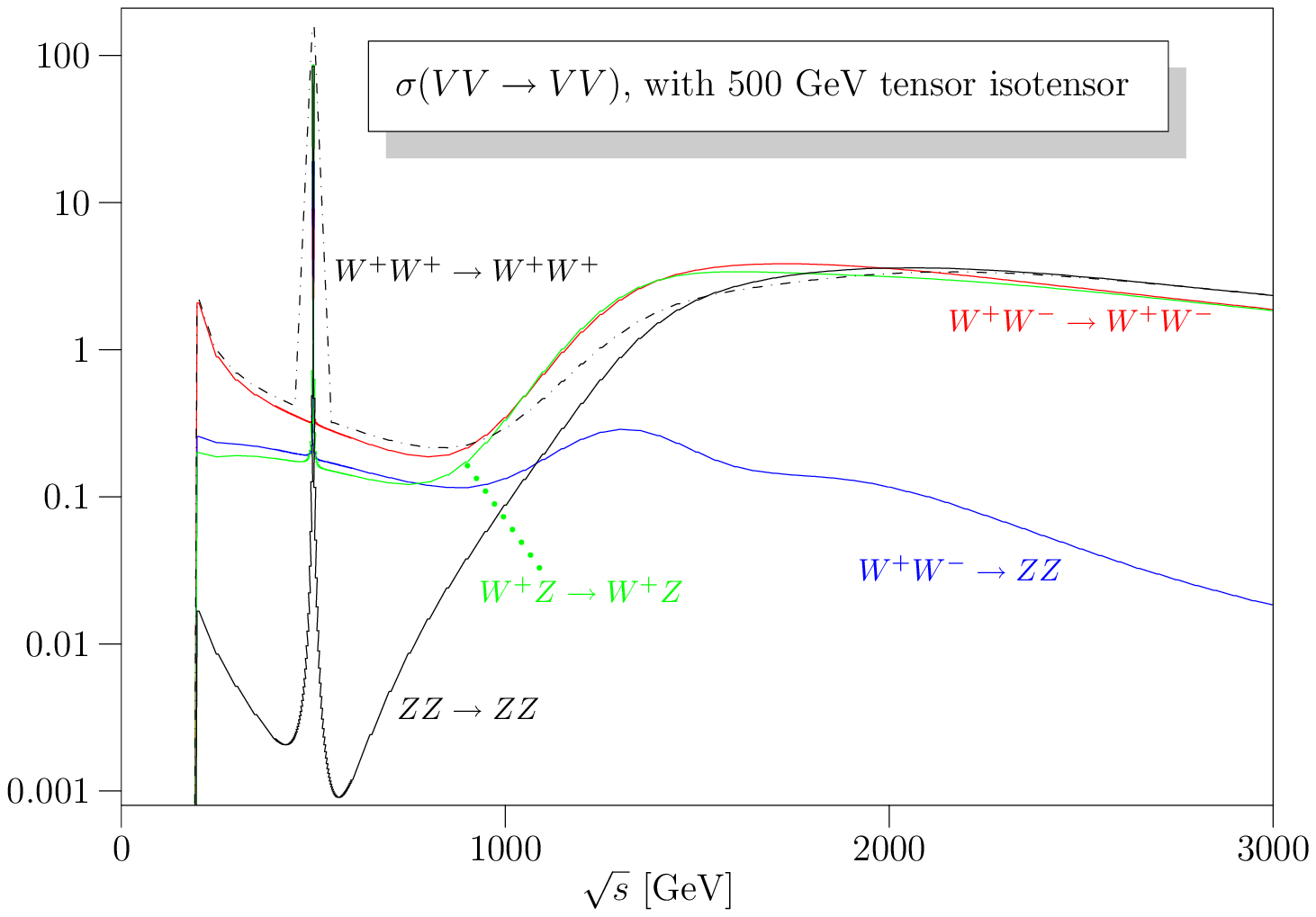} 
\end{center}
  \caption{
    Cross sections for $VV\to VV$ scattering in nanobarns, with
    the presence of resonances: scalars (isoscalar $\sigma$ and isotensor
    $\phi$) in the upper line, vector isovector $\rho$ in the middle,
    and tensors (isoscalar $f$ and isotensor $a$) in the lower
    line, respectively. All amplitudes have been unitarized according
    to the K-matrix method. The resonance mass is set to 500 GeV in
    each case. Again, the contribution from the forward region is cut
    out by a 15 degree cut around the beam axis.
    \label{fig:amp_resonance}}
\end{figure}

In Fig.~\ref{fig:amp_resonance} we show the cross sections for the
five different vector boson scattering processes with the presence of
the five isospin-allowed resonances mentioned in the text. $ZZ \to ZZ$
and $WW \to ZZ$ contain all three isospin eigenamplitudes, hence show
a resonance in all channels except for the vector resonance case where
it is forbidden by the Yang-Landau theorem. The amplitude $WZ \to WZ$ does
not have isoscalar resonances, while the amplitude $W^+W^+ \to W^+W^+$
allows only isotensor resonances. Finally, $W^+ W^- \to W^+ W^-$
contains all resonances. 

%%%%%%%%%%%%%%%%%%%%%%%%%%%%%%%%%%%%%%%%%%%%%%%%%%%%%%%%%%%%%%%%%%%%%%%%
%%% References
%%%%%%%%%%%%%%%%%%%%%%%%%%%%%%%%%%%%%%%%%%%%%%%%%%%%%%%%%%%%%%%%%%%%%%%%

%%% \end{fmffile}
\end{document}